\documentclass[12pt]{article}

\usepackage{color}
\definecolor{darkblue}{rgb}{0.1,0.1,.7}
\usepackage[colorlinks, linkcolor=darkblue, citecolor=darkblue, urlcolor=darkblue, linktocpage]{hyperref}
\usepackage[]{amsmath}
\usepackage[]{graphicx}
\usepackage[]{latexsym}
\usepackage{slashed,graphicx,color,amsmath,amssymb}
\usepackage[mathscr]{eucal}
\usepackage{mathrsfs}
\usepackage{geometry}
\usepackage[margin=10pt,font=small,labelfont=bf]{caption}
\usepackage{amscd}
\usepackage{bm}
\usepackage{xcolor}
\usepackage{upgreek}
\usepackage[square, comma, sort&compress,numbers]{natbib}
\usepackage[all,cmtip]{xy}
\numberwithin{equation}{section}
\newcommand{\tr}{\mathrm{Tr}\,}

\newcommand{\ra}{\mathrm{a}}
\newcommand{\rb}{\mathrm{b}}
\newcommand{\rc}{\mathrm{c}}

\newcommand{\rs}{\mathrm{s}}
\newcommand{\rt}{\mathrm{t}}

\newcommand{\rz}{\mathrm{z}}

\newcommand{\rA}{\mathrm{A}}
\newcommand{\rB}{\mathrm{B}}
\newcommand{\rC}{\mathrm{C}}
\newcommand{\rD}{\mathrm{D}}

\newcommand{\rG}{\mathrm{G}}

\newcommand{\rI}{\mathrm{I}}

\newcommand{\rN}{\mathrm{N}}

\newcommand{\rP}{\mathrm{P}}
\newcommand{\rQ}{\mathrm{Q}}

\begin{document}
\begin{flushright}
\begin{tabular}{r}
MPP-2018-291
\end{tabular}
\end{flushright}
%\vspace{-14.1mm}
%\vspace*{-.6in}

\thispagestyle{empty}
\vspace{.2in} {\Large
\begin{center}
{\bf Torus partition function of the six-vertex model\\ from algebraic geometry}
\end{center}}
\vspace{.2in}
\begin{center}
Jesper Lykke Jacobsen$^{a,b,c}$, Yunfeng Jiang$^{d,e}$,  Yang Zhang$^{d,f,g}$
\\
\vspace{.3in}
\small{
$^a$ \textit{Laboratoire de Physique Th\'eorique, D\'epartement de Physique de l'ENS,\\
\'Ecole Normale Sup\'erieure, Sorbonne Universit\'e, CNRS,\\ PSL Research University, 75005 Paris, France}\\\vspace{.3cm}
$^b$ \textit{Sorbonne Universit\'e, \'Ecole Normale Sup\'erieure, CNRS,\\
Laboratoire de Physique Th\'eorique (LPT ENS), 75005 Paris, France}\\\vspace{.3cm}
$^c$ \textit{Institut de Physique Th\'eorique, Paris Saclay, CEA, CNRS, 91191 Gif-sur-Yvette, France}\\\vspace{.3cm}
$^d$ \textit{Institut f{\"u}r Theoretische Physik,
ETH Z{\"u}rich}},\\
\small{\textit{
Wolfgang Pauli Strasse 27,
CH-8093 Z{\"u}rich, Switzerland}\\\vspace{.3cm}
$^e$ \textit{CERN Theory Department, Geneva, Switzerland}\\\vspace{.3cm}
$^f$ \textit{Max Planck Institut f\"ur Physik Werner Heisenberg Institut\\ 80805 M\"unchen, Germany}
}\\\vspace{.3cm}
$^g$ \textit{PRISMA Cluster of Excellence, Johannes Gutenberg University,\\ 55128 Mainz, Germany}

\end{center}

\vspace{.3in}

\newpage
\begin{abstract}
We develop an efficient method to compute the torus partition function of the six-vertex model exactly for finite lattice size. The method is based on the algebro-geometric approach to the resolution of Bethe ansatz equations initiated in a previous work, and on further ingredients introduced in the present paper. The latter include rational $Q$-system, primary decomposition, algebraic extension and Galois theory. Using this approach, we probe new structures in the solution space of the Bethe ansatz equations which enable us to boost the efficiency of the computation. As an application, we study the zeros of the partition function in a partial thermodynamic limit of $M \times N$ tori with $N \gg M$. We observe that for $N \to \infty$ the zeros accumulate on some curves and give a numerical method to generate the curves of accumulation points.
\end{abstract}

\vskip 1cm \hspace{0.7cm}

\newpage

\setcounter{page}{1}
\begingroup
\hypersetup{linkcolor=black}
\tableofcontents
\endgroup

%%%%%%%%%%%%%%%%%%%%%%%%%%%%%%%%%%%%%%%%%%%%%%%%%%%%%%%%%%%%%%%%%%%%%%%%
\section{Introduction}
\label{sec:intro}
%%%%%%%%%%%%%%%%%%%%%%%%%%%%%%%%%%%%%%%%%%%%%%%%%%%%%%%%%%%%%%%%%%%%%%%%
Computing partition functions of two-dimensional lattice models is one of the central pro\-blems in statistical mechanics. When the lattice model is integrable, one can often compute the partition function exactly. However, even for integrable models, exact results are typically only available in two limits, \emph{i.e.}, when the lattice size is very small or in the thermodynamic limit where the lattice size is infinite. For the intermediate case, obtaining exact results for the partition function is actually a hard task.\par

For definiteness, we consider the torus partition function of the six-vertex model at its isotropic point, with lattice size $M\times N$. This is a well-known integrable model which is equivalent to the Heisenberg XXX spin chain \cite{Baxter:book}. The partition function can be computed by the transfer matrix method. Usual folklore of integrability tells us that we can diagonalize the transfer matrix by Bethe ansatz. The partition function can be written in terms of the eigenvalues of the transfer matrix. This works for any lattice size. However, the eigenvalues of the transfer matrix obtained in this way are only \emph{formal}, since they are written in terms of Bethe roots, which are solutions of Bethe ansatz equations (BAE). To compute the partition function explicitly, we need to actually solve the BAE and find \emph{all} physical solutions and then plug in the eigenvalues of the transfer matrix. What is usually not stressed in the literature is that it is in fact a highly non-trivial task to find all the solutions of the BAE, even numerically.
%For example, for the $SU(2)$ Heisenberg XXX spin chain, one can find all the solutions of BAE up to length 14 [NW] using advanced numerical %approach and clusters.\par

Morever, even if one finds all the solutions of BAE numerically, the result is not exact. Although numerical results are sufficient for many purposes, our goal here is to find exact results. In this work, we propose an efficient method to compute the partition function of integrable lattice model for finite-size system \emph{exactly} and \emph{analytically} without using any numerics. This method involves several recent developments in integrability, such as rational $Q$-systems and the algebro-geometric approach to Bethe ansatz equations. In this sense, the current work is a continuation of the work \cite{Jiang:2017phk} initiated by two of us.\par

The general goal of the program started in \cite{Jiang:2017phk} is to study the solution space of Bethe ansatz equations systematically by algebraic geometry and develop new analytical methods for physical applications in different contexts. In this paper, we extend the results of our previous work by introducing two new ingredients from algebraic geometry, namely \emph{primary decomposition} and \emph{algebraic extension}. Using these methods, we can probe structures in the solution space of the BAE. We find that the solution space can be decomposed naturally into non-intersecting subspaces upon primary decomposition on $\mathbb{Q}$. The physical interpretation of such decomposition is related to the decomposition of the transfer matrix with respect to the total momentum. This decomposition can be performed more thoroughly on an algebraically extended field. The structure we find is interesting on its own right. Furthermore, as an application, by using this decomposition and Galois theory, we gain a huge boost in the efficiency of the partition function computation.\par

Let us summarize the main results of this paper.
\begin{enumerate}
\item We perform the decomposition with respect to total momentum on an extended field $\mathbb{F}_M$ (defined in section~\ref{sec:primaryDec}). We denote each subspace by $\rI_{M,K,\ell}$ where $M$ is the length of the spin chain, or equivalently the size of one direction of the lattice; $K$ is the number of magnons and $\ell$ denotes the momentum sector with total momentum $2\pi \ell/M$. We compute the Gr\"obner basis and construct the quotient ring for each $\rI_{K,M,\ell}$. The dimensions of the quotient rings, or equivalently the number of physical solutions of the BAE are given in Tables~\ref{BAE_decomposition_M6}--\ref{BAE_decomposition_M18} for length $6\le M\le 18$.
\item We compute the companion matrices of the transfer matrix $\mathbf{T}_{M,K,\ell}(z)$ and Baxter's $Q$-polynomials $\mathbf{Q}_{M,K,\ell}(z)$ within each subspace $\rI_{M,K,\ell}$ up to $M=14$. Using this data, one can compute the exact torus partition function for any number of $N$. Of course, as $N$ grows, the results becomes quickly large (meaning here high-degree polynomials with huge rational coefficients). In practice, we can compute the partition function up to rather large $N\sim 100$ for all $M$ up to $M=14$. In addition, we can also diagonalize these matrices numerically and find Bethe roots and eigenvalues of the transfer matrix up to $M=14$ efficiently.
\item Our results for the partition function with $N \gg M$ are close to the partial thermodynamic limit of fixed $M$ and $N \to \infty$. Powerful techniques for studying this limit have been developed in the framework of the (antiferromagnetic) Potts model \cite{Pottszeros1,Pottszeros2,Pottszeros3,Pottszeros4,Pottszerostorus,Berahanonplanar,Trilattpotts}. Building on elements of these works, we here study the partition function zeros for the isotropic six-vertex model in this limit. As $N$ grows, the zeros accumulate on curves in accordance with the Beraha-Kahane-Weiss theorem. We give a numerical method to determine these limiting curves of accumulation points and discuss the universal behaviors of the curves for different values of $M$.
\end{enumerate}
All the results mentioned above can be downloaded from the github repository,
\begin{quotation}
  \url{https://github.com/yzhphy/BAE_AG/tree/master/Results/Exact_Partition_Function/partition_function_M14}
\end{quotation}
For the future reference, we also present the Gr\"obner basis results
for the six-vertex model,
\begin{quotation}
  \url{https://github.com/yzhphy/BAE_AG/tree/master/Results/XXX_Groebner_basis}
\end{quotation}

The rest of this paper is organized as follows. We introduce the six-vertex model and its torus partition function in section~\ref{sec:PFreview}. This part is standard and can be skipped by experienced readers. In section~\ref{sec:AG}, we discuss the algebro-geometric approach to compute the torus partition function. We present explicit results for the partition function in section~\ref{sec:sum_over_BAE}. For $M\le 6$, we have closed-form expressions for any $N$. For larger $M$, we give some partial results for fixed large $N$ to convey an idea about the exact results. The full results can be found in the github repository. The partition function zeros in the partial thermodynamic limit are discussed in section~\ref{sec:zero}, where we also compute the corresponding limiting curves. In section~\ref{sec:primaryDec}, we show how the algebro-geometric decomposition of the solution space of the BAE can be refined by working with an algebraic extension of $\mathbb{Q}$. Using primary decomposition over this extension and some Galois theory provides us with a finer and computationally more efficient decomposition, which we can physically relate to the construction of momentum sectors for the transfer matrix. We state our conclusions and perspectives for further work in section~\ref{sec:concDisc}. Appendices~\ref{app:Q-system}--\ref{sec:codes} contain details on more technical aspects, and the tables giving the number of physical solutions of the BAE for $6\le M\le 18$ are relegated to Appendix~\ref{sec:resultDcom9to18}.

%%%%%%%%%%%%%%%%%%%%%%%%%%%%%%%%%%%%%%%%%%%%%%%%%%%%%%%%%%%%%%%%%%%%%%%%
\section{Torus partition function of the six-vertex model}
\label{sec:PFreview}
%%%%%%%%%%%%%%%%%%%%%%%%%%%%%%%%%%%%%%%%%%%%%%%%%%%%%%%%%%%%%%%%%%%%%%%%
In this section, we review some basic facts about the six-vertex model, which also serve to fix our notations. The six-vertex model is a prototype of integrable lattice models. It is well known that it can be mapped to the Heisenberg XXZ spin chain and solved by Bethe ansatz. Throughout this work, we consider the isotropic point of the six-vertex model, which can be mapped to the Heisenberg XXX spin chain. We leave the more general case for future investigation.\par

\begin{figure}[h!]
\begin{center}
\includegraphics[scale=0.4]{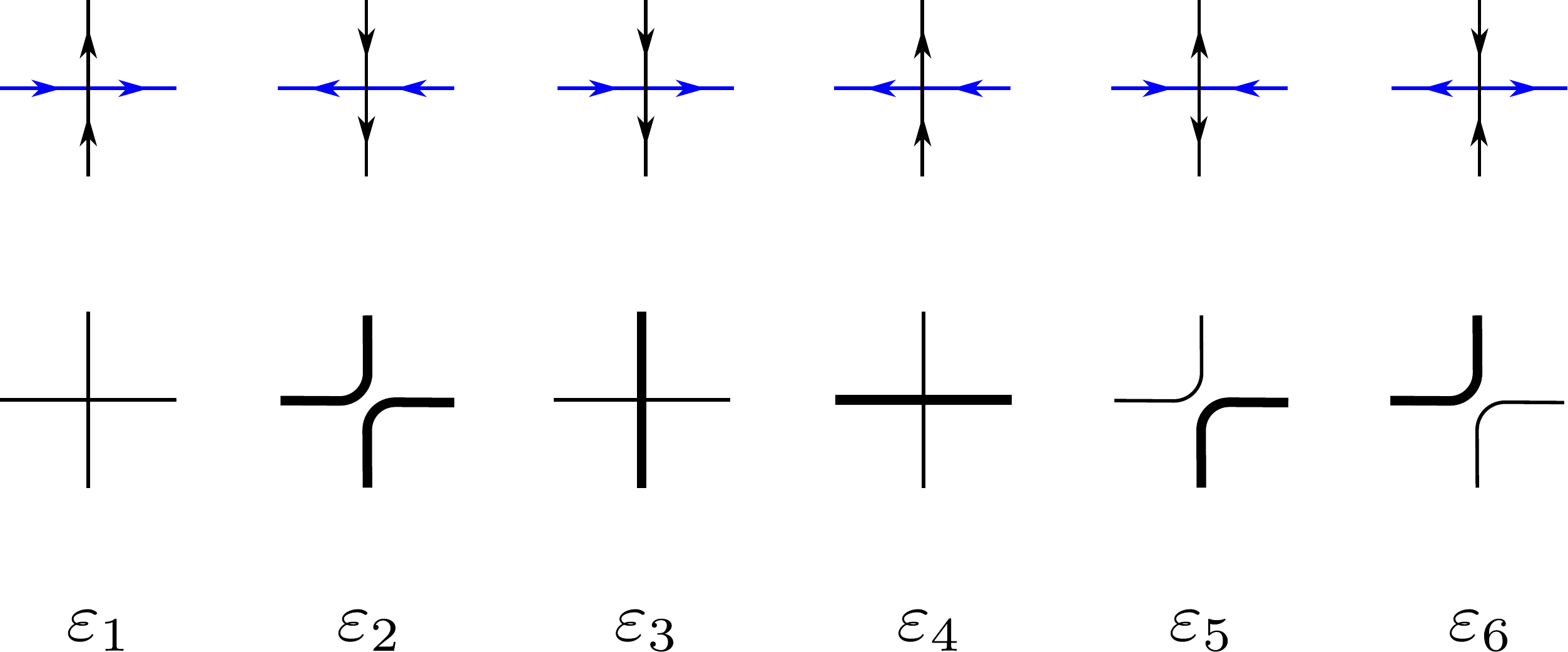}
\caption{Configurations of 6-vertex model.}
\label{fig:6vertex}
\end{center}
\end{figure}

\begin{figure}[h!]
\begin{center}
\includegraphics[scale=0.4]{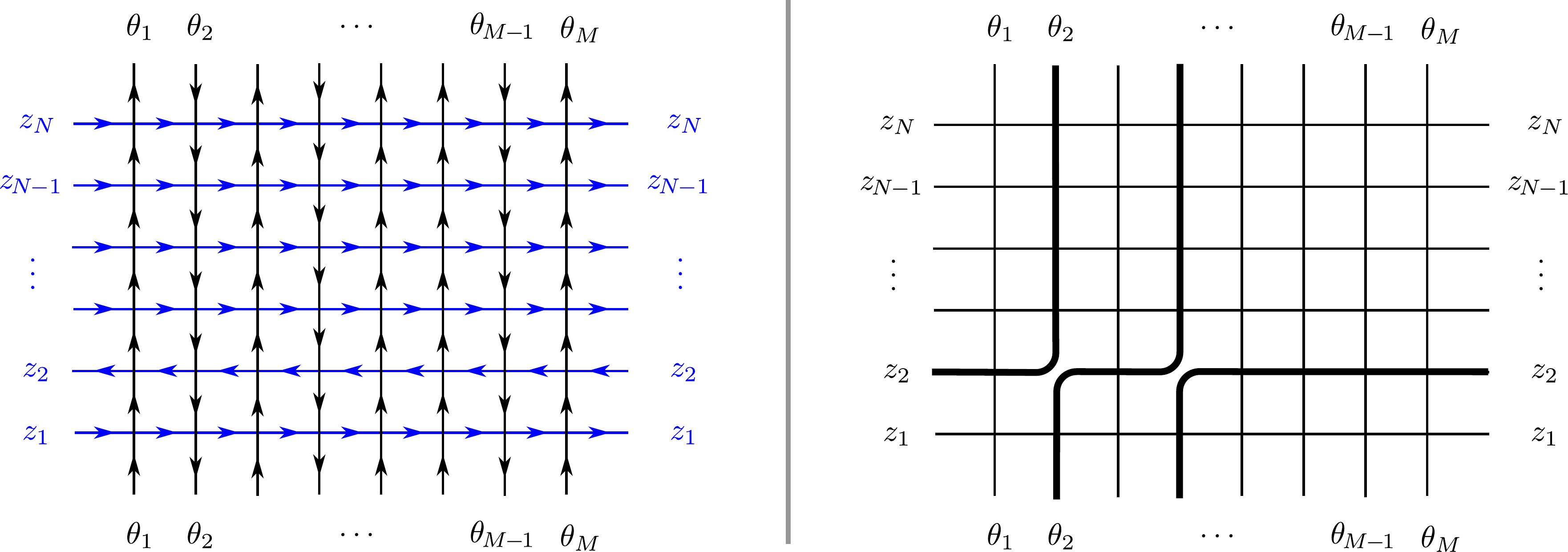}
\caption{One possible configuration for the six-vertex model of size $M\times N$.}
\label{fig:lattice}
\end{center}
\end{figure}

The six-vertex model is a two-dimensional lattice model. At each site there are six possible configurations obeying the so-called ice rule, namely the number of incoming arrows should equal the number of outgoing arrows. The six possible configurations are depicted in figure\,\ref{fig:6vertex}.
Each configuration can be represented in two ways. One is by putting arrows on each edge and the other is by using thin and thick lines. Each configuration is associated with an interaction energy $\varepsilon_i(z,\theta)$, $(i=1,\cdots,6)$ subject to the following constraints
\begin{align}
\varepsilon_1=\varepsilon_2\,,\qquad  \varepsilon_3=\varepsilon_4\,,\qquad \varepsilon_5=\varepsilon_6\,.
\end{align}
Following Baxter \cite{Baxter:book}, we denote the corresponding Boltzmann weights by $\omega_j=\exp(-\beta\varepsilon_j)$ and define
\begin{align}
\ra=\omega_1=\omega_2\,,\qquad \rb=\omega_3=\omega_4\,,\qquad \rc=\omega_5=\omega_6\,,
\end{align}
where in the isotropic case
\begin{align}
\ra(z,\theta)=z-\theta+i/2\,,\qquad \rb(z,\theta)=z-\theta-i/2\,,\qquad \rc(z,\theta)=i\,.
\label{weights_abc}
\end{align}
Indeed, the anisotropy parameter is then $\Delta=\frac{\ra^2+\rb^2-\rc^2}{2\ra \rb} = 1$, so the corresponding spin chain is the XXX one.
We consider a lattice of $M$ columns and $N$ rows with periodic boundary condition in both directions, as is shown in figure\,\ref{fig:lattice}. The partition function is given by summing the product of local Boltzmann weights over all possible configurations.\par
The partition function of the six-vertex model can be computed by the transfer matrix method (see for example \cite{Baxter:book}). To define the transfer matrix, we start from the $R$-matrix
\begin{align}
R_{an}(z,\theta)=\left(
                                       \begin{array}{cccc}
                                         \ra(z,\theta) & 0 & 0 & 0 \\
                                         0 & \rb(z,\theta) & \rc(z,\theta) & 0 \\
                                         0 & \rc(z,\theta) & \rb(z,\theta) & 0 \\
                                         0 & 0 & 0 & \ra(z,\theta) \\
                                       \end{array}
                                     \right) \,.
\end{align}
Integrability of the six-vertex model is guaranteed by the fact that the $R$-matrix satisfies the Yang-Baxter equation
\begin{align}
R_{an}(u)R_{bn}(v)R_{ab}(u-v)=R_{ab}(u-v)R_{bn}(v)R_{an}(u) \,.
\end{align}
The transfer matrix is defined as
\begin{align}
\label{eq:TMz_6v}
T_M(z)=\tr_a\left(\prod_{n=1}^M R_{an}(z,\theta_n)\right) \,.
\end{align}
The partition function can be written in terms of the transfer matrix
\begin{align}
Z_{M,N}=\tr \left[T_M(z_1)T_M(z_2)\cdots T_M(z_N)\right] \,,
\end{align}
where the parameters $z_j$ and $\theta_k$ characterize the Boltzman weights at each site of the lattice. We consider the homogeneous case $\theta_k=0$ and $z_j=z$. Then the partition function is simply given by
\begin{align}
\label{eq:PDtrace}
Z_{M,N}=\tr \left[T_M(z)^N\right] \,.
\end{align}
It is clear from the above definitions that it is a polynomial of degree $M N$ in the variable $z$ with rational coefficients.

\paragraph{Transfer matrix.} Our task is to compute the trace of $T_M(z)^N$. The transfer matrix $T_M(z)$ is a matrix of dimension $2^M$. The most straightforward way to compute the trace in (\ref{eq:PDtrace}) is by first constructing the matrix explicitly, performing the matrix multiplication $N$ times and then taking the trace. To simplify this task, it is useful to notice that the transfer matrix is block-diagonal and we can construct the transfer matrix in each spin sector separately. The spin sectors are labelled by $K$ where $K=0,1,\ldots,M$ is the number of vertical thick black lines in the right panel of figure\,\ref{fig:lattice}. In the spin chain language, $K$ corresponds to the number of flipped spins, or the number of magnons with respect to the pseudo-vacuum state $|\uparrow^M\rangle$. The dimension of spin sector $K$ is
\begin{align}
d_{M,K}={M\choose K} \,.
\label{spindim}
\end{align}
We denote the transfer matrix of the spin sector labelled by $K$ as $T_{M,K}(z)$. The partition function is then given by
\begin{align}
\label{eq:ZMN-bf}
Z_{M,N}=\sum_{K=0}^{M}\tr\left[T_{M,K}(z)^N\right] \,.
\end{align}
In what follows, we shall call this approach the \emph{brute-force method}. It serves as a check for other approaches. This approach becomes cumbersome very quickly since the dimension $d_{M,K}$ grows rapidly with $M$ and $K$. For example, $d_{12,6}=924$ which implies that for $M=12$ we already need to deal with matrices of dimension about $10^3$.

\paragraph{Bethe ansatz.} A better method which makes use of the integrability of the model is the Bethe ansatz. One can diagonalize the transfer matrix by directly constructing its eigenvectors using Bethe ansatz. The construction can be done in each spin sector. Let us denote the eigenvectors by $|\mathbf{u}\rangle$, parameterized by a set of parameters $\mathbf{u}=\{u_1,\ldots,u_K\}$ called \emph{rapidities}. We then have
\begin{align}
T_{M,K}(z)|\mathbf{u}\rangle=t_{\mathbf{u}}(z)|\mathbf{u}\rangle\,.
\end{align}
Notice that for the spin sector $K$, the number of rapidities that characterize the state is $K$. The eigenvalue $t_{\mathbf{u}}(z)$ is given by
\begin{align}
\label{eq:eigenT}
t_{\mathbf{u}}(z)=a(z)\frac{Q_{\mathbf{u}}(z-i)}{Q_{\mathbf{u}}(z)}+d(z)\frac{Q_{\mathbf{u}}(z+i)}{Q_{\mathbf{u}}(z)} \,,
\end{align}
where
\begin{align}
\label{eq:ad}
a(z)=\left(z+\frac{i}{2}\right)^M,\qquad d(z)=\left(z-\frac{i}{2}\right)^M \,,
\end{align}
and
\begin{align}
Q_{\mathbf{u}}(z)=\prod_{j=1}^K(z-u_j) \label{Baxter-polynomial}
\end{align}
is called the \emph{Baxter polynomial}. The rapidities are constrained by the Bethe ansatz equations (BAE)
\begin{align}
\label{eq:BAE}
\left(\frac{u_j+i/2}{u_j-i/2}\right)^M=\prod_{k\ne j}^K\frac{u_j-u_k+i}{u_j-u_k-i}\,,\qquad j=1,\ldots,K\,.
\end{align}

Bethe ansatz is a very powerful analytical method and it leads to the solution of the six-vertex model in the thermodynamic limit, when $M,N\to\infty$. However, for fixed and finite $M$ and $N$, the expression for $t_{\mathbf{u}}(z)$ is only formal since the parameters $\{u_1,\ldots,u_K\}$ are not known explicitly. To find them, we need to solve the system of algebraic equations (\ref{eq:BAE}).

This raises some serious problems if we want to compute the partition function exactly and explicitly as a polynomial in $z$. First of all, the BAE for generic $M$ and $K$ are impossible to solve analytically and can only be solved numerically. Even finding the numerical solutions turns out to be a highly non-trivial task. Worse, not all the solutions of the the BAE lead to true eigenvectors and eigenvalues of the transfer matrix. We know that within each spin sector labelled by $K$, the dimension of the transfer matrix is $d_{M,K}$. One therefore needs to show that there are exactly $d_{M,K}$ \emph{physical solutions} of the BAE for fixed $M$ and $K$. This is intimately related to the \emph{completeness} problem of Bethe ansatz and is quite subtle.\par

\paragraph{Physical solutions of the BAE.} In general, the number of solutions to the BAE is more than the number of physical states. It is a non-trivial problem to characterize physical solutions of BAE among all the solutions. Fortunately, this problem has been studied systematically in \cite{Hao:2013jqa}. The conclusion is that the physical solutions can be classified as pairwise distinct non-singular solutions and singular physical solutions. For a detailed discussion of these solutions, we also refer to \cite{Jiang:2017phk}. To single out these solutions, one needs to impose further constraints in addition to the original set of BAE \cite{Hao:2013jqa,Jiang:2017phk}.\par

To find the physical solutions, it is actually more convenient to work with other formulations of the BAE, in particular Baxter's $TQ$-relation and the rational $Q$-system. Baxter's $Q$-operator provides a powerful method for solving integrable lattice models \cite{Baxter:book}. The central equation of this method is an operator equation called the $TQ$-relation. In terms of the eigenvalues of the $T$ and $Q$ operators, the $TQ$ relation becomes the following functional equation
\begin{align}
\label{eq:tQ}
t(z)Q(z)=a(z)Q(z-i)+d(z)Q(z+i) \,.
\end{align}
Note that this equation is basically equivalent to (\ref{eq:eigenT}) if we multiply both sides of the latter by $Q_{\mathbf{u}}(z)$. For a state of length $M$ and magnon number $K$, $t(z)$ and $Q(z)$ are polynomials in the variable $z$ of degree $M$ and $K$, respectively. To solve the $TQ$-relation (\ref{eq:tQ}), one first writes $t(z)$ and $Q(z)$ in the explicit polynomial form
\begin{align}
\label{eq:sumTQ}
t(z)=\sum_{j=0}^M\rt_j z^j,\qquad Q(z)=z^K+\sum_{k=0}^{K-1}\rs_k z^k \,.
\end{align}
The unknown variables that we are solving for are $\{\rt_0,\ldots,\rt_M,\rs_0,\ldots,\rs_{K-1}\}$. Plugging (\ref{eq:sumTQ}) and the explicit form (\ref{eq:ad}) of $a(z)$, $d(z)$ into the equation (\ref{eq:tQ}) and demanding that it is satisfied for any value of $z$, one obtains a system of algebraic equations for the unknown variables. These equations are linear in both sets of variables $\{\rt_0,\ldots,\rt_M\}$ and $\{\rs_0,\ldots,\rs_{K-1}\}$ and are easier to solve than the original set of BAE. Solving the $TQ$-relation gives at the same time both polynomials $t(z)$ and $Q(z)$. The zeros of $Q(z)$ in turn provide the solution of the BAE. Another advantage of using the $TQ$-relation is that it automatically eliminates the solutions with coinciding rapidities, \emph{i.e.}, the solutions of the $TQ$-relation lead to pairwise distinct solutions of the BAE.\par

Working with the $TQ$-relation instead of the original BAE is already more efficient for our purpose. However, the use of the $TQ$-relation does not eliminate all the unphysical solutions. Very recently, an even more efficient method has been developed for solving the BAE which is the rational $Q$-system approach \cite{Marboe:2016yyn,Marboe:2017dmb}. In this method, one defines a rational $Q$-system associated with a Young tableaux with specific boundary conditions. The configuration of the Young tableaux is related to the length and magnon number of the Bethe state. By requiring all the $Q$-functions at the vertices of the Young tableaux to be polynomials, one ends up with a set of algebraic equations called zero-remainder conditions (ZRC) where the unknown coefficients are the coefficients of the $Q$-functions, \emph{i.e.}, the analogues of the $\{s_0,\ldots,s_{K-1}\}$ defined above. Solving the $Q$-system amounts to finding all the $Q$-polynomials on the Young tableaux. One specific $Q$-polynomial coincides with $Q(z)$ defined above, so its zeros give the Bethe roots. For more details of this approach and some explicit examples, we refer to appendix~\ref{app:Q-system}.

It turns out that solving the ZRC is even more efficient than solving the $TQ$-relation. Moreover, the solutions of the ZRC are in one-to-one correspondence with the physical solutions of the BAE, so no further constraints need to be imposed. One minor disadvantage of this method is that the equations themselves are not known explicitly for any length and magnon numbers and have to be derived case by case. This is not a big issue in practice since the equations can be generated rather efficiently for not too large $M$ and $K$.\par

In what follows, to find all the physical solutions of the BAE, we will work with rational $Q$-systems. So far, the $Q$-system approach for BAE has only been developed for the isotropic limit ($\Delta = 1$) corresponding to the XXX spin chain. To treat the XXZ spin chain at generic $\Delta$, we still need to rely on the $TQ$-relation, together with additional constraints to select physical solutions. The strategy is to first find the $Q$-polynomial, and next use the $TQ$-relation to find the $t$-polynomials. After finding all the $t$-polynomials, it is straightforward to write down the torus partition function that we are after.\par

However, it is clear that if we want to find all the $t$-polynomials explicitly, no matter which approach (BAE, $TQ$-relation, $Q$-system) we are using, we have to solve the system of algebraic equations numerically. The results are thus approximate and not exact. To avoid solving equations and obtain instead exact and analytical results for the partition function, we can however apply methods in computational algebraic geometry. We discuss these approaches in the next section.

%%%%%%%%%%%%%%%%%%%%%%%%%%%%%%%%%%%%%%%%%%%%%%%%%%%%%%%%%%%%%%%%%%%%%%%%
\section{Algebro-geometric approach to the partition function}
\label{sec:AG}
%%%%%%%%%%%%%%%%%%%%%%%%%%%%%%%%%%%%%%%%%%%%%%%%%%%%%%%%%%%%%%%%%%%%%%%%
In this section, we describe our method for computing the torus
partition function using an algebro-geometric approach. The main tools
that we are going to use are \emph{Gr\"obner basis}, \emph{quotient ring}
and \emph{companion matrix}. See \cite{MR3330490} for a textbook reference to the corresponding
mathematics. For a detailed introduction to these notions in the context of Bethe ansatz, we refer to \cite{Jiang:2017phk}.\par

As discussed in the previous section, in order to select all the physical solutions, we work with the rational $Q$-system. For the $\mathfrak{su}(2)$ invariant XXX spin chain which is equivalent to the six-vertex model, the corresponding Young tableaux have two rows with the number of boxes being $(M-K,K)$. The $Q$-polynomial that we are interested is $Q_{0,1}$. The computation of the $Q$-polynomials relies on the definition of certain paths on the Young tableau (for more details, see appendix~\ref{app:Q-system}). For the $\mathfrak{su}(2)$ case, we can choose the path such that the unknown coefficients are precisely the coefficients of $Q_{0,1}(z)$, and we have
\begin{align}
Q(z) = Q_{0,1}(z)=z^K+\sum_{k=0}^{K-1}\rs_k\,z^k \,.
\end{align}
\paragraph{Ideal.} The zero-remainder conditions (ZRC) then give a set of algebraic equations for the $K$ variables $\{\rs_{0},\ldots,\rs_{K-1}\}$,
\begin{align}
\label{eq:idealf}
f_1(\rs_0,\rs_1,\cdots,\rs_{K-1})=f_2(\rs_0,\rs_1,\cdots,\rs_{K-1})=f_S(\rs_0,\rs_1,\cdots,\rs_{K-1})=0\,,
\end{align}
where $f_k(\rs_0,\rs_1,\cdots,\rs_{K-1})$ are polynomials in the variables $\{\rs_{0},\cdots,\rs_{K-1}\}$. Here $S$ is the number of equations and it depends on the path we choose. The polynomials $f_1,\cdots,f_S$ define an ideal in the polynomial ring $\mathbb{C}[\rs_0,\rs_1,\cdots,\rs_{K-1}]$, denoted
\begin{align}
\rI_{M,K}=\langle f_1,\cdots,f_S\rangle.
\end{align}
A given ideal can be generated by different bases, among which the so-called Gr\"obner basis is particularly useful for us. We denote the Gr\"obner basis by $\rG_k$
\begin{align}
\rI_{M,K}=\langle f_1,\cdots,f_S\rangle=\langle \rG_1,\cdots,\rG_{S'}\rangle\,
\end{align}
where in general $S$ and $S'$ are different. Notice that when computing the Gr\"obner basis we need to choose a \emph{partial ordering} of the monomials formed of the variables $\{\rs_0,\rs_1,\cdots,\rs_{K-1}\}$. For different orderings, the corresponding Gr\"obner basis can look quite different.

\paragraph{Quotient ring.} The quotient ring is defined as
\begin{align}
\mathcal{Q}_{M,K}=\mathbb{C}[\rs_0,\rs_1,\cdots,\rs_{K-1}]/\rI_{M,K}\,,
\end{align}
and it is a finite-dimensional linear space. The dimension of this linear space equals the number of physical solutions of the BAE for given $M$ and $K$, which is given by \cite{Hao:2013jqa}
\begin{align}
\mathcal{N}_{M,K}={M\choose K}-{M\choose K-1}\,.
\label{binomialdiff}
\end{align}
Since $\mathcal{Q}_{M,K}$ is a linear space, it can be spanned by a
basis set. The standard basis of the quotient ring $\mathcal{Q}_{M,K}$
is given by all the monomials of $\{\rs_0,\cdots,\rs_{K-1}\}$ that
cannot be divided by $\text{LT}[\rG_k]$ ($k=1,\cdots,S'$) where ``LT''
stands for the leading term in some given partial ordering. In order
to construct the standard basis, one needs to calculate the Gr\"obner
basis of the ideal $\langle f_1,\cdots,f_S\rangle$. This is one of the
main calculations of the current work which can be done by standard
algorithms such as Buchberger's algorithm or the F4/F5 algorithm of
Faug\`ere \cite{Faugere199961, Faugere:2002:NEA:780506.780516}. These
algorithms are implemented in several packages for algebraic geometry,
such as \texttt{Singular} \cite{DGPS} .\par

\paragraph{Companion matrix.} Let us denote the basis of the quotient ring by $\{e_{1},e_2,\cdots,e_{\mathcal{N}_{M,K}}\}$. Any polynomial $P(\rs_0,\rs_1,\ldots,\rs_{K-1})$ can be mapped to a numerical matrix $\mathbf{M}_P$ called the \emph{companion matrix} of dimension $\mathcal{N}_{M,K}$. The algorithm for doing so is as follows. We multiply the polynomial $P$ by one of the standard basis elements $e_j$ and then find the remainder of the polynomial reduction with respect to the Gr\"obner basis,
\begin{align}
P\,e_j=\sum_{k=1}^{S'}a_k\,\rG_k+r_j\,.
\end{align}
The remainder $r_j(\rs_0,\cdots,\rs_{K-1})$ can be expanded in terms of the standard basis where the coefficients of the expansion are the elements of the companion matrix, \emph{i.e.},
\begin{align}
r_j= \sum_{k=1}^{{\cal N}_{M,K}} M_{jk}\,e_k\,,\qquad \mathbf{M}_P=\left(M_{ij}\right)\,.
\end{align}
The companion matrix satisfies the following homomorphism properties
\begin{align}
\mathbf{M}_{P_1\pm P_2}=&\,\mathbf{M}_{P_1}\pm\mathbf{M}_{P_2}\,, \nonumber \\
\mathbf{M}_{P_1  P_2}=&\,\mathbf{M}_{P_1}\cdot\mathbf{M}_{P_2}\,, \label{eq:propertyM} \\
\mathbf{M}_{P_1/P_2}=&\,\mathbf{M}_{P_1}\cdot\mathbf{M}_{P_2}^{-1}\,. \nonumber
\end{align}
The main result from algebraic geometry which we are going to use is
\begin{align}
\label{eq:trace}
\sum_{\text{sol}}P(\rs_0,\cdots,\rs_{K-1})=\tr\,\mathbf{M}_{P}\,,
\end{align}
where the sum `sol' is over all solutions of the system of equations (\ref{eq:idealf}). It is straightforward to see that if we start with equations whose coefficients are in $\mathbb{Q}$, the right-hand side of (\ref{eq:trace}) is rational (viz., it is a polynomial in $z$ with rational coefficients), even though individual terms appearing on the left-hand side may be irrational.\par

Using the procedure described above, after the construction of quotient ring, we can map the function $Q_{0,1}(z)$ into a companion matrix by mapping each coefficient $\rs_k\mapsto\mathbf{S}_k$, so that
\begin{align}
Q_{0,1}(z)\mapsto \mathbf{Q}_{M,K}(z)=z^K\,\mathbf{I}+\sum_{k=1}^{K-1}\mathbf{S}_k z^k\,.
\end{align}
where $\mathbf{I}$ is the identity matrix of dimension $\mathcal{N}_{M,K}$. Our goal is to find the companion matrix of the transfer matrix $t_{\mathbf{u}}(z)$, since this will permit us to access the partition function. This can be done by using the explicit expression of the transfer matrix (\ref{eq:eigenT}) and the properties of the companion matrix (\ref{eq:propertyM}). More explicitly, we have
\begin{align}
\label{compmatrix_tT}
t_{\mathbf{u}}(z)\mapsto \mathbf{T}_{M,K}(z) \,,
\end{align}
where $\mathbf{T}_{M,K}(z)$ can be computed using the ideal generated by the $TQ$-relations\footnote{More precisely, to select physical solutions, we need to combine the equations coming from the $TQ$-relations and the rational $Q$-system and then eliminate the variables $\mathbf{s}$.} or from the companion matrix $\mathbf{Q}_{M,K}(z)$ by the following relation
\begin{align}
\mathbf{T}_{M,K}(z)=\left[a(z)\mathbf{Q}_{M,K}(z-i)+d(z)\mathbf{Q}_{M,K}(z+i)\right]\cdot\mathbf{Q}_{M,K}(z)^{-1} \,,
\end{align}
valid whenever $\mathbf{Q}_{M,K}(z)$ is non-singular. The partition function is given by
\begin{align}
\label{eq:ZMN-ag}
Z_{M,N}=\sum_{K=0}^{[M/2]}(M-2K+1)\,\tr\!\left(\mathbf{T}_{M,K}(z)^N\right)\,.
\end{align}

Several comments are in order. The multiplicities $M-2K+1$, for $K\le M/2$, take into account the descendant states in the Bethe ansatz. These states are obtained by sending some of the Bethe roots to infinity. It is easy to see that adding a root at infinity does not change the eigenvalue of the transfer matrix. When solving the BAE or the ZRC, we only find regular solutions that do not have roots at infinity. Since the descendant states are indeed part of the spectrum, we need to take them into account in the computation of the partition function by the proper multiplicity.

The expression in (\ref{eq:ZMN-ag}) takes a very similar form to the one in (\ref{eq:ZMN-bf}). However, there are important differences. Firstly, the result in (\ref{eq:ZMN-ag}) makes use of the full $\mathfrak{su}(2)$ symmetry, and hence the dimensions of the transfer matrices $\mathbf{T}_{M,K}(z)$ are smaller than those of $T_{M,K}(z)$. The dimensions of $\mathbf{T}_{M,K}(z)$ and $T_{M,K}(z)$ are $\mathcal{N}_{K,M}$ and $d_{K,M}$ respectively, see (\ref{binomialdiff}) and (\ref{spindim}). For example, for $M=14$, $K=7$, we have $\mathcal{N}_{14,7}=429$ and $d_{14,7}=3432$. Therefore (\ref{eq:ZMN-ag}) is computationally more efficient than (\ref{eq:ZMN-bf}).

Secondly, one can impose further constraints on the solution space of the BAE and decompose the solution space into even smaller subspaces for (\ref{eq:ZMN-ag}). This means that we can make the matrices $\mathbf{T}_{M,K}(z)$ in (\ref{eq:ZMN-ag}) block-diagonal and hence work with even smaller matrices, which improves the efficiency further. This point will be discussed in detail in section~\ref{sec:primaryDec}.

%%%%%%%%%%%%%%%%%%%%%%%%%%%%%%%%%%%%%%%%%%%%%%%%%%%%%%%%%%%%%%%%%%%%%%%%
\section{Explicit results}
\label{sec:sum_over_BAE}
%%%%%%%%%%%%%%%%%%%%%%%%%%%%%%%%%%%%%%%%%%%%%%%%%%%%%%%%%%%%%%%%%%%%%%%%
In this section, we give results of partition function for different values of $M$ and $N$. We obtain closed-form results up to $M=6$ for arbitrary $N$. The reason we stop at $M=6$ is that the ZRC can be solved analytically up to this length in $\mathbb{Q}$. For $M=7$ and $M=8$, analytical solutions of the ZRC can also be found by working in extended fields; see section~\ref{sec:primaryDec} for more details. For $M\ge9$, we give an efficient algorithm for computing the partition function for fixed $M$ and $N$ (which can be large). The results are polynomials in $z$ of high degrees with rational coefficients (typically large) and it does not make sense to write them down in this paper. Instead, interested readers can find the results on the repository which we mentioned in the introduction.

\subsection{Closed-form expressions for $M\le 6$}
In this section, we list the results for $M\le 6$. We denote the partition function of an $M\times N$ toroidal lattice by $Z_{M,N}$.

\paragraph{$M=1.$} This is the simplest case. The sum in (\ref{eq:ZMN-ag}) contains only $K=0$, and the result is given by
\begin{align}
Z_{1,N}=2\times(2z)^N.
\end{align}

\paragraph{$M=2.$} We have to sum over $K=0$ and $K=1$. The partition function is given by
\begin{align}
Z_{2,N}=3\left(2z^2-\frac{1}{2}\right)^N+\left(2z^2+\frac{3}{2}\right)^N.
\end{align}

\paragraph{$M=3.$} We again have to sum over $K=0$ and $K=1$. The result is
\begin{align}
Z_{3,N}=4\left(2z^3-\frac{3}{2}z \right)^N+2\left[\left(2z^3+\frac{3}{2}z+\frac{\sqrt{3}}{2}\right)^N+\left(2z^3+\frac{3}{2}z-\frac{\sqrt{3}}{2}\right)^N\right] \,.
\end{align}
We see that irrational numbers start to show up within some eigenvalues of the transfer matrix. However, for any $N \in \mathbb{N}$ the result for
$Z_{3,N}$ is a rational-coefficient polynomial in $z$ after simplification.

\paragraph{$M=4.$} We have to sum over $K=0,1,2$. The result is
\begin{align}
\label{eq:Z4N}
Z_{4,N}=&\,5\left(2z^4-3z^2+\frac{1}{8}\right)^N \\\nonumber
+&3\left(2z^4+z^2+2z+\frac{1}{8}\right)^N
+ 3\left(2z^4+z^2-2z+\frac{1}{8}\right)^N +3\left(2z^4+z^2-\frac{7}{8}\right)^N\\\nonumber
+&\left(2z^4+3z^2+\frac{13}{8}\right)^N+\left(2z^4+3z^2-\frac{3}{8}\right)^N.
\end{align}

\paragraph{$M=5.$} We have to sum over $K=0,1,2$. The final result reads
\begin{align}
\label{eq:Z5N}
Z_{5,N}=&\,6\left(2 z^5- 5 z^3+\frac{5}{8}z\right)^N  \\\nonumber
+&\,4\left(2z^5-\frac{1}{2}\sqrt{25+10\sqrt{5}}\,z^2-\frac{1}{8}(5-4\sqrt{5})z+\frac{1}{8}\sqrt{5-2\sqrt{5}}\right)^N\\\nonumber
+&\,4\left(2z^5+\frac{1}{2}\sqrt{25+10\sqrt{5}}\,z^2-\frac{1}{8}(5-4\sqrt{5})z-\frac{1}{8}\sqrt{5-2\sqrt{5}}\right)^N\\\nonumber
+&\,4\left(2z^5-\frac{1}{2}\sqrt{25-10\sqrt{5}}\,z^2-\frac{1}{8}(5+4\sqrt{5})z+\frac{1}{8}\sqrt{5+2\sqrt{5}}\right)^N\\\nonumber
+&\,4\left(2z^5+\frac{1}{2}\sqrt{25-10\sqrt{5}}\,z^2-\frac{1}{8}(5+4\sqrt{5})z-\frac{1}{8}\sqrt{5+2\sqrt{5}}\right)^N\\\nonumber
+&\,2\left(2z^5+3z^3+\frac{21}{8}z\right)^N\\\nonumber
+&\,2\left(2z^5+3z^3-\frac{1}{2}\sqrt{10-2\sqrt{5}}\,z^2+\frac{1}{8}(1+4\sqrt{5})z-\frac{1}{8}\sqrt{50+22\sqrt{5}}\right)^N\\\nonumber
+&\,2\left(2z^5+3z^3+\frac{1}{2}\sqrt{10-2\sqrt{5}}\,z^2+\frac{1}{8}(1+4\sqrt{5})z+\frac{1}{8}\sqrt{50+22\sqrt{5}}\right)^N\\\nonumber
+&\,2\left(2z^5+3z^3-\frac{1}{2}\sqrt{10+2\sqrt{5}}\,z^2+\frac{1}{8}(1-4\sqrt{5})z+\frac{1}{8}\sqrt{50-22\sqrt{5}}\right)^N\\\nonumber
+&\,2\left(2z^5+3z^3+\frac{1}{2}\sqrt{10+2\sqrt{5}}\,z^2+\frac{1}{8}(1-4\sqrt{5})z-\frac{1}{8}\sqrt{50-22\sqrt{5}}\right)^N.\\\nonumber
\end{align}
We see that the eigenvalues of the transfer matrix now become more complicated, with double square roots showing up in the coefficients.

\paragraph{$M=6.$} We have to sum over $K=0,1,2,3$. The final result reads
\small{
\begin{align}
\label{eq:Z6N}
Z_{6,N}=&\,7\left(2z^6-\frac{15}{2}z^4+\frac{15}{8}z^2-\frac{1}{32}\right)^N \\\nonumber
 +&\,5\left(2z^6-\frac{3}{2}z^4-\frac{25}{8}z^2+\frac{11}{32}\right)^N\\\nonumber
 +&\,5\left(2z^6-\frac{3}{2}z^4+\sqrt{3}z^3-\frac{21}{8}z^2-\frac{3\sqrt{3}}{4}z-\frac{1}{32}\right)^N+5\left(2z^6-\frac{3}{2}z^4-\sqrt{3}z^3-\frac{21}{8}z^2+\frac{3\sqrt{3}}{4}z-\frac{1}{32}\right)^N\\\nonumber
 +&\,5\left(2z^6-\frac{3}{2}z^4+3\sqrt{3}z^3+\frac{11}{8}z^2-\frac{\sqrt{3}}{4}z-\frac{1}{32}\right)^N +5\left(2z^6-\frac{3}{2}z^4-3\sqrt{3}z^3+\frac{11}{8}z^2+\frac{\sqrt{3}}{4}z-\frac{1}{32}\right)^N\\\nonumber
 +&\,3\left(2z^6+\frac{5}{2}z^4-\frac{25}{8}z^2+\frac{3}{32}\right)^N\\\nonumber
 +&\,3\left(2z^6+\frac{5}{2}z^4+\frac{1}{8}(15-8\sqrt{5})z^2-\frac{1}{32}(21+8\sqrt{5})\right)^N\\\nonumber
 +&\,3\left(2z^6+\frac{5}{2}z^4+\frac{1}{8}(15+8\sqrt{5})z^2-\frac{1}{32}(21-8\sqrt{5})\right)^N\\\nonumber
 +&\,3\left(2z^6+\frac{5}{2}z^4-\sqrt{3}z^3+\frac{11}{8}z^2-\frac{5\sqrt{3}}{4}z -\frac{9}{32}\right)^N\\\nonumber
 +&\,3\left(2z^6+\frac{5}{2}z^4+\sqrt{3}z^3+\frac{11}{8}z^2+\frac{5\sqrt{3}}{4}z -\frac{9}{32}\right)^N\\\nonumber
 +&\,3\left(2z^6+\frac{5}{2}z^4-\frac{\sqrt{54-6\sqrt{17}}}{2}z^3+\frac{(2\sqrt{17}-3)}{8}z^2-\frac{\sqrt{54-6\sqrt{17}}(3+\sqrt{17})}{16}z
 +\frac{(9+2\sqrt{17})}{32} \right)^N\\\nonumber
 +&\,3\left(2z^6+\frac{5}{2}z^4+\frac{\sqrt{54-6\sqrt{17}}}{2}z^3+\frac{(2\sqrt{17}-3)}{8}z^2+\frac{\sqrt{54-6\sqrt{17}}(3+\sqrt{17})}{16}z
 +\frac{(9+2\sqrt{17})}{32} \right)^N\\\nonumber
 +&\,3\left(2z^6+\frac{5}{2}z^4-\frac{\sqrt{54+6\sqrt{17}}}{2}z^3-\frac{(2\sqrt{17}+3)}{8}z^2-\frac{\sqrt{54+6\sqrt{17}}(\sqrt{17}-3)}{16}z
 +\frac{(9-2\sqrt{17})}{32} \right)^N\\\nonumber
 +&\,3\left(2z^6+\frac{5}{2}z^4+\frac{\sqrt{54+6\sqrt{17}}}{2}z^3-\frac{(2\sqrt{17}+3)}{8}z^2-\frac{\sqrt{54+6\sqrt{17}}(\sqrt{17}-3)}{16}z
 +\frac{(9-2\sqrt{17})}{32} \right)^N\\\nonumber
 +&\,\left(2z^6+\frac{9}{2}z^4+\frac{23}{8}z^2-\sqrt{3}z-\frac{1}{32}\right)^N
 +\left(2z^6+\frac{9}{2}z^4+\frac{23}{8}z^2+\sqrt{3}z-\frac{1}{32}\right)^N\\\nonumber
 +&\,\left(2z^6+\frac{9}{2}z^4+\frac{7-8\sqrt{13}}{8}+\frac{31-8\sqrt{13}}{32}\right)^N
 +\left(2z^6+\frac{9}{2}z^4+\frac{7+8\sqrt{13}}{8}+\frac{31+8\sqrt{13}}{32}\right)^N\\\nonumber
 +&\,\left(2z^6+\frac{9}{2}z^4+\frac{15}{8}z^2-\frac{25}{32}\right)^N.
\end{align}
}
\normalsize
Again, we see that some of the terms in (\ref{eq:Z5N})--(\ref{eq:Z6N}) contain multiple square roots. However, once we sum over all terms for any $N\in\mathbb{N}$, we obtain polynomials whose coefficients are rational numbers, as expected.

\subsection{Partition functions for higher $M$ and $N$}
For $M=7$ and $M=8$, we can also work out the analytical results. However, the closed-form results involve complicated multi-square roots and are not very illuminating to write down explicitly here. For $M\ge 9$, we are not able to find analytical solutions anymore. For these cases, we give an efficient approach to compute the partition function for fixed $M$ and $N$ based on the companion matrix.\par

Although our method is much more efficient than the brute-force approach, the complexity still grows exponentially with $M$. On a laptop, we are able to compute the Gr\"obner basis and companion matrices of the $Q$-polynomial $\mathbf{Q}(z)$ and the transfer matrix $\mathbf{T}_{M,K}(z)$ up to $M=14$ after algebraic extension (see more details in section~6). The dimensions of the companion matrices with fixed $M,K,\ell$ are given in tables~\ref{BAE_decomposition_M6}--\ref{BAE_decomposition_M18}. In general, the partition function $Z_{M,N}(z)$ is a polynomial of order $MN$. Let us consider one example and take $M=14$, $N=100$. We have
\begin{align}
Z_{14,100}(z)=\sum_{k=0}^{700}c_{n} z^{2n}
\end{align}
where $c_n$ are rational numbers. The full result is too large to show, we present simply one coefficient, say $c_{50}$, here just to give an idea about the result. It can be written as $c_{50}=\rN_{50}/\rD_{50}$ where $\rN_{50}$ and $\rD_{50}$ are integers and are given by
\begin{align*}
\rN_{50}=&\,3549714199509718765414261648405948375346908444631449814070999015959834\\\nonumber
&\,5566548305771114077497148041577082024237243782436360433278999500001467\\\nonumber
&\,7415020115130416461041186374238411444900889750187515469354178173296042\\\nonumber
&\,6263542784005444527370571255655399436082869031810512249627664031748092\\\nonumber
&\,4072971596196276059412147685707041129531252668420023067545577282800096\\\nonumber
&\,9295580666703059631065928833005677223301695252356346794040665489002919\\\nonumber
&\,252969420827675164522054378930548720509758103112303817657569605
\end{align*}
and
\begin{align*}
\rD_{50}=&\,5254662920039596746236382353531688074822975081008511328653941899536358\\\nonumber
&\,8535283647520133924979358051065937411251928593712733718650179494204788\\\nonumber
&\,1824032065695719746688948719970424831685787857667697873122163831147965\\\nonumber
&\,2780562037740551931521573661427623891169657954033123606560887017391291\\\nonumber
&\,1075991202517005852196882800252419269626961951377918107052351183818088\\\nonumber
&\,1989632 \,.
\end{align*}
Approximately this number is $c_{50}\approx 6.75536\times 10^{125}$. Using numerical methods such as the function \texttt{Rationalize} in {\sc{Mathematica}} to guess such a large number will be quite difficult in practice since it requires working with floating point numbers with very high accuracy.\par

Since the results for partition functions for large $N$ are typically large, we find it more useful to give the companion matrices. These companion matrices contain all the information we need. To find the explicit eigenvalues of $Q_{\mathbf{u}}(z)$ and $t_{\mathbf{u}}(z)$, we can diagonalize the companion matrices. In general this diagonalization can only be done numerically, but the matrices that we need to diagonalize are much smaller and can be handled much more efficiently. The zeros of $Q_{\mathbf{u}}(z)$ give the Bethe roots. That is to say, we can straightforwardly find \emph{all} physical solutions of the BAE up to length $M=14$ using our results. The eigenvalues $t_{\mathbf{u}}(z)$ contain all the information about conserved charges of the system, \emph{i.e.}, momentum, energy and higher conserved charges of the Bethe states.

To find the exact partition function $Z_{M,N}$, we need to take matrix powers of the compa\-nion matrices $\mathbf{T}_{M,K}(z)^N$ and then take the trace. We will be interested in the case of large $N$. Naively, we would need to perform $N$ matrix multiplications involving the companion matrix before taking the trace. When the size of the matrix is large, the analytic computation of matrix multiplication become time consuming. To reach a high value of $N$, we actually need a better way to do the computation. This is described in appendix~\ref{sec:largematrix}.

\subsection{Consistency check}
\begin{figure}[h!]
\begin{center}
\includegraphics[scale=0.5]{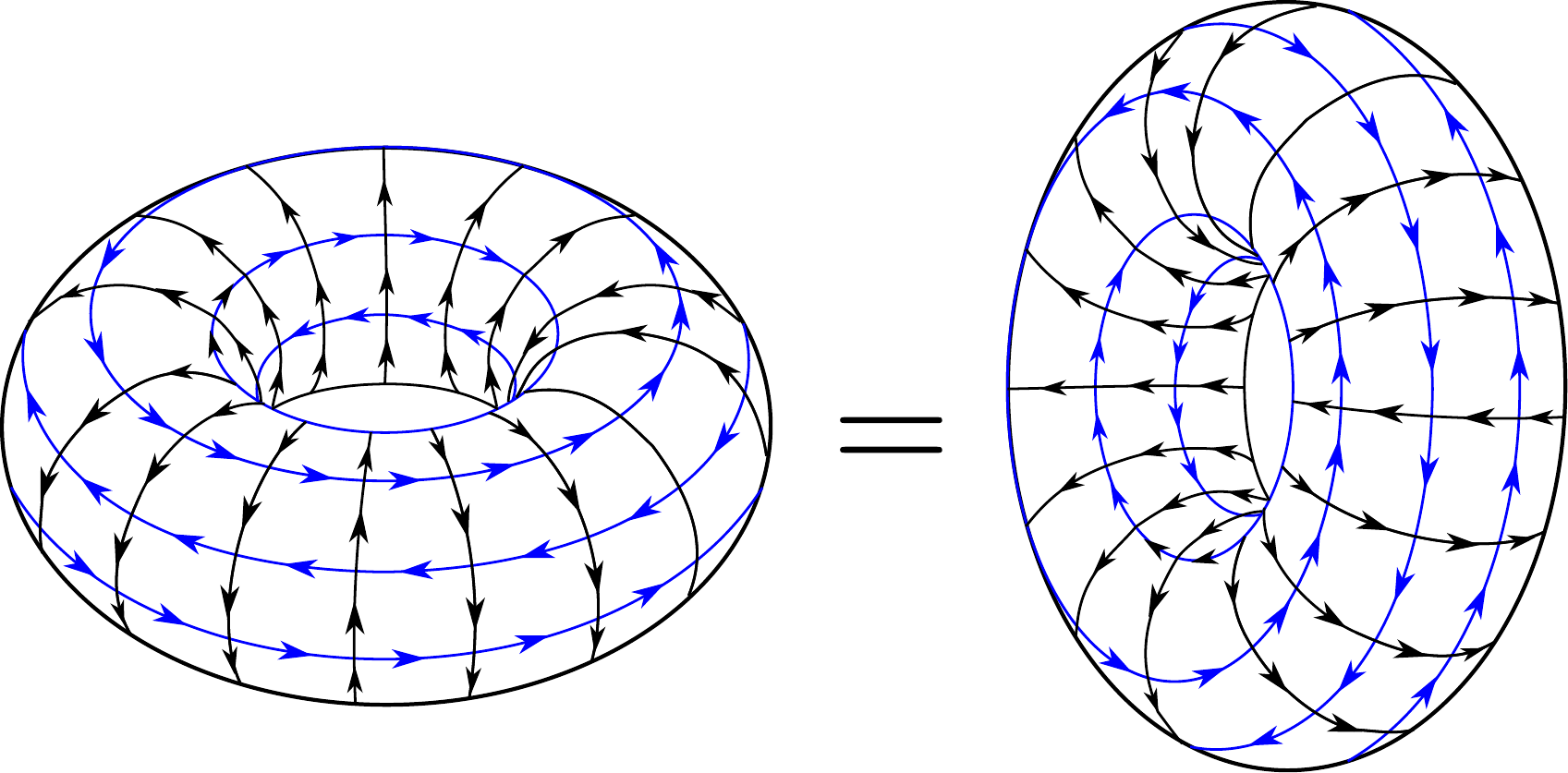}
\caption{Lattice modular invariance of the lattice partition function.}
\label{fig:torus}
\end{center}
\end{figure}

Since our results are usually large polynomials, it is important to perform some checks for their correctness. One important consistency check is the following. We can compute the lattice partition function in two ways, corresponding to two different choices of the transfer direction, and the result should be the same, as is shown in figure~\ref{fig:torus}.
Specifically, we consider the transformation of the partition function where we rotate the lattice by $\pi/2$. To analyze the effect of this transformation, we consider the $M\times N$ lattice with $M$ vertical lines and $N$ horizontal lines. We associate to each vertical line the same spectral parameter $\theta$, and to each horizontal line the same spectral parameter $z$. We denote this partition function by $Z_{M,N}(z-\theta)$. There are six possible configurations at each site with three different Boltzmann weights given by
\begin{align}
\ra(z,\theta)=z-\theta+i/2\,,\qquad \rb(z,\theta)=u-\theta-i/2\,,\qquad \rc(z,\theta)=i\,.
\end{align}
The partition function can be written as
\begin{align}
\label{eq:fmnk}
Z_{M,N}(z-\theta)=\sum_{m,n,k\ge 0\atop m+n+k=MN}f_{m,n,k}\,\ra(z,\theta)^m\,\rb(z,\theta)^n\,\rc(z,\theta)^k\,,
\end{align}
where $f_{m,n,k}$ denotes the multiplicity of configurations with $m,n,k$ vertices with Boltzmann weight $\ra,\rb,\rc$ respectively. Rotating the lattice by $\pi/2$, it is clear, from the first line of figure~\ref{fig:6vertex}, that we have the following transformation
\begin{align}
&\ra(u,\theta)\mapsto \rb(\theta,u)=-\ra(u,\theta)\,,\\\nonumber
&\rb(u,\theta)\mapsto \ra(\theta,u)=-\rb(u,\theta)\,,\\\nonumber
&\rc(u,\theta)\mapsto \rc(\theta,u)=\rc(u,\theta)\,.
\end{align}
Therefore, under this transformation, which we denote by $\mathsf{R}$, the partition function transforms as
\begin{align}
Z^{\mathsf{R}}_{M,N}(u-\theta)=Z_{N,M}(\theta-u)=\sum_{m,n,k\ge 0\atop m+n+k=MN}(-1)^{m+n}\,f_{m,n,k}\,\ra(u,\theta)^m\,\rb(u,\theta)^n\,\rc(u,\theta)^k.
\end{align}
The number of type-$\rc$ vertices $k$ is even due to the ice rule. Therefore, if $M N$ is even, the transformation leaves the partition function invariant, otherwise it gives an additional minus sign. This invariance is a strong consistency check.

We use this relation to check the correctness of the companion matrices as follows. For a given length $M$, we construct the corresponding companion matrices and compute the partition function $Z_{M,N}(z-\theta)$ for all $N\le M$. These partition functions can also be computed as $Z_{N,M}(\theta-z)$ using the companion matrices constructed for length $N$. If the results are correct, the two calculations should give the same result. We have checked the companion matrix in this way up to $M=14$.

%%%%%%%%%%%%%%%%%%%%%%%%%%%%%%%%%%%%%%%%%%%%%%%%%%%%%%%%%%%%%%%%%%%%%%%
\section{Zeros of partition functions}
\label{sec:zero}
%%%%%%%%%%%%%%%%%%%%%%%%%%%%%%%%%%%%%%%%%%%%%%%%%%%%%%%%%%%%%%%%%%%%%%%%
The torus partition function $Z_{M,N}(z)$ is a polynomial of order $MN$ in $z$. It is instructive to find the zeros of this partition function in the complex $z$-plane.

Partition function zeros for statistical models with one complex parameter have been studied in a variety of contexts---including
Lee-Yang zeros (complex magnetic field) \cite{Lee-Yang}, Fisher zeros (complex temperature) \cite{Fisher:zeros}, and graph polynomials such as the $Q$-colour
chromatic polynomial \cite{Pottszeros1,Pottszeros2,Pottszeros3,Pottszeros4,Pottszerostorus,Berahanonplanar,Trilattpotts}---and has given rise to an immense literature (see, \emph{e.g.}, references in \cite{Pottszeros1}).
The zeros of partition functions in the thermodynamic limit contain information about the phases and critical behavior of the model at hand.
In many cases the zeros will accumulate on curves, for $M,N \to \infty$, which ``pinch'' the real axis at one or more critical points.
Isolated accumulation points provide another possible scenario. In the simplest cases---such as the Ising model with suitable boundary conditions---the
curves of accumulation points can be proved to form circles, giving rise to so-called circle theorems. More generally, the density of zeros near a critical point
obey scaling laws that can be related to critical exponents.

The true thermodynamic limit of an $M \times N$ system is obtained by letting $M,N \to \infty$ with a fixed and finite ratio, $0 < M/N < \infty$.
But another means of obtaining relevant information is to fix a finite value of $M$, and study the accumulation points of zeros as $N \to \infty$.
For a partition function of the form (\ref{eq:ZMN-ag}), and supposing a mild non-degeneracy condition,
the Beraha-Kahane-Weiss theorem \cite{Beraha4209} states that the accumulation points will form curves.
By standard analyticity theorems, any closed region delimited by such curves will constitute a thermodynamical phase (for $N \to \infty$).
Under reasonable (but not entirely innocuous) assumptions about the commutativity of limits, the phase diagram in the thermodynamic
limit can then be inferred by studying the convergence of these curves upon taking $M \to \infty$.

\subsection{Partition functions for different $M$ and $N$}
In this subsection, we give the partition function zeros for different $M$ and $N$. We fix the value of $M$ and increase $N$ to see how the distribution of the zeros change. In this way, we try to extrapolate the behaviors of the zeros to the (partial) thermodynamic limit $N\to\infty$ (with $M$ being fixed and finite). One might wonder what is the benefit of knowing the partition function exactly for finding the zeros. Naively one might expect that numerical approximations will be sufficient for finding the zeros of the partition function. However, it is known that the locations of zeros of a polynomial can be very sensitive to perturbations of coefficients, especially when the degree of the polynomial is large. One famous example is the so-called Wilkinson's polynomial where a change of one of the coefficients by $10^{-7}$ leads to significant changes in the locations of the zeros. Having exact results eliminates this potential subtlety.

\begin{figure}[h!]
\begin{center}
\includegraphics[scale=0.3]{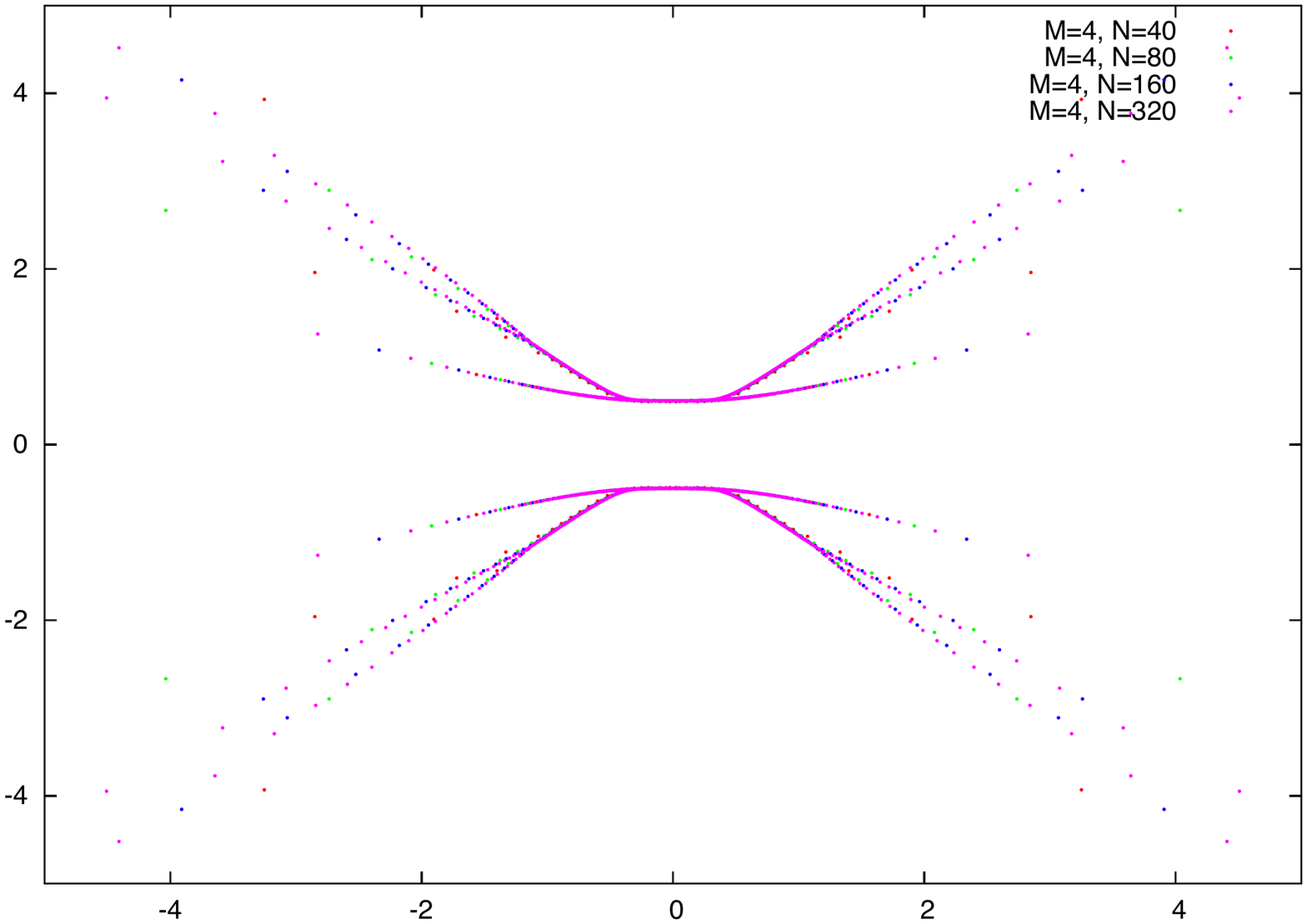}\includegraphics[scale=0.3]{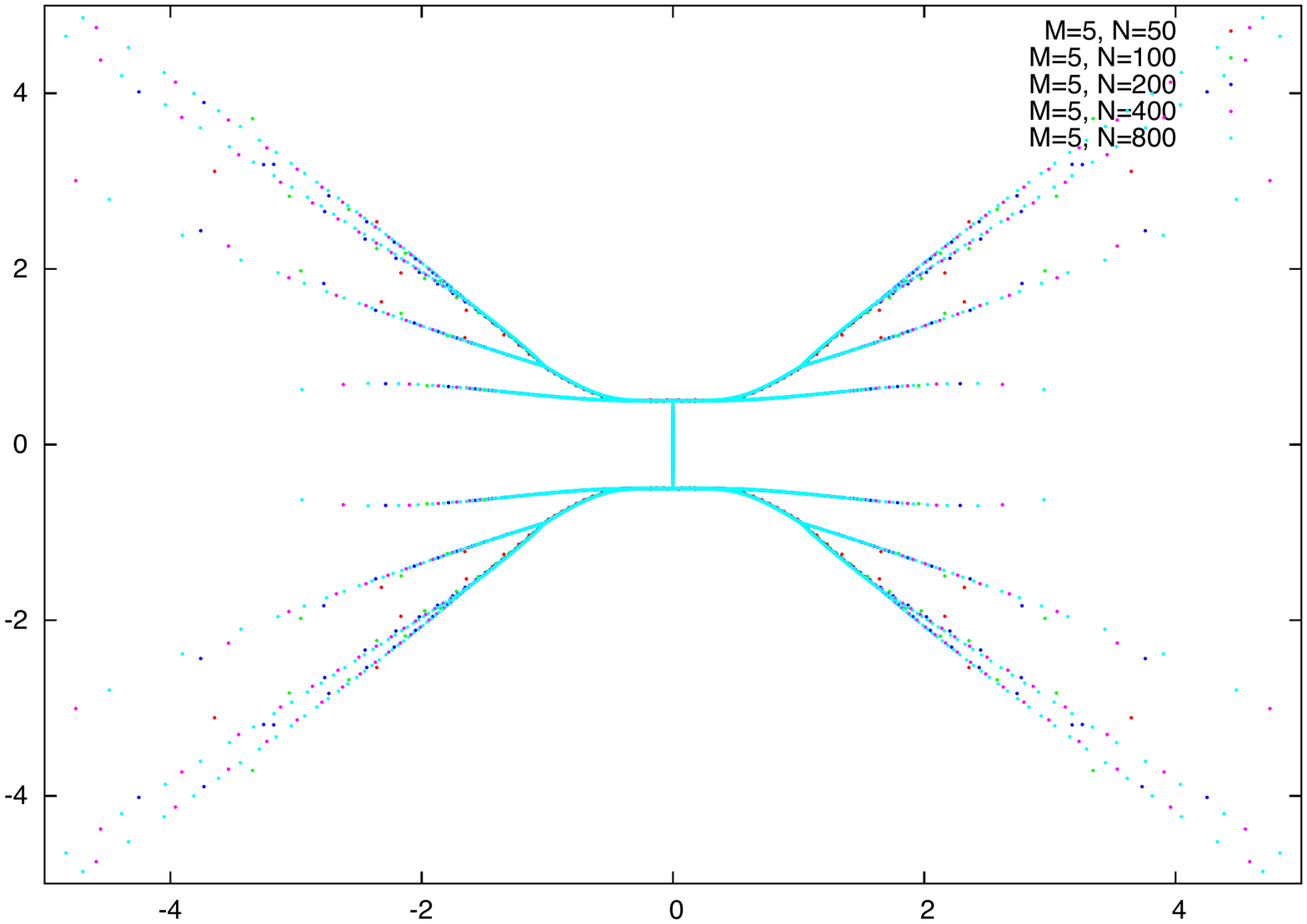}
\includegraphics[scale=0.3]{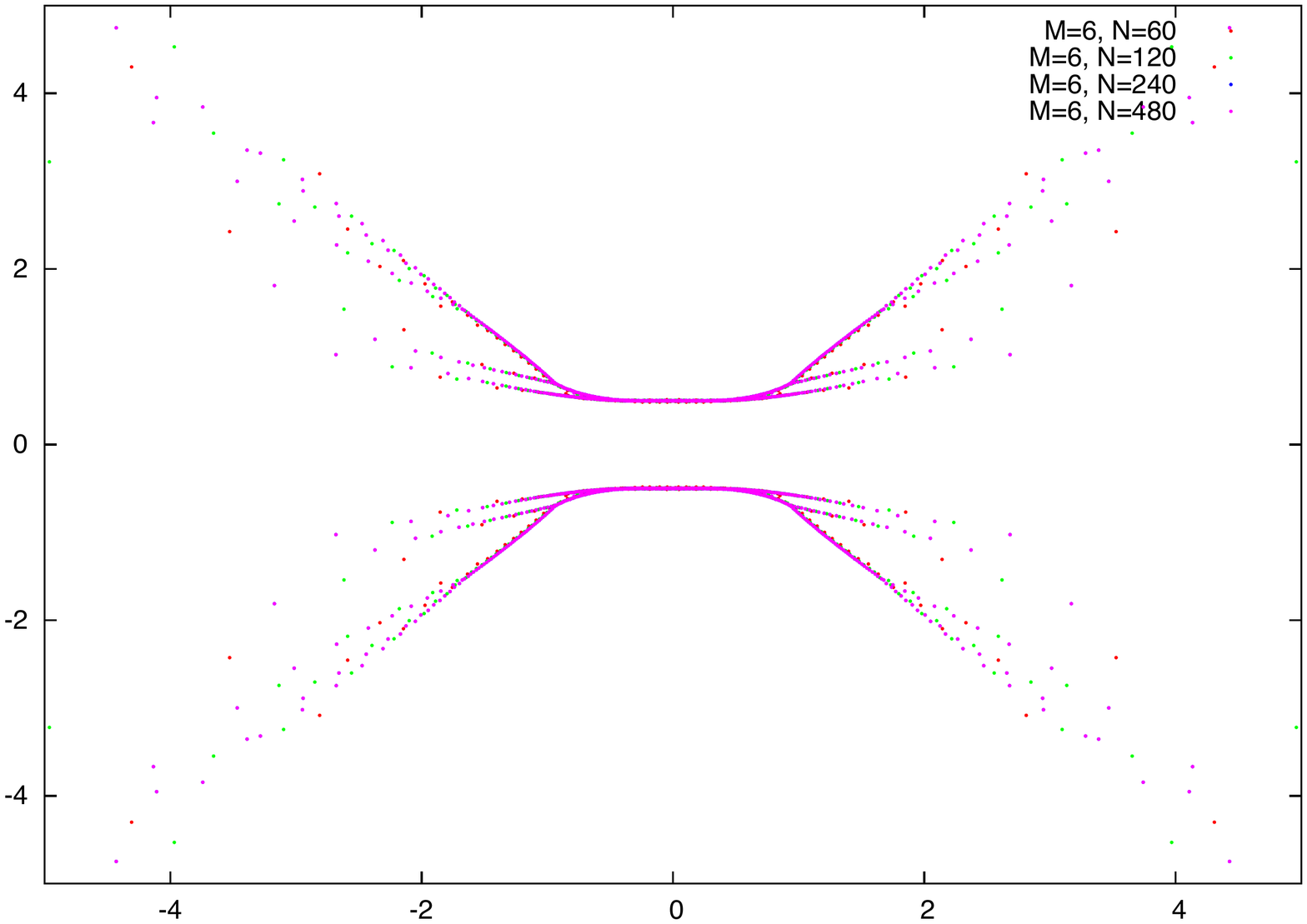}\includegraphics[scale=0.3]{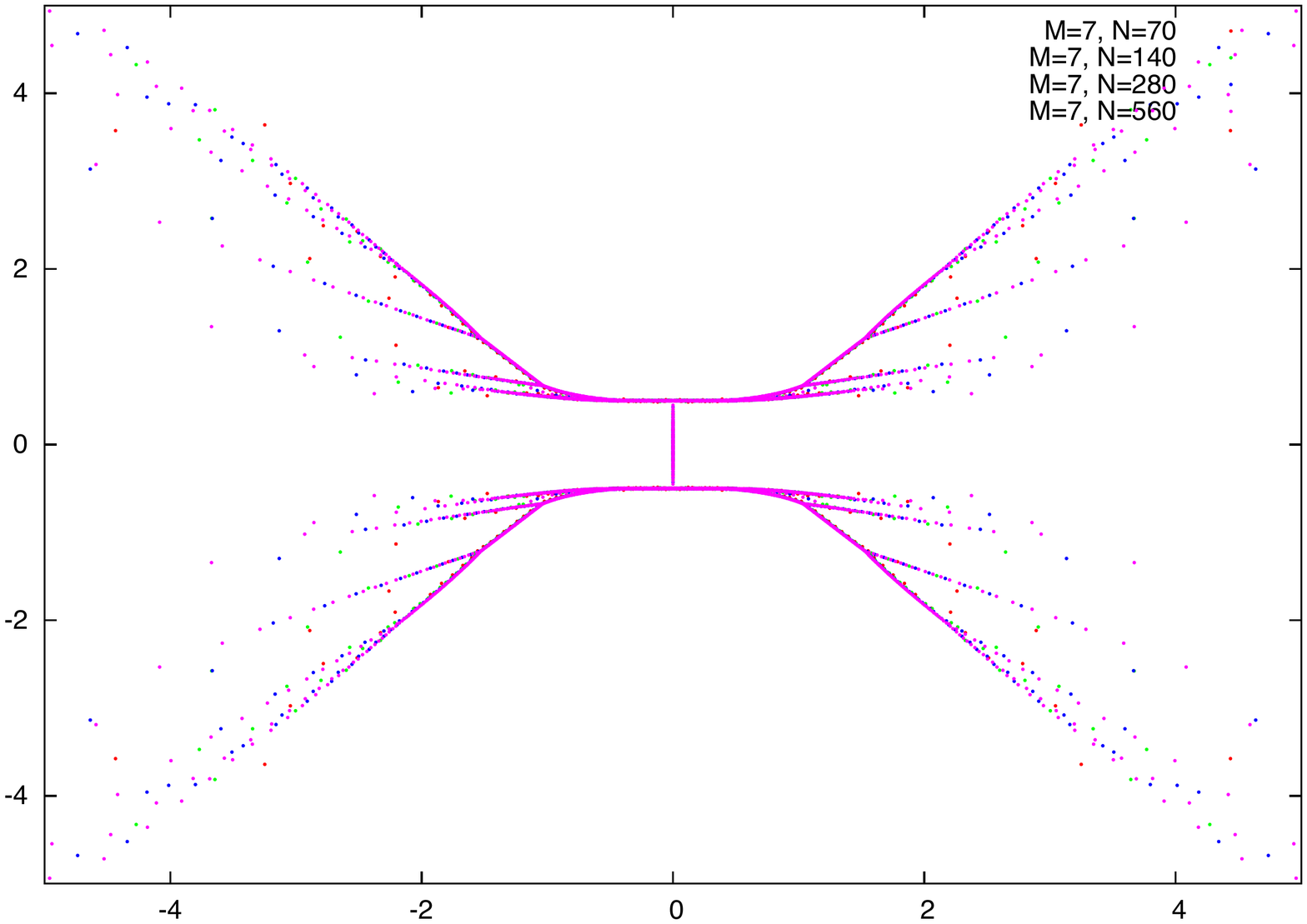}
\includegraphics[scale=0.3]{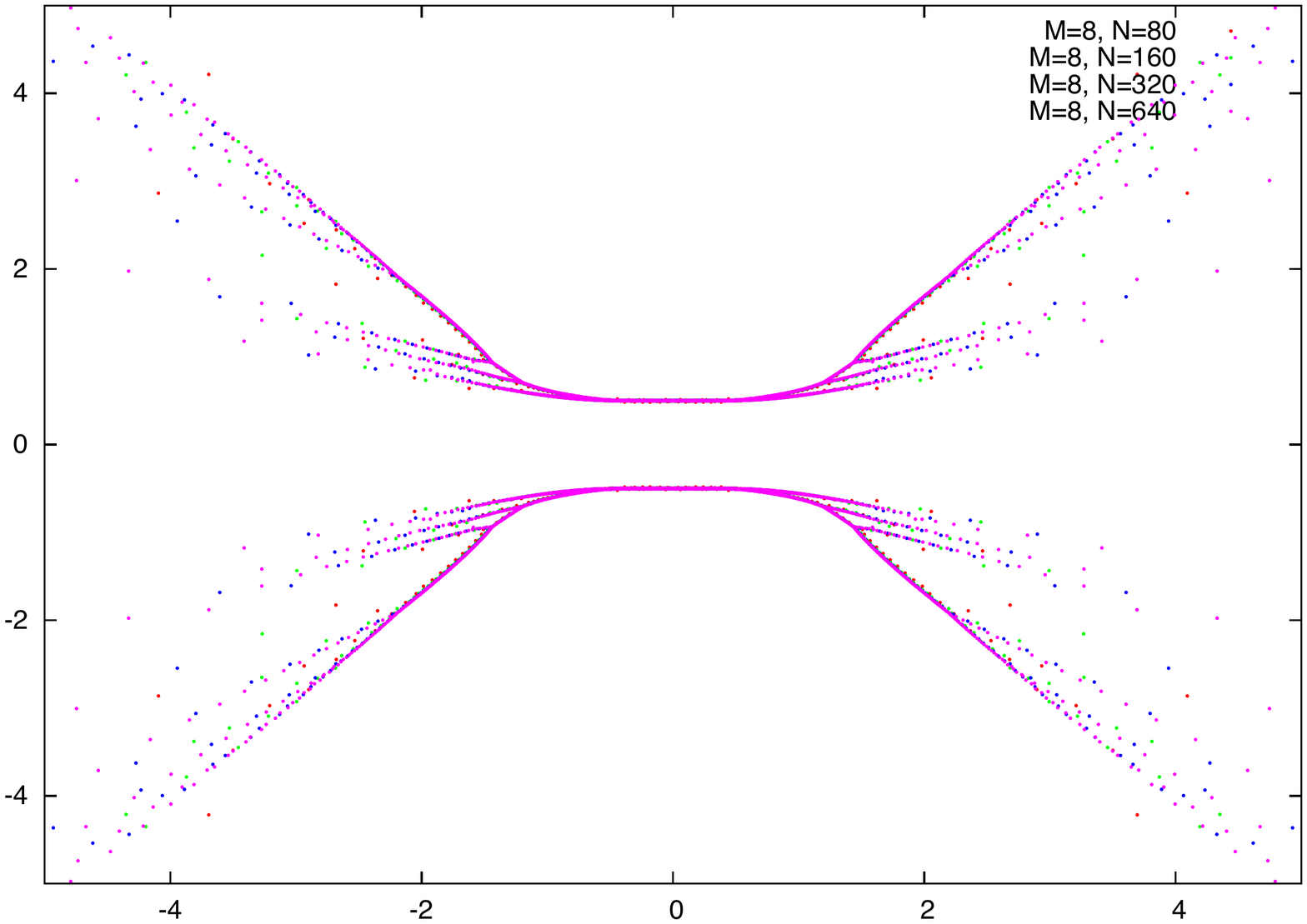}\includegraphics[scale=0.3]{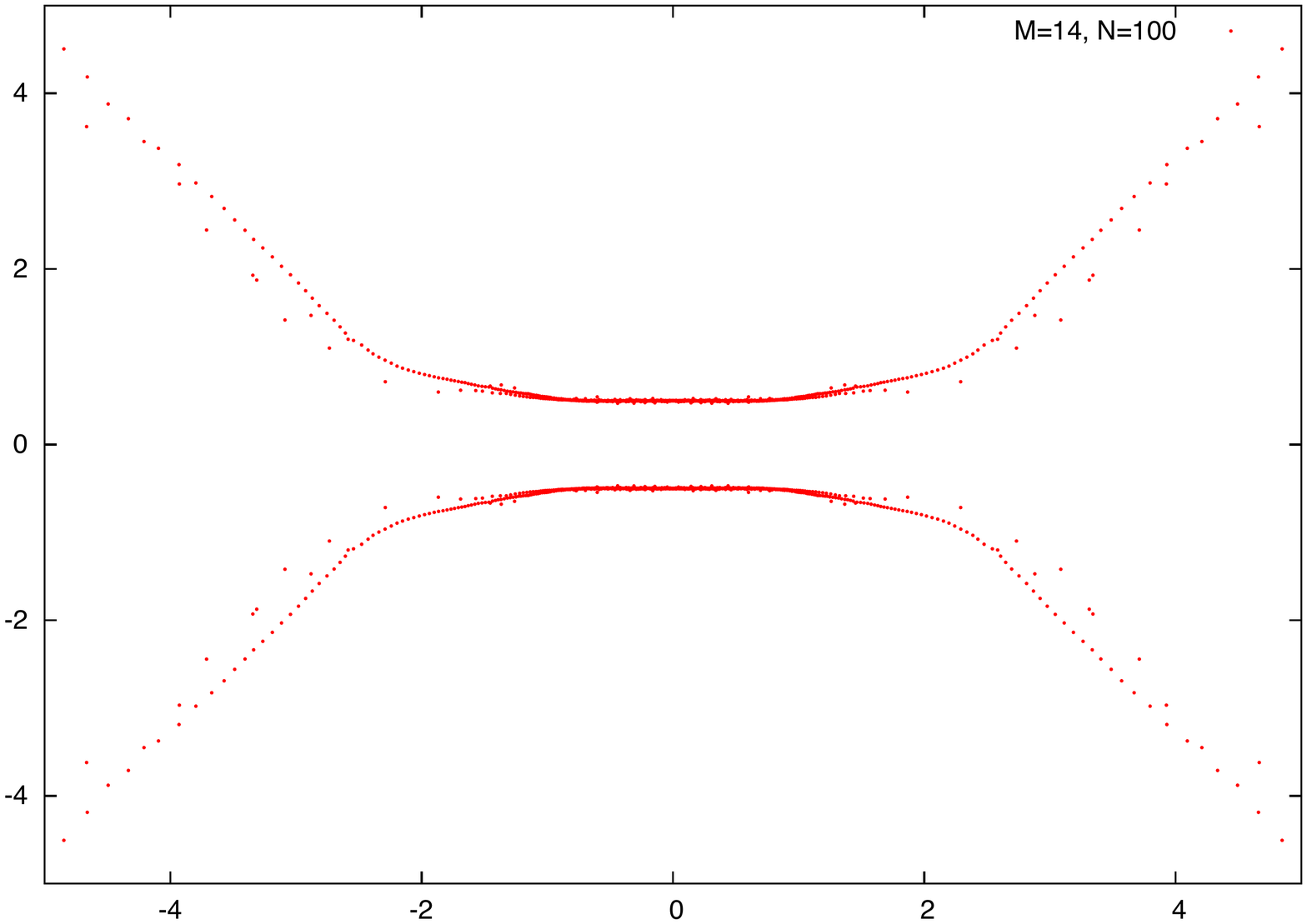}
\caption{Partition function zeros in the complex $z$-plane, shown in reading direction for $M=4,5,\cdots,8$ and $M=14$.}
\label{fig:zeros4-8}
\end{center}
\end{figure}

In figure~\ref{fig:zeros4-8} we show the partition function zeros for $M=4,5,\cdots,8$ with various aspect ratios, namely $N = \rho M$ for $\rho = 10, 20, 40, 80$
(and in one case $\rho = 160$). The lower right corner shows the result of out largest computation with $M=14$ and $N=100$. The partition functions for $M=4,5,6$ have
been computed by the direct application of the explicit formulae (\ref{eq:Z4N})--(\ref{eq:Z6N}), and the remaining results by the algebro-geometric method.
These are polynomials of degree $M N$, and the coefficients can be normalized to be integers by multiplying $Z_{M,N}(z)$ by the overall factor $2^{(M-1)N}$.
Given the very large degree and the size of their coefficients, it is actually a non-trivial task to compute the zeros of these polynomials.
These difficulties are however efficiently overcome by the application of the software {\sc MPSolve} \cite{MPSolve},
which is a multiprecision implementation of the Aberth method \cite{Aberth}. The main advantage of the latter method is that it approximates
all the roots of a univariate polynomial simultaneously.

The zero plots in figure~\ref{fig:zeros4-8} reveal several interesting features. As the aspect ratio $\rho$ grows, the zeros tend to settle on certain curves---the
limiting curves of accumulation points to be discussed further in the next subsection. In the regions close to the origin the finite-$\rho$ effects are small, but
further away their importance increases, and the fine structure of the limiting curves are barely visible even at the largest $\rho$ shown. Moreover, far from
the origin the density of zeros is very scarce. Regarding the case $M=14$, it seems likely that it would develop rich details as those seen in the other plots, provided large
$\rho$ could be accessed. In particular, there are ``stray'' zeros around the central almost-horizontal branches that appear as precursors of multiple branches
and T-points. While all these features could certainly be analyzed at length, we instead move on to the direct determination
of the limiting curves as $\rho \to \infty$.

\subsection{Limiting curves}
\label{sec:condcurv}
In the previous subsection, we have seen that as $N$ increases, the zeros of the partition function accumulate on some curves. Following \cite{Pottszeros1}
we shall refer to these as {\em limiting curves}.
By the Beraha-Kahane-Weiss (BKW) theorem \cite{Beraha4209}, this is a consequence of the form (\ref{eq:ZMN-bf}), or equivalently of (\ref{eq:ZMN-ag}),
that relates the partition function $Z_{M,N}$ to sum over traces of the $N$'th power of the transfer matrix $T_{M,K}(z)$, or of the
corresponding companion matrix $\mathbf{T}_{M,K}(z)$ given by (\ref{compmatrix_tT}).

More precisely, the BKW theorem applies to an expression of the form
\begin{align}
 Z_{M,N}(z) = \sum_i \alpha_i(z) \Lambda_i(z)^N \,,
\end{align}
where we shall refer to the $\Lambda_i(z)$ as eigenvalues, and the $\alpha_i(z)$ as the corresponding multiplicities.
For a given $z$, let us order the eigenvalues by norm, so that $|\Lambda_1(z)| \ge |\Lambda_2(z)| \ge \cdots$, and we call
an eigenvalue $\Lambda_i(z)$ dominant (at $z$) if its norm is maximal, $|\Lambda_i(z)| \le |\Lambda_1(z)|$.
Supposing a mild non-degeneracy condition, the BKW theorem then states that the accumulation set of zeros, as $N \to \infty$,
will form either isolated points or curves. An isolated accumulation point occurs for $z = z_0$, when there is a unique dominant eigenvalue
(i.e., $|\Lambda_1(z_0)| > |\Lambda_2(z_0)|$) and the corresponding multiplicity vanishes (\emph{i.e.}, $\alpha_1(z_0) = 0$).
A curve of accumulation points occurs when there are at least two dominant eigenvalues (\emph{i.e.}, $|\Lambda_1(z)| = |\Lambda_2(z)|$), and the relative
phase $\phi(z) \in \mathbb{R}$ defined by $\Lambda_2(z) = {\rm e}^{i \phi(z)} \Lambda_1(z)$ varies along the curve. The speed of variation
of $\phi(z)$ along the curves can be related to the density of partition function zeros \cite{Pottszeros1}.
Note also that the limiting curves may have T-points or higher-order bifurcations at a point $z_0$ where more than two eigenvalues
are equimodular. We refer to \cite{Pottszeros1} for more details on the BKW theorem and the detailed analysis of the generic setup.

In our context, $\alpha_i(z)$ and $\Lambda_i(z)$ depend on $M$ and $K$, and moreover $\alpha_i(z) = M-2K+1$ are simply constants.
Therefore all accumulation points form curves, and not isolated points, in agreement with the observations of the preceding subsection.
To trace these curves for a given $M$, we use an approach for identifying the loci of equimodularity that is described in appendix~\ref{sec:codes}.
This consists in two steps: first we identify some points of equimodularity by a direct search (\emph{e.g.}, along suitably chosen straight lines),
and second we trace the equimodular curves starting from each of those points, using a procedure explained in the appendix.
While this approach may fail to detect very small curves of accumulation points, we believe to have obtained complete results for $M \le 8$.

\begin{figure}[h!]
\begin{center}
\includegraphics[scale=0.3]{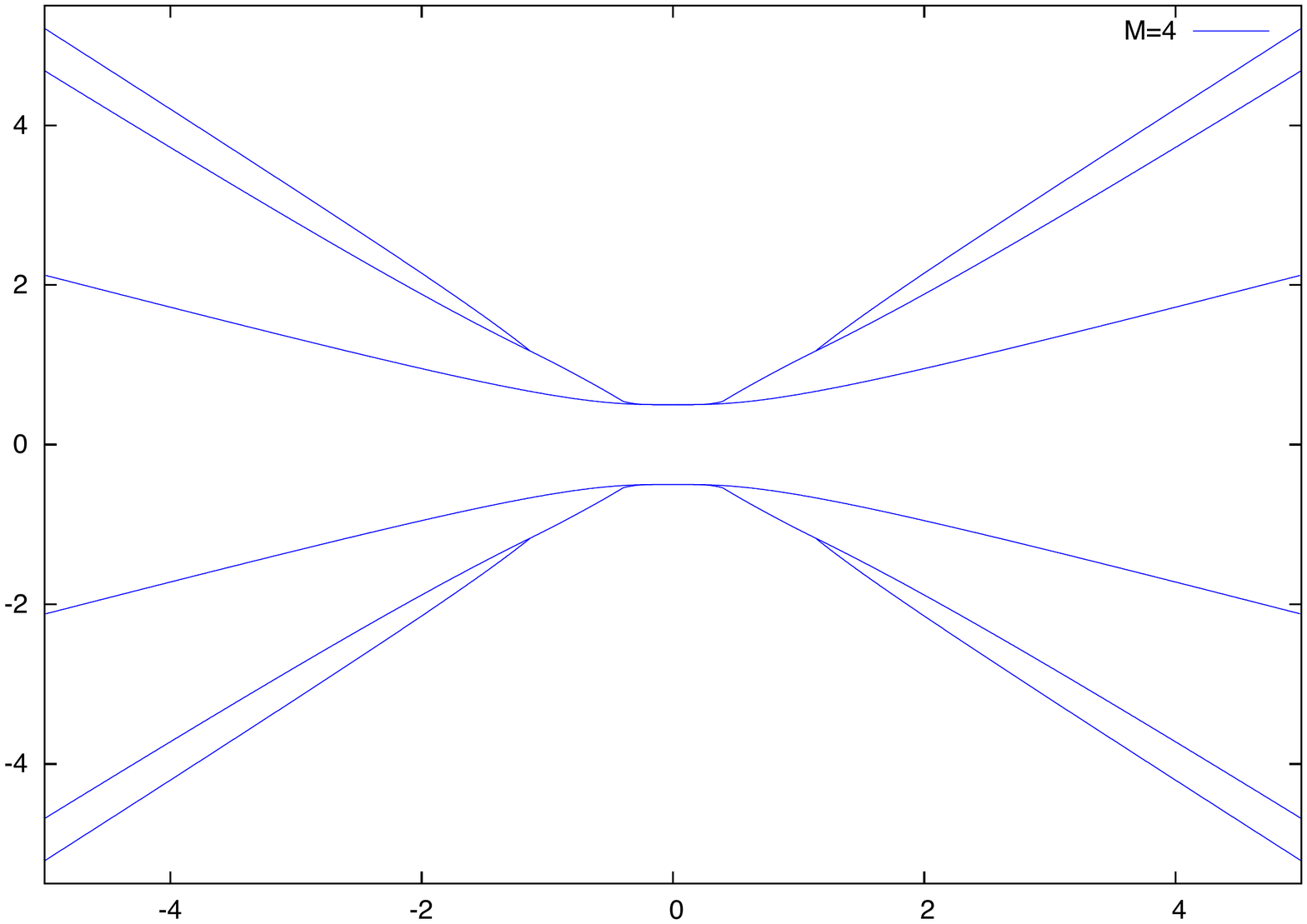}\includegraphics[scale=0.3]{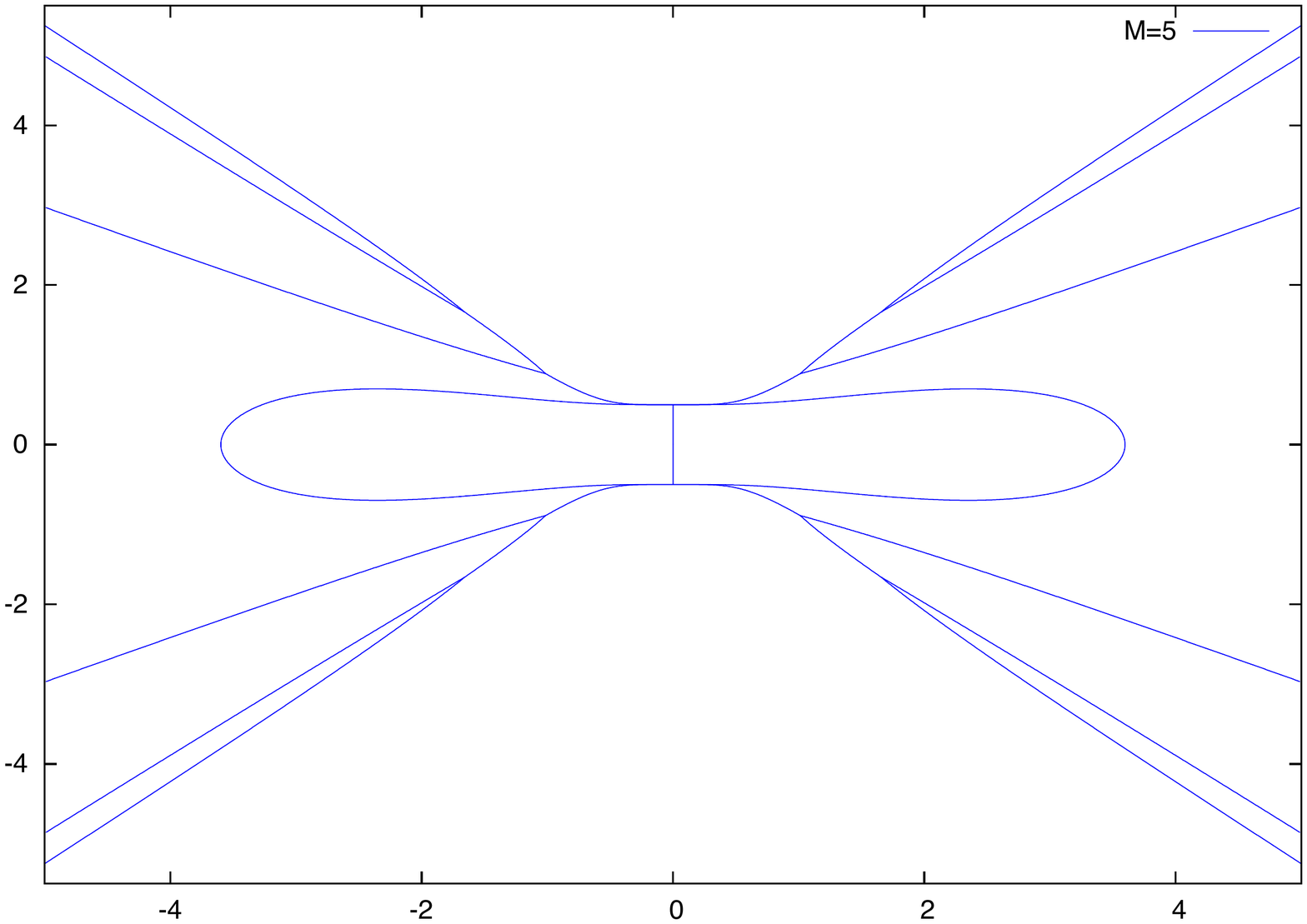}
\includegraphics[scale=0.3]{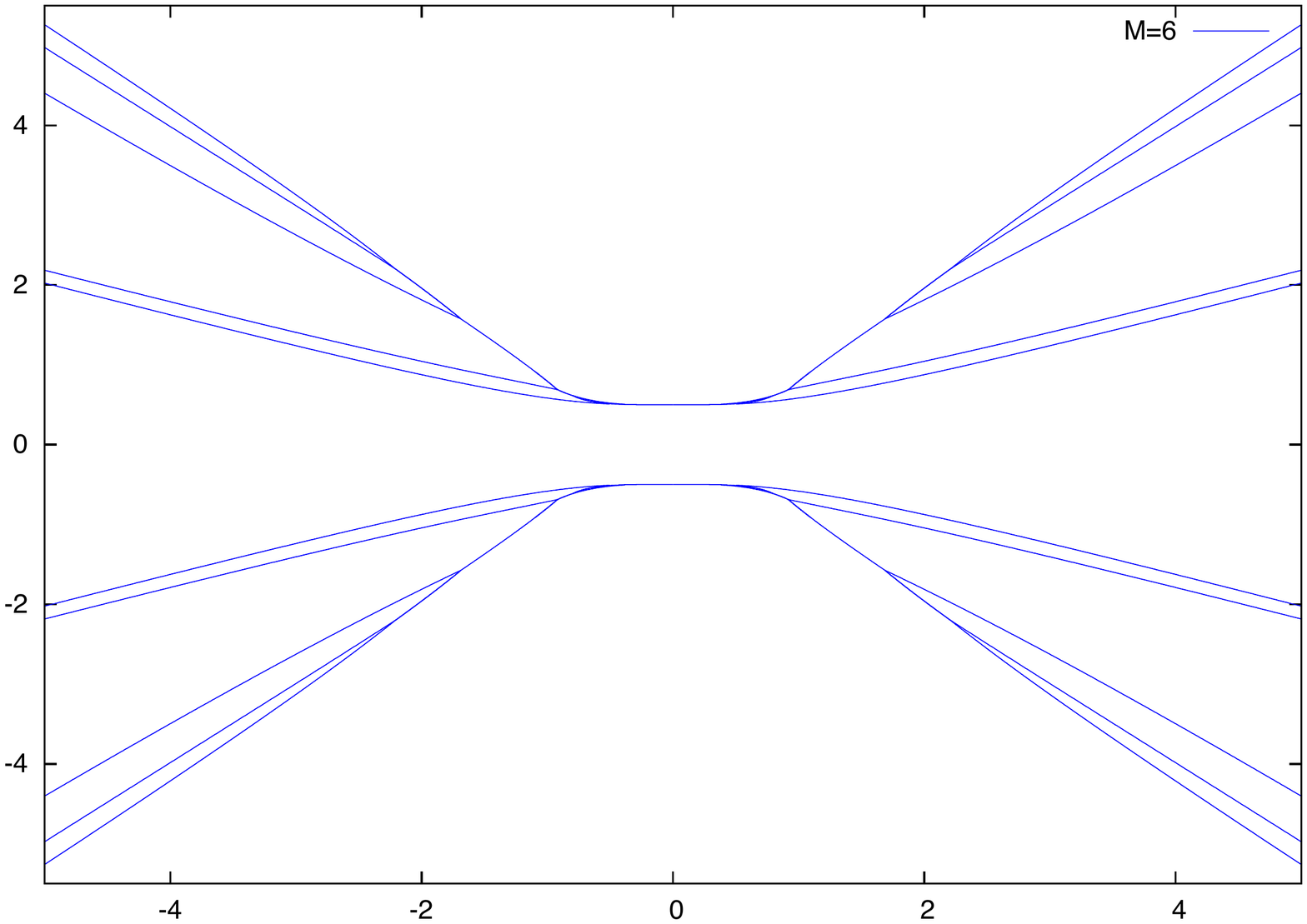}\includegraphics[scale=0.3]{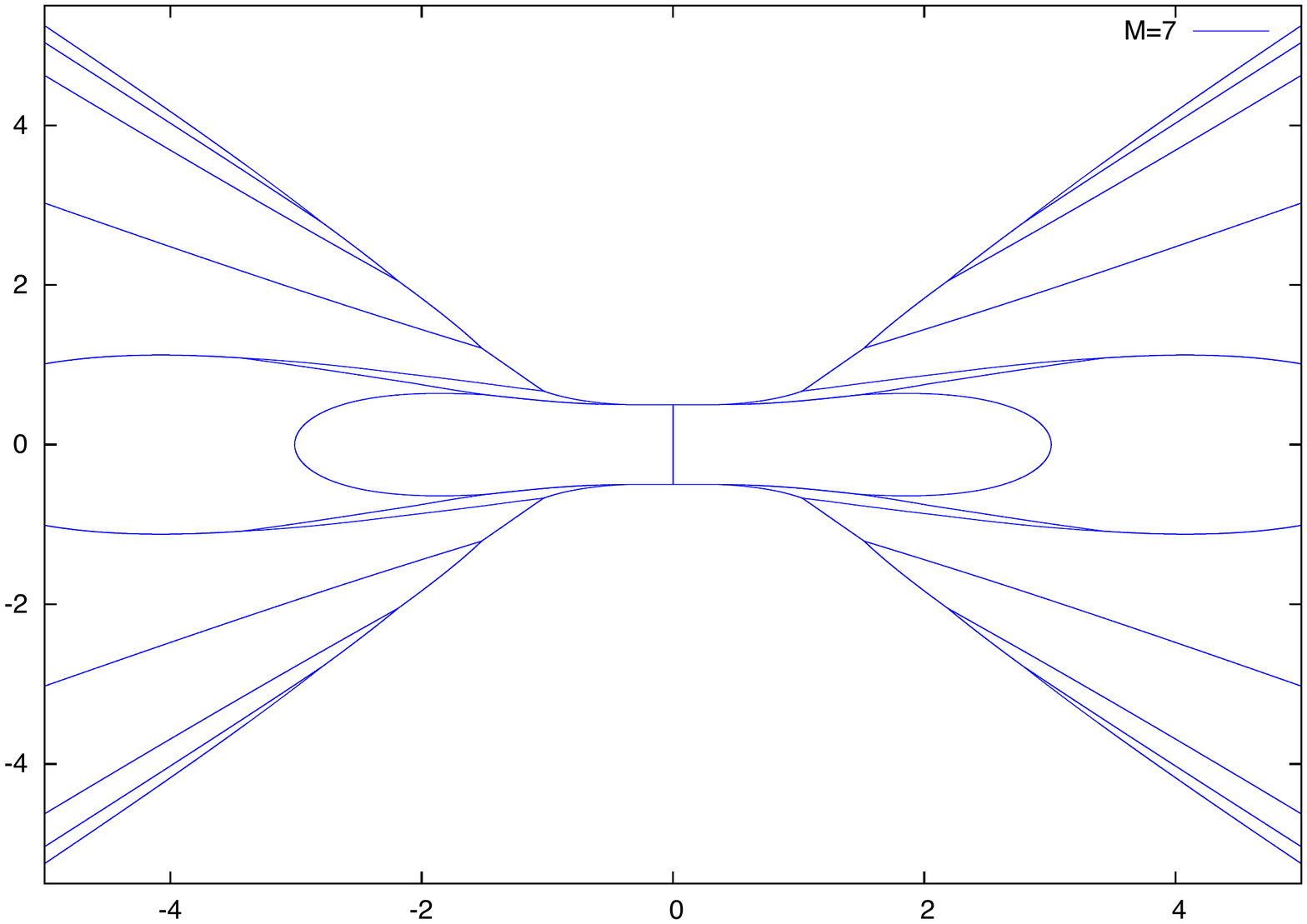}
\includegraphics[scale=0.3]{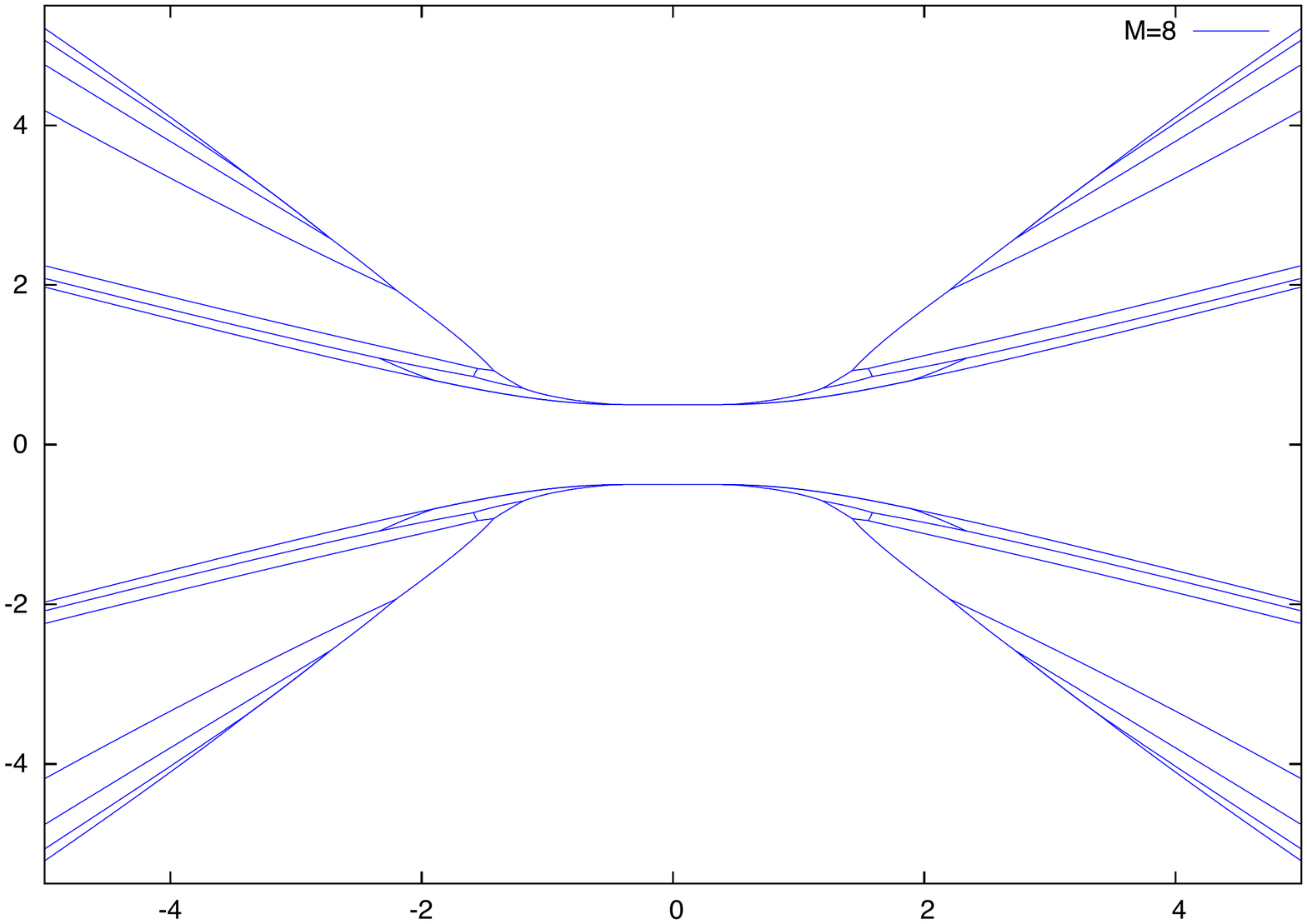}\hphantom{\includegraphics[scale=0.3]{8.pdf}}
\caption{Limiting curves of accumulation points of partition function zeros in the complex $z$-plane, shown in reading direction for $M=4,5,\cdots,8$.}
\label{fig:curves4-8}
\end{center}
\end{figure}

The resulting limiting curves for $M=4,5,\cdots,8$ are shown in figure~\ref{fig:curves4-8}. A number of qualitative features can be read off from these examples.
First, the curves are invariant under the independent sign changes of ${\rm Re}\, z$ and ${\rm Im}\, z$. Second, they all
contain the point $z = i/2$. Third, they contain a number of branches extending to infinity; the number of such branches
within each quadrant appears to be $3, 3, 5, 4, 7$ for the sizes considered. Fourth, for even $M$ the curves do not intersect the real axis, while
for odd $M$ they contain an exact vertical ray $z \in [-i/2,i/2]$. For odd $M$, there are further intersections with the real axis, namely
$z \simeq \pm 3.5970$ for $M=5$, as well as $z \simeq \pm 3.0096$ and $z \simeq \pm 6.0139$ for $M=7$.
Fifth, we only find T-points and now higher-order bifurcations.

\begin{figure}[h!]
\begin{center}
\includegraphics[scale=0.3]{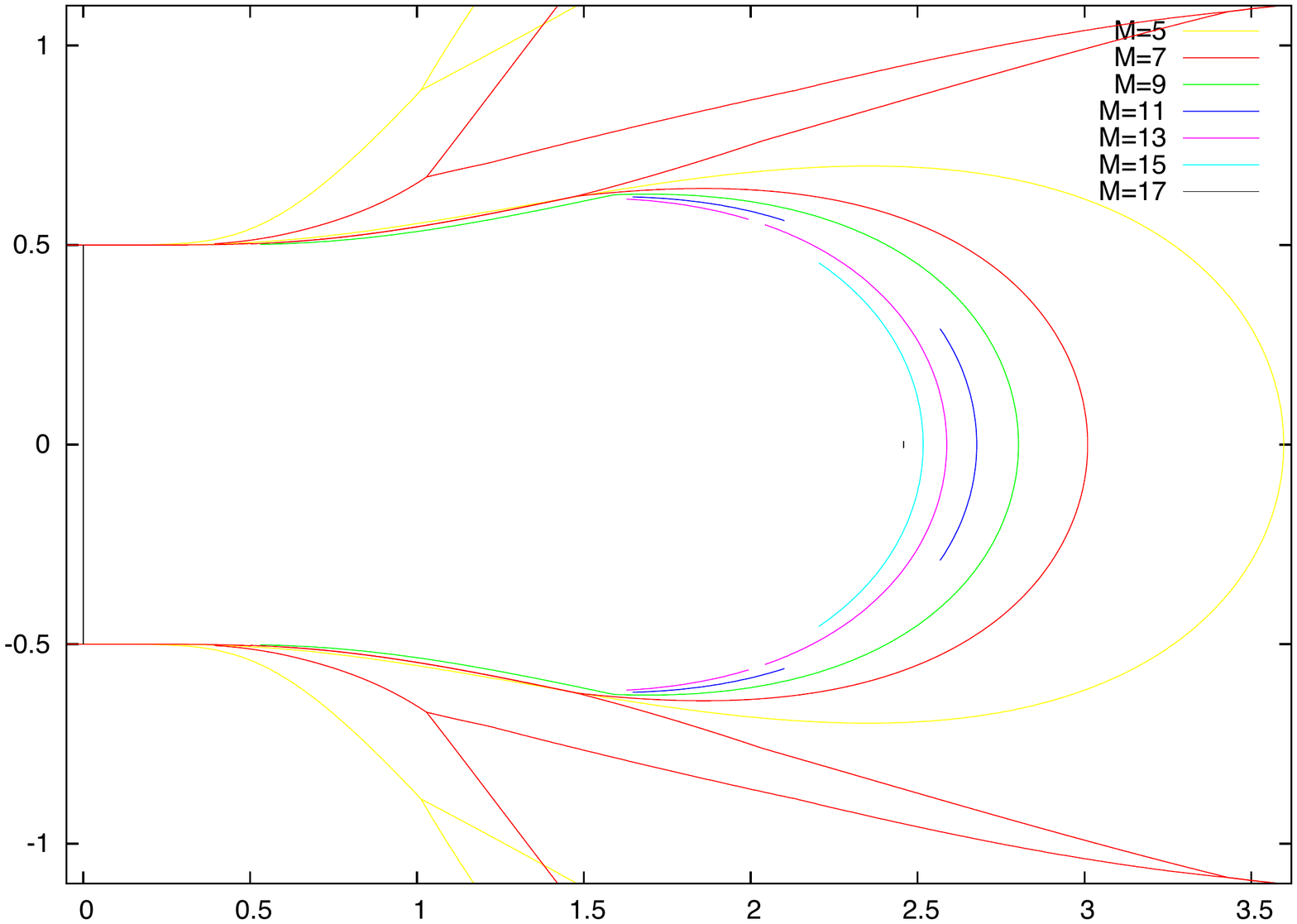} \qquad \includegraphics[scale=0.5]{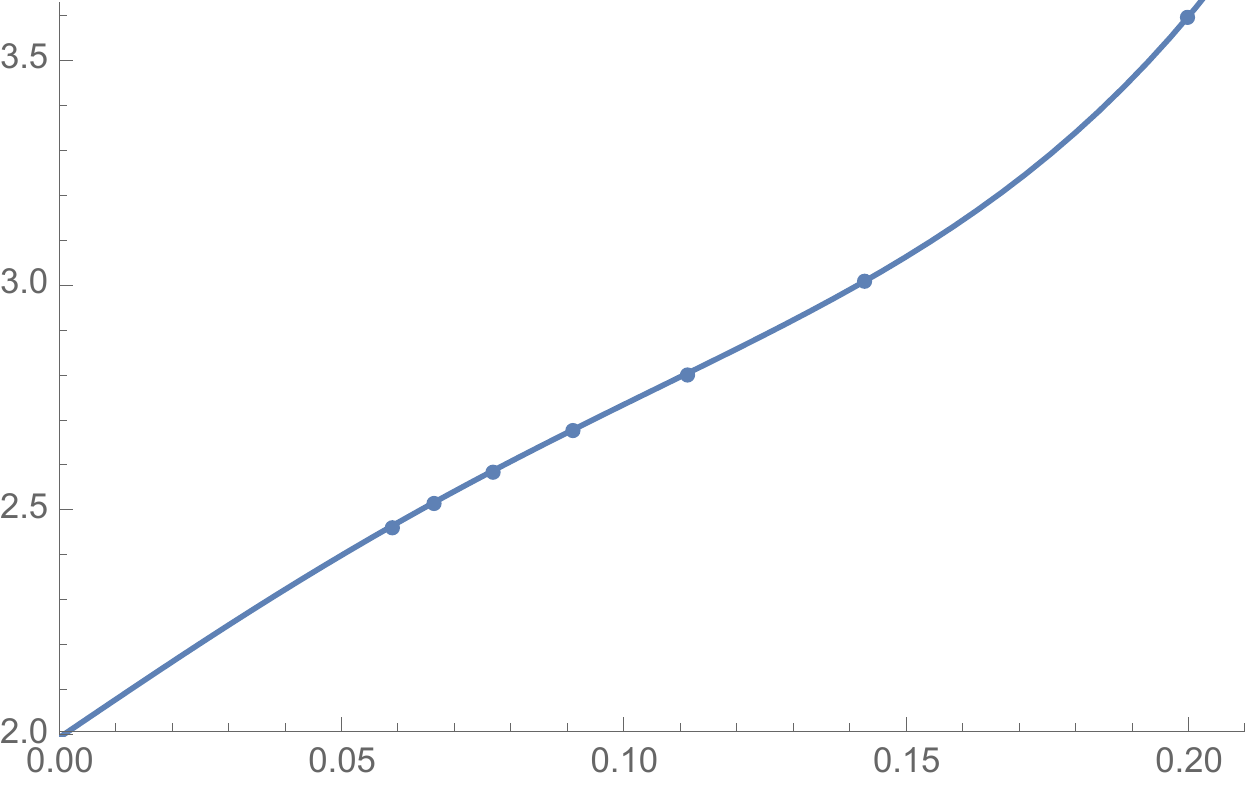}
\caption{Left panel: Partial results for the inner parts of the curves of accumulation points for odd $M=5,7,\cdots,17$.
Right panel: Extrapolation of the first real intersection, using a polynomial fit in the variable $1/M$.}
\label{fig:curves-odd-M}
\end{center}
\end{figure}

A close scrutiny of figure~\ref{fig:curves4-8} reveals that the limiting curves some tantalizingly tiny features for $M \ge 6$, including almost-parallel curves
and short stems linking the various branches. We have taken great case to represent (what we believe to be) all such features.

From a numerical point of view, the diagonalization computations become increasingly difficult as we approach the highly degenerate points
$z = \pm i/2$. Practical details about the computational approach to limiting curves can be found in appendix~\ref{sec:codes}.

\begin{figure}[h!]
\begin{center}
\includegraphics[scale=0.6]{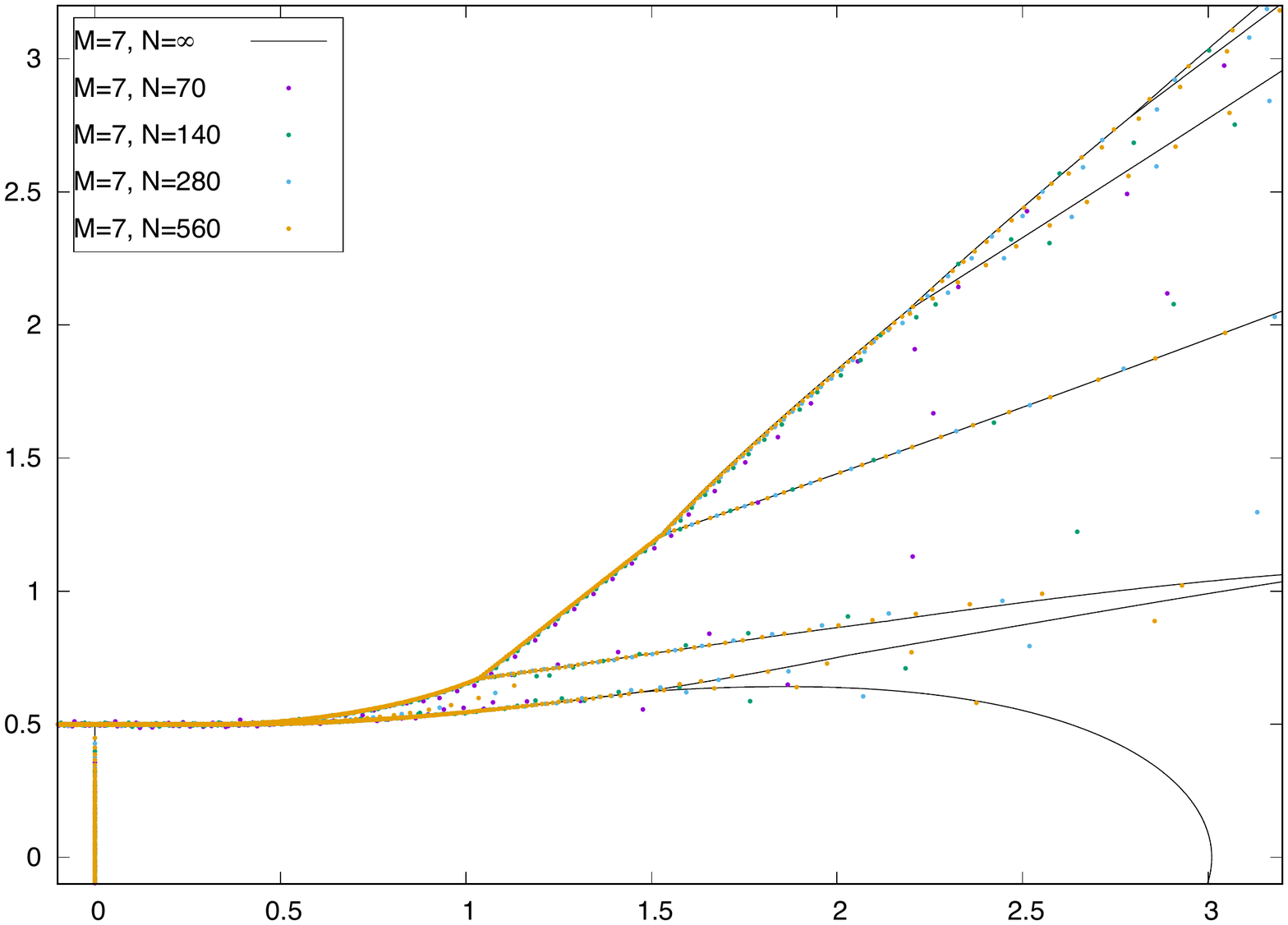}
\caption{Comparison between the limiting curves and the partition function zeros for $M=7$.}
\label{fig:curves-M7-compare}
\end{center}
\end{figure}

Comparing figures~\ref{fig:zeros4-8}--\ref{fig:curves4-8} gives convincing evidence that the partition function zeros indeed accumulate on the limiting curves,
in the limit of large aspect ratio $\rho \to \infty$. That this is indeed the case is proved by the BKW theorem. However, it is also clear that some parts of the limiting curves are
very scarcely populated by the zeros, even for the large values of $\rho$ shown in figure~\ref{fig:zeros4-8}. Moreover, some of the fine details of
the limiting curves are hardly discernable on the plots of zeros, such as the short stem-like pieces connecting the almost-parallel branches for $M=8$ or the
(barely visible) sliver-shaped enclosed region for $M=6$. Figure~\ref{fig:curves-M7-compare} shows a comparison between the limiting curves and the $M=7$ partition function zeros for various aspect ratios.

We have examined the curves delimiting the region close to the origin in some more detail. For even $M$, it has a corridor-like aspect, with the
branches containing $z = \pm i/2$ appearing to become more horizontal as $M$ increases. For odd $M$, it forms an elongated bubble, with the
above-mentioned vertical ray in the middle, whose upper and lower boundaries tend as well to become more horizontal as $M$ increases. To illustrate
the size-dependence of this bubble, we have produced {\em partial} results for this part of the curves for higher, odd values of $M$ (up to $M=17$),
as shown in figure~\ref{fig:curves-odd-M}. The first intersection with the real axis appears to extrapolate to $z_\star = 2$ as $M \to \infty$ (see the
right panel of figure~\ref{fig:curves-odd-M}). Based on
this, we conjecture that the enclosed region tends to a rectangle, given by $0 < {\rm Re}\, z < 2$ and $-1/2 < {\rm Im}\, z < 1/2$, in the thermodynamic limit,
for $M$ odd.

%%%%%%%%%%%%%%%%%%%%%%%%%%%%%%%%%%%%%%%%%%%%%%%%%%%%%%%%%%%%%%%%%%%%%%%%
\section{Primary decomposition}
\label{sec:primaryDec}
%%%%%%%%%%%%%%%%%%%%%%%%%%%%%%%%%%%%%%%%%%%%%%%%%%%%%%%%%%%%%%%%%%%%%%%%
Let us summarize what we have achieved so far in computing the torus partition function of the six-vertex model. Computing the partition function by brute force, we need to work with matrices of dimension $d_{M,K}$ within the spin sector of $K$ magnons. Using Bethe ansatz and the algebro-geometric method, we are able to reduce the problem to the computation of \emph{companion matrices} of dimension $\mathcal{N}_{M,d}=d_{M,K}-d_{M,K-1}$. This reduction in the dimensions of the matrices makes use of the full $\mathfrak{su}(2)$ symmetry of the theory. Recall that we classify the Bethe states as primary states and their descendants with respect to the $\mathfrak{su}(2)$ algebra. Since all the descendant states of a given primary state have the same eigenvalue of the transfer matrix, we can focus on the primary states only. The number of primary states is much less than the total number of states in a given spin sector.\par

Can we do better? Notice that we have not yet exploited all the symmetries of the model. For example, the model is also invariant under a lattice translation. This symmetry leads to the the total momentum of the Bethe state being quantized, taking only a finite number of possible values. Therefore, apart from decomposing the Hilbert space according to spin sectors, we can also decompose the Hilbert space according to \emph{momentum sectors}, \emph{i.e.}, states with different values of the lattice momentum. These two decompositions can be performed simultaneously and leads to even smaller companion matrices. This will greatly enhance the efficiency of our approach.\par

Mathematically, the decomposition with respect to momentum sectors is intimately related to \emph{primary decomposition} and \emph{algebraic extension} in algebraic geometry. Physically, this decomposition also allows us to probe much deeper into the solution space of BAE and find new structures that have not been studied in the literature. In this section, we discuss the decomposition of the solution space with respect to momentum sectors. We first introduce the notion of primary decomposition on $\mathbb{Q}$ from some interesting observations about the partition function. We will see that it is useful to perform the decomposition on a larger field $\mathbb{Q}(i,\xi_M)$ obtained by an algebraic extension. In addition, exploiting Galois theory in the current context, we will show that many of the subspaces after the decomposition are actually related by the Galois group, and it is thus sufficient to perform the computation for a representative. The decomposition together with Galois theory lead to a huge boost in the efficiency of our computation. More details are given in appendix~\ref{sec:Gr} and an upcoming publication \cite{Jiang:AG_BAE2}.

%%%%%%%%%%%%%%%%%%%%%%%%%%%%%%%%%%%%%%%%%%%%%%%%%%%%%%%%%%%%%%%%%%%%%%
\subsection{Primary decomposition over $\mathbb{Q}$}
\label{sec:primaryDecQ}
%%%%%%%%%%%%%%%%%%%%%%%%%%%%%%%%%%%%%%%%%%%%%%%%%%%%%%%%%%%%%%%%%%%%%%
To see that the solution space of the BAE has more structure, we take a careful look at the closed-form results of the partition function in (\ref{eq:Z6N}). We can see that it is natural to group some of the terms together since they take very similar forms. For example, we can group the following four eigenvalues in (\ref{eq:Z6N}):

\begin{footnotesize}
\begin{align}
\label{eq:Lambda1234}
\Lambda_1=2z^6+\frac{5}{2}z^4-\frac{\sqrt{54-6\sqrt{17}}}{2}z^3+\frac{(2\sqrt{17}-3)}{8}z^2-\frac{\sqrt{54-6\sqrt{17}}(3+\sqrt{17})}{16}z
 +\frac{(9+2\sqrt{17})}{32} \,, \\\nonumber
\Lambda_2=2z^6+\frac{5}{2}z^4+\frac{\sqrt{54-6\sqrt{17}}}{2}z^3+\frac{(2\sqrt{17}-3)}{8}z^2+\frac{\sqrt{54-6\sqrt{17}}(3+\sqrt{17})}{16}z
 +\frac{(9+2\sqrt{17})}{32} \,, \\\nonumber
\Lambda_3=2z^6+\frac{5}{2}z^4-\frac{\sqrt{54+6\sqrt{17}}}{2}z^3-\frac{(2\sqrt{17}+3)}{8}z^2-\frac{\sqrt{54+6\sqrt{17}}(\sqrt{17}-3)}{16}z
 +\frac{(9-2\sqrt{17})}{32} \,, \\\nonumber
\Lambda_4=2z^6+\frac{5}{2}z^4+\frac{\sqrt{54+6\sqrt{17}}}{2}z^3-\frac{(2\sqrt{17}+3)}{8}z^2-\frac{\sqrt{54+6\sqrt{17}}(\sqrt{17}-3)}{16}z
 +\frac{(9-2\sqrt{17})}{32} \,. \nonumber
\end{align}
\end{footnotesize}
One can check that although each $\Lambda_1,\cdots,\Lambda_4$ is complicated and has irrational coefficients for generic $z$, their symmetric power sums
\begin{align}
\label{eq:symm-pow-sum}
\Lambda_1^n+\Lambda_2^n+\Lambda_3^n+\Lambda_4^n
\end{align}
are always polynomials whose coefficients are rational numbers! Since each $\Lambda_i$ corresponds to a solution of the BAE or the $TQ$-relation, this implies that we can group the four corresponding solutions of the BAE. Notice that we cannot make the decomposition further on $\mathbb{Q}$. If we further divide the four solutions into two groups, say $\Lambda_1,\Lambda_2$ and $\Lambda_3,\Lambda_4$, then the coefficients of the symmetric power sums $\Lambda_1^n+\Lambda_2^n$ and $\Lambda_3^n+\Lambda_4^n$ are no longer rational. This implies that these four solutions form an \emph{irreducible} or \emph{primary} block on $\mathbb{Q}$. Similarly, the remaining terms in (\ref{eq:Z6N}) can be divided into such primary blocks. In geometrical terms, this grouping is equivalent to decomposing an affine variety into independent components. Such an operation is called \emph{primary decomposition} in algebraic geometry. We refer to appendix~\ref{sec:Gr} for more details.\par

Given an ideal, it is straightforward to compute the primary decomposition using standard algorithms. To understand the physical meaning of primary decomposition, we now analyze the example $M=6$ carefully.

\paragraph{An example: $M=6$.} The result of the primary decomposition is given in table~\ref{PD_M6}.
\begin{table}[h!]
\begin{center}
\begin{tabular}{c|l}
  \hline
   $K$ & $M=6$ \\
    \hline
  1 & 5=1+2+2 \\
  2 & 9=1+2+2+4 \\
  3 & 5=1+2+2 \\
  \hline
\end{tabular}
\caption{Primary decomposition of solution space of BAE with $M=6$. The numbers on the right-hand sides of each line represent the dimension of each subspace.}
\label{PD_M6}
\end{center}
\end{table}
Let us consider the spin sector $K=2$. From table~\ref{PD_M6} we see that there are 9 physical solutions for $M=6,K=2$, and that these solutions can be divided into four groups, with dimensions 1,2,2,4. In particular, the dimension-4 subspace corresponds to the four eigenvalues given in (\ref{eq:Lambda1234}).\par

As we alluded to before, this decomposition is related to the lattice translational invariance which is generated by the shift operator $U=e^{iP}$. As an operator, it is related to the transfer matrix as
\begin{align}
U_M=(-i)^M T_M(i/2).
\end{align}
For a closed spin chain of length $M$, the allowed eigenvalues of the shift operator are
\begin{align}
\exp\left(\frac{2\pi i\ell}{M}\right),\qquad \ell=1,\cdots,M.
\end{align}
Let us denote the four subspaces as $\rA,\rB,\rC,\rD$; we can then compute the values of $\ell$ for $M=6$ for each subspace. The result is shown in table~\ref{tab:ABCD}.
\begin{table}[h!]
\begin{center}
\begin{tabular}{|c|c|l|c|}
  \hline
    & dimensions & values of $\ell$ & eigenvalues of $U^3$ \\
  \hline
  $\rA$ & 1 & $\{3\}$ & $-1$ \\
  $\rB$ & 2 & $\{6,6\}$ & $+1$ \\
  $\rC$ & 2 & $\{1,5\}$ & $-1$ \\
  $\rD$ & 4 & $\{2,2,4,4\}$ & $+1$ \\
  \hline
\end{tabular}
\caption{The values of $\ell$ and eigenvalues of $U^3$ for physical solutions of BAE with $M=6,K=2$ in the four subspaces under primary decomposition. All the solutions in the same subspace have the same eigenvalues of $U^3$.}
\label{tab:ABCD}
\end{center}
\end{table}
We see that the value of $\ell$ is not the same within each subspace. However, if we compute the eigenvalues of the operator $U^3$, we find that they are the same within each subspace, as is shown in the last column of table~\ref{tab:ABCD}. This is due to the fact that we work on the field $\mathbb{Q}$. Let us denote $\xi_M=\exp(2\pi i/M)$. It is clear that $\xi_M^{\ell}$ is not always rational for all $\ell=1,\cdots,M$. Therefore one cannot perform the decomposition over momentum sector completely on $\mathbb{Q}$. For each $M$, we can find the smallest integer $1\le m\le M$ such that the all the eigenvalues of $U^m$ are rational. Then we can perform the decomposition with respect to the eigenvalues of $U^m$. As a result, we can restrict ourselves to each subspace by imposing an additional constraint on the original BAE. In our example, for $M=6,K=2$, the additional constraints for the four subspaces are
\begin{align}
&\rA:\qquad U=(-i)^6 t(i/2)=-1,\\\nonumber
&\rB:\qquad U=(-i)^6 t(i/2)=+1,\\\nonumber
&\rC:\qquad U^3=\left[(-i)^6 t(i/2) \right]^3=-1,\quad U=(-i)^6 t(i/2)\ne-1,\\\nonumber
&\rD:\qquad U^3=\left[(-i)^6 t(i/2) \right]^3=+1,\quad U=(-i)^6 t(i/2)\ne+1.
\end{align}
Notice that we need to include the constraints $U\ne\pm 1$ in the cases $\rC$ and $\rD$ because otherwise they will include cases $\rA$ and $\rB$.

\paragraph{Algebraic extension.}
As we see in the previous discussion, the primary decomposition is related to the decomposition with respect to the lattice momentum. Due to the fact that $\xi_M^{\ell}$ is not always a rational number, we cannot perform the decomposition completely. However, we are not constrained to work on the field $\mathbb{Q}$. If we extend the field slightly to include $\xi_M$ and perform the primary decomposition on the extended field, then the decomposition with respect to the lattice momentum can be performed \emph{completely}. More precisely, the extended field will turn out to be $\mathbb{F}_M=\mathbb{Q}(i,\xi_M)$ where $i$ is the imaginary unit.\par

After the decomposition into momentum sectors, we have $M$ subspaces\footnote{Some of the subspaces might not exist for some values of $M$ and $K$, as witnessed by the tables in the next subsection.} (corresponding to $\ell=1,\cdots,M$) in each spin sector $K$.
In principle, we need to compute the Gr\"obner basis and companion matrices of all subsystems. However, we will show that by making use of {\it the Galois group} of the algebraic extension, we just need to calculate a very few BAE subsystems. We get the contribution from all subsystems by the Galois group actions.

\subsection{Primary decomposition over $\mathbb{F}_M$}
\label{sec:primaryDecFM}
We explain in detail how to implement the decomposition over $\mathbb{F}_M$ in practice. Since we work with the rational $Q$-system, it is most convenient to express the momentum condition in terms of Baxter polynomials $Q(z)$ as
\begin{align}
\label{eq:momentumcond}
\prod_{j=1}^K\frac{u_j+i/2}{u_j-i/2}=\frac{Q_{\mathbf{s}}(-i/2)}{Q_{\mathbf{s}}(+i/2)}=\xi_M^{\ell}\,,\qquad \ell=1,\cdots,M \,.
\end{align}
Here $\mathbf{s}=\{\rs_0,\cdots,\rs_{K-1}\}$ are the unknown coefficients that we solve for in the rational $Q$-system. Alternatively, we can write $Q_{\mathbf{s}}(z)$ as
\begin{align}
Q_{\mathbf{s}}(z)=\prod_{j=1}^K (z-u_j) \,,
\end{align}
where $\{u_1,\cdots,u_j\}$ are the Bethe roots. When the solution of BAE is \emph{singular}, \emph{i.e.}, two of the Bethe roots are $\pm i/2$, we have $Q_{\mathbf{s}}(\pm i/2)=0$, whence the left-hand side of (\ref{eq:momentumcond}) is singular. This singularity can be eliminated by using the $TQ$-relations, as we will comment on below.

%This case need to be treated separately by careful regularization or using the $TQ$-relations.\par

Let $\rI_{M,K}$ be the ideal of the $Q$-system, for a spin-chain state of length $M$ and magnon number $K$, in the variables $\mathbf{s}$. When $Q_{\mathbf{s}}(\pm i/2)$ is non-singular, we can write the momentum condition (\ref{eq:momentumcond}) in the following polynomial form
\begin{align}
\label{momentum}
\rP_{M,K,\ell}=Q_{\mathbf{s}}(-i/2)-\xi_M^{\ell}\,Q_{\mathbf{s}}(i/2)=0\,.
\end{align}
Consider the polynomial ring $\mathbb{A}_{M,K}=\mathbb{F}_M[\rs_0,\cdots,\rs_{K-1}]$, and by abuse of notation we denote by $\rI_{M,K}$ the ideal in $\mathbb{A}_{M,K}$ generated by the rational $Q$-system. We define $M$ ideals in $\mathbb{A}_{M,K}$ as
\begin{align}
\label{regular_sub_ideal}
\rI_{M,K,\ell}\equiv\left(\rI_{M,K}+\langle \rP_{M,K,\ell} ,w Q_{\mathbf{s}}(i/2)-1\rangle \right)\cap\mathbb{F}_M[\rs_0,\cdots,\rs_{K-1}] \,,
\end{align}
where $\ell=1,\cdots,M$, and ``$+$'' means the sum of two ideals. Here $w$
is an auxiliary variable to remove the singular Bethe roots.

We further define an additional ideal for the singular case,
\begin{align}
\rI_{M,K,\infty}\equiv \rI_{M,K}+\langle Q_{\mathbf{s}}(i/2)\rangle.
\end{align}
Let $\mathcal Z( \ldots)$ be the common solution of a set of equations, or equivalently the algebraic set of the corresponding ideal, in the algebraic closure $ \bar{\mathbb Q}$ of rational numbers. We
claim that,
\begin{equation}
  \label{root_decomposition}
  \mathcal{Z}(\rI_{M,K}) =\bigg( \bigcup_{\ell=1}^M \mathcal{Z}(\rI_{M,K,\ell})\bigg) \bigcup  \mathcal{Z}(\rI_{M,K,\infty})\,.
\end{equation}
and thus the BAE roots are classified into $M+1$ subsets.

From the construction of these ideals, we see that for any point ${\bf x} \in \mathcal
  Z(\rI_{M,K,\ell})$,
  \begin{equation}
    \label{eq:7}
    Q_{\bf x}(i/2) \not =0, \quad \frac{Q_{\bf x}(-i/2)}{Q_{\bf x}(+i/2)} =\xi_M^{\ell}
  \end{equation}
and it is clear that for ${\bf x} \in \mathcal
  Z(\rI_{M,K,\infty})$,
\begin{equation}
  \label{eq:8}
   Q_{\bf x}(i/2)=0\,.
\end{equation}
Hence, for different $\ell \in \{1, \ldots M,\infty\}$,  the algebraic sets ${\cal Z}(\rI_{M,K,\ell})$
have no intersection, so the union (\ref{root_decomposition}) is disjoint. Since the $Q$-system equation has no coinciding Bethe roots by construction, $\rI_{M,K,\ell}$ are all radical ideals. By Hilbert's
Nullstellensatz,
\begin{equation}
\label{decomposition}
\rI_{M,K} =\bigg( \bigcap_{\ell=1}^M
\rI_{M,K,\ell}\bigg) \bigcap \rI_{M,K,\infty}\,.
\end{equation}
This is the ideal decomposition which is crucial for the efficient computation of
exact partition function via Gr\"obner bases. Note that in
this paper, we do not prove that for $\ell \in \{1,\ldots,M,\infty\}$,
each $\rI_{M,K,\ell}$ is primary over the field $\mathbb{F}_{M}$, \emph{i.e.},
that there exists no further decomposition beyond the computation in this
paper. This discussion is left for future work. Another comment is that one may well expect that in addition to the lattice translation symmetry, there can be other discrete symmetries such as reflection symmetry that may play a similar role. Namely, we can further decompose the solution space with respect to these symmetries. This interesting possibility is also left for future work.
%\jesper{One may well expect that the reflection symmetry (in addition to translation symmetry) could play some role. Did you have any thoughts about this? I know some models (beyond XXX) where this extra symmetry actually diminishes the number of distinct eigenvalues.} \par

With the decomposition \eqref{root_decomposition}, the exact partition function is presented as a sum over the contributions from the ideals in \eqref{decomposition},
%\jesper{Consider writing $\lfloor x \rfloor$ for integer part, throughout the paper. I think this is standard notation.}
\begin{gather}
  \label{partition_function_decomposition}
  Z_{M,N}(z) =\sum_{K=0}^{\lfloor M/2 \rfloor}(M-2K+1)\left(\sum_{\ell \in \{1,\ldots M,\infty\}} \tr({\bf T}_{M,K,\ell}(z)^N)\right) \,,
\end{gather}
where the companion matrix ${\bf T}_{M,K,\ell}(z)$ is the companion
matrix for
\begin{gather}
  \label{eq:1}
  \big(a(z) Q_{M,K}(z-i)+d(z) Q_{M,K}(z+i)\big) Q_{M,K}(z)^{-1}
\end{gather}
in the ideal $\rI_{M,K,\ell}$. Note that ${\bf T}_{M,K,\ell}(z)$
contains polynomials in $i$ and $\xi_M$ but no other algebraic
numbers. As we will see, each ${\bf T}_{M,K,l}(z)$ has a much smaller
size than the original companion matrix ${\bf{T}}_{M,K}(z)$, hence
\eqref{partition_function_decomposition} provides a highly efficient
way of computing the partition function.

Finally, we comment on the Bethe roots in $\rI_{M,K,\infty}$, \emph{i.e.}, singular
roots. The singularity of the eigenvalue of $U$ in terms of $Q_{\mathbf{s}}(\pm i/2)$ is actually spurious. They can be eliminated by using the $TQ$-relation. We can combine the equations from the rational $Q$-system and the $TQ$-relation and then eliminate the variables $\mathbf{s}$. The elimination procedure is actually quite simple, due to the structure of equations from $TQ$-relations. This gives us a set of equations that only involve the variables $\mathbf{t}=\{\rt_0,\rt_1,\cdots,\rt_M\}$. The momentum conditions in terms of $\mathbf{t}$ variables are simply
\begin{align}
(-i)^M t_{\mathbf{t}}(i/2)=\xi_M^{\ell},\qquad\ell=1,2,\cdots,M
\end{align}
and are free of singularities. We can of course directly work with the equations involving only $\mathbf{t}$ variables and there will not be the spurious class $\rI_{M,K,\infty}$.\par

The reason that we also work with equations involving the $\mathbf{s}$ variables is that we can separate the \emph{regular} and \emph{singular} solutions in this case. Singular physical solutions are special among the solutions of the BAE, since naively plugging them into the eigenvalues and eigenstates lead to divergences and one needs to perform judicious regularizations \cite{Nepomechie:2013mua}. According to the conjecture of \cite{Hao:2013jqa}, all the physical solutions of BAE can be divided into regular and singular physical solutions. These authors worked out the number of these two kinds of solutions up to $M=14,K=7$ and checked the validity of the conjecture. As a by-product of the current paper, we can actually provide more data points up to $M=18,K=9$ and find that the conjecture still holds up to these values.

\paragraph{Galois theory.} The decomposition
\eqref{partition_function_decomposition} is a very convenient
expression. Moreover, there is a further short-cut for the
computation. The equations in different  $\rI_{M,K,\ell}$ are related
by Galois group actions; therefore, instead of exhaustively going through the sum over all
$\ell$'s in \eqref{partition_function_decomposition}, we just need to
compute a few $\ell$'s, namely one for each orbit.

Note that for two different
decomposed BAE with $1\leq \ell_1, \ell_2 \leq M$ described in the previous
subsection, if we replace $\xi_M^{\ell_1}$ by $\xi_M^{\ell_2}$ in the
generators from $\rI_{M,K,\ell_1}$ \eqref{regular_sub_ideal}, without changing the imaginary unit $i$ or any rational
coefficient, then the ideal
$\rI_{M,K,\ell_2}$ is obtained. This implies that we need to consider the
field automorphisms of $\mathbb F_M$ which keeps $i$ and the rational numbers
invariant, or the Galois group $G\equiv \textrm{Gal}({\mathbb
  F}_{M}/\mathbb Q(i))$ where
\begin{align}
\mathbb{Q}(i)=\{a+b\,i:a,b\in\mathbb{Q}\}.
\end{align}
Specifically, if there exists an element $g\in
G$ such that,
\begin{equation}
  \label{eq:10}
  g(\xi_M^{\ell_1}) =\xi_M^{\ell_2}\,,
\end{equation}
then by the field automorphism of Gr\"obner basis computation,
\begin{equation}
  \label{galois_action_partition_function}
 g( {\bf T}_{M,K,\ell_1}(z)) ={\bf T}_{M,K,\ell_2}(z)\,,
\end{equation}
so the computation for ${\bf T}_{M,K,\ell_2}$ is no longer needed. Here $g$
acts on each element of the companion matrix. Then,
instead of taking the sum over $\ell=1,\ldots, M$, we just need to find
the orbits, under the $G$-action, of the set $\{\xi_M,\xi_M^2, \ldots ,1\}$ and compute only
one companion matrix for each orbit.

Hence it is important to analyze the structure of the algebraic extension
$[{\mathbb F}_{M}:\mathbb Q(i)]$.  We consider three different cases
for $M$. The analysis is based on elementary Galois theory. Here we just
list the classification results, and the proof will be presented in
the future work.
\begin{enumerate}
\item $M$ is odd. In this case, the field $ {\mathbb F}_{M}$ is the {\it cyclotomic field},
  \begin{equation}
    \label{eq:6}
    {\mathbb F}_{M} =\mathbb Q(\xi_{M},i) =\mathbb Q(\xi_{4M}) =\mathbb Q\left(e^\frac{2 \pi i}{4M}\right).
  \end{equation}
Note that $i= \xi_{4M}^M$. The large Galois group, $\textrm{Gal}({\mathbb F}_{M}/\mathbb Q)$, is the
multiplication group $(\mathbb Z/(4M) \mathbb Z)^\times$ with the size
$\phi(4M)=\phi(4)\phi(M)=2\phi(M)$. Here $\phi(\ldots)$ is the Euler totient function. We have
\begin{equation}
  \label{eq:13}
  |G|=|\textrm{Gal}({\mathbb F}_{M}/\mathbb Q(i))|=\phi(M)
\end{equation}

From elementary number theory, there exists a $g\in G$ such that,
\begin{equation}
  \label{eq:12}
  g( \xi_M^{\ell_1})=\xi_M^{\ell_2}, \quad g(i)=i\,.
\end{equation}
Hence $\rI_{M,K,\ell_1}$ and $\rI_{M,K,\ell_2}$ are equivalent if and
only if $\gcd(\ell_1,M)=\gcd(\ell_2,M)$. We conclude that in this
case, the ideals \eqref{regular_sub_ideal} are classified by the
greatest common divisors. Under the Galois group $G$, the number of orbits is $\sigma_0(M)$, where $\sigma_0(\ldots)$ denotes the divisor function which counts the number of divisors of $M$. Furthermore, with the ideal $\rI_{M,K,\infty}$, we need to compute $\sigma_0(M)+1$ companion matrices when $M$ is odd.

\item $M$ is even and $4 \not | M$. In this case, the field $ {\mathbb
    F}_{M}$ is the cyclotomic field
 \begin{equation}
    \label{eq:15}
    {\mathbb F}_{M} =\mathbb Q(\xi_{M},i) =\mathbb Q(\xi_{2M}) =\mathbb Q\left(e^\frac{2 \pi i}{2M}\right).
  \end{equation}
$\textrm{Gal}({\mathbb F}_{M}/\mathbb Q)$ is the
multiplication group $(\mathbb Z/(2M) \mathbb Z)^\times$ with the size
$\phi(2M)=\phi(4)\phi(M/2)=2\phi(M)$. We have
\begin{equation}
  \label{eq:16}
  |\textrm{Gal}({\mathbb F}_{M}/\mathbb Q(i))|= \phi(M).
\end{equation}
As in the previous case, $\rI_{M,K,\ell_1}$ and $\rI_{M,K,\ell_2}$ are equivalent if and
only if $\gcd(\ell_1,M)=\gcd(\ell_2,M)$. We need to compute $\sigma_0(M)+1$ companion matrices, as in the preceding case.

\item $4|M$. This case is different from the previous ones.  The field $ {\mathbb
    F}_{M}$ is the cyclotomic field
 \begin{equation}
    \label{eq:15}
    {\mathbb F}_{M} =\mathbb Q(\xi_{M},i) =\mathbb Q(\xi_{M}) =\mathbb Q\left(e^\frac{2 \pi i}{M}\right).
  \end{equation}
Note that $i$ is a power of $\xi_M$. $G= \textrm{Gal}({\mathbb
  F}_{M}/\mathbb Q(i))$ is the subgroup of $\textrm{Gal}({\mathbb
  F}_{M}/\mathbb Q)$ which keeps $i$ invariant.

The classification of $\ell$'s is more complicated in this case, since
$i \in Q(\xi_M)$. From detailed Galois theory analysis, $\rI_{M,K,\ell_1}$ and $\rI_{M,K,\ell_2}$ are equivalent if and
only if
\begin{equation}
  \label{M4pre}
  \gcd(M,\ell_1)=\gcd(M,\ell_2),\quad \frac{\ell_2}{\ell_1}=1 \mod
  \gcd\left(\frac{M}{\gcd(M,\ell_1)},4\right) \,.
\end{equation}
Note that the denominator of the reduced fraction $\ell_2/\ell_1$ is
relatively prime to $M/\gcd(M,\ell_1)$, so the congruence condition for
$\ell_2/\ell_1$ is meaningful.

The condition \eqref{M4pre} for $4|M$ is complicated. However, it is possible to
simplify it and get a similar condition as in the previous two cases. We
notice that there is an enhanced symmetry for the $Q$-system,
\begin{equation}
  \label{enhanced_symmetry}
  s_i \mapsto (-1)^{k-i} s_i, \quad i=0,\ldots, K \,.
\end{equation}
Under this transformation, we find that up to $M\leq 16$ the ideal
$I_{M,K}$ is invariant. Furthermore, the polynomial for the momentum
condition transforms as
\begin{equation}
  \label{eq:3}
    \rP_{M,K,\ell}(i) \mapsto (-1)^K \rP_{M,K,\ell}(-i)\,.
\end{equation}
This means that the imaginary unit $i$ in $\rP_{M,K,\ell}$ transforms to
$-i$. Hence, for two integers $\ell_1$ and $\ell_2$, such that
$\gcd(M,\ell_1)=\gcd(M,\ell_2)$ but which do not satisfy  \eqref{M4pre},
when the enhanced symmetry \eqref{enhanced_symmetry} exists,
$\rI_{M,K,\ell_1}$ and $\rI_{M,K,\ell_2}$ are still equivalent.
\end{enumerate}

In summary, with the enhanced symmetry \eqref{enhanced_symmetry},
for any positive integer $M$, $\rI_{M,K,\ell_1}$ and
$\rI_{M,K,\ell_2}$ are equivalent if and only if
$\gcd(M,\ell_1)=\gcd(M,\ell_2)$. Hence there are always $\sigma_0(M)+1$ classes.

\subsection{Results on decomposition over $\mathbb{F}_M$}
In this subsection, we give some results of BAE decomposition over the extended field $\mathbb{F}_M$.

\subsubsection*{$M=6$}
In this case $ {\mathbb F}_{6} =\mathbb Q\big(e^\frac{2 \pi i}{12}\big)$. From
the discussion of Galois theory in the previous section, the
decomposed BAE sub-systems are classified by the value of $\ell$:
\begin{equation}
  \label{eq:15}
  \{1,5\},\{2,4\},\{3\},\{6\}, \{\infty\} \,.
\end{equation}
Two sub-systems, whose $\ell$-values live in the same class, are
equivalent by a Galois group
action. Hence the number of solutions to the two sub-systems must be equal.

We compute the Gr\"obner basis of
$\rI_{6,K,\ell}$ with $\ell=1,2,3,4,5,6,\infty$, and get the
Bethe root counting for the decomposed BAE with $M=6$ in
Table~\ref{BAE_decomposition_M6}. For the singular Bethe roots, via
the $TQ$-relation, we can find the values of their
regularized momenta. These regularized values are indicated
by the numbers between brackets. For example, in
Table~\ref{BAE_decomposition_M6}, the entry ``$0\ (1)$'' for $K=2$ and $\ell=3$ means
that $I_{6,2,3}$ has no solution but there is one singular Bethe root
whose regularized momentum value is $e^{3\times 2\pi i/6}=e^{\pi i}$.

\begin{table}[h!]
  \centering
\begin{tabular}{c|c|c|c|c|c}
\hline
  & $\{1,5\}$ & $\{2,4\}$ & $\{3\}$ & $\{6\}$ & $\{\infty\}$ \\
\hline
  $K=1$ & $1$ & $1$ & $1$ & $0$ & $0$ \\
\hline
$K=2$ & $1$ & $2$ & $0\ (1)$ & $2$ & $1$ \\
\hline
$K=3$ & $1$ & $0$ & $2$ & $0\ (1)$ & $1$\\
\hline
\end{tabular}
\caption{Number of Bethe roots for decomposed BAE with $M=6$ . There are $\sigma_0(6)+1=5$
  classes, which correspond to $\{1,5\},\{2,4\},\{3\},\{6\},
  \{\infty\}$. The numbers in brackets are the singular Bethe roots.  }
\label{BAE_decomposition_M6}
\end{table}

\subsubsection*{$M=7$}
In this case ${\mathbb F}_{7} =\mathbb Q\big(e^\frac{2 \pi i}{28}\big)$. The
decomposed BAE sub-systems are classified by the value of $\ell$:
\begin{equation}
  \label{eq:15}
  \{1,2,3,4,5,6\}, \{7\}, \{\infty\} \,.
\end{equation}
Note that since $7$ is a prime number, the classification of BAE
sub-systems is very simple. The root counting for the decomposed BAE with $M=7$ is given in Table~\ref{BAE_decomposition_M7}.
\begin{table}[h!]
  \centering
\begin{tabular}{c|c|c|c}
\hline
  & $\{1,2,3,4,5,6\}$ & $\{7\}$ & $\{\infty\}$ \\
\hline
  $K=1$ & $1$ & $0$ & $0$ \\
\hline
$K=2$ & $2$ & $2$ & $0$ \\
\hline
$K=3$ & $2$ & $2$ & $0$ \\
\hline
\end{tabular}
\caption{Number of Bethe roots for decomposed BAE with $M=7$. }
\label{BAE_decomposition_M7}
\end{table}

\subsubsection*{$M=8$}
In this case ${\mathbb F}_{8} =\mathbb Q\big(e^\frac{2 \pi i}{8}\big)$. The
decomposed BAE sub-systems are classified by the value of $\ell$, via the
condition \eqref{M4pre}:
\begin{equation}
  \label{eq:15}
  \{1,5\}, \{2\}, \{3, 7\},\{4\},\{6\}, \{8\}, \{\infty\} \,.
\end{equation}
However, due to the enhanced symmetry \eqref{enhanced_symmetry}, classes like $\{1,5\}$ and
$\{3,7\}$ are equivalent, and similarly $\{2\}$ and $\{6\}$ are also equivalent. Therefore we can use the new classification,
\begin{equation}
  \label{eq:17}
  \{1,3,5,7\},\{2,6\}, \{4\},\{8\}, \{\infty\}
\end{equation}
The root counting for the decomposed BAE with $M=8$ is given in Table~\ref{BAE_decomposition_M8}.
\begin{table}[h!]
  \centering
\begin{tabular}{c|c|c|c|c|c}
\hline
  & $\{1,3,5,7\}$ & $\{2,6\}$ & $\{4\}$ & $\{8\} $ & $\{\infty\}$ \\
\hline
  $K=1$ & $1$ & $1$ & $1$ & $0$ & $0$ \\
\hline
$K=2$ & $2$ & $3$ & $2\ (1)$ & $3$ & $1$\\
\hline
$K=3$ & $4$ & $3$ & $3$ & $2\ (1)$ & $1$ \\
\hline
$K=4$ & $1$ & $2$ & $0\ (3)$ & $3$ & $3$ \\
\hline
\end{tabular}
\caption{Number of Bethe roots for decomposed BAE with $M=8$. }
\label{BAE_decomposition_M8}
\end{table}
Note that for the cases $M\leq 8$, each BAE subsystem contains at most
four roots. Hence the Bethe roots form a {\it solvable algebraic
  extension} over $\mathbb F_M$ by Galois theory, which means that each Bethe root can be
expressed as radicals of rational numbers for $M\leq 8$. The
closed-form expressions for the partition function with $M=7,8$ will
be presented in the future work.

%\subsection{$9\leq M \leq 18$}
Similar results for $9\leq M \leq 18$ are presented in appendix~\ref{sec:resultDcom9to18}
(see Tables~\ref{BAE_decomposition_M9}--\ref{BAE_decomposition_M18}). We use
the Gr\"obner basis method with coefficients in the algebraic
extension $\mathbb F_{M}$ to get the Bethe roots classification. Our code is powered
by {\sc Singular} \cite{DGPS}.

%%%%%%%%%%%%%%%%%%%%%%%%%%%%%%%%%%%%%%%%%%%%%%%%%%%%%%%%%%%%%%%%%
\subsection{Relation to naive momentum-sector diagonalization}
\label{sec:naiveMSD}
%%%%%%%%%%%%%%%%%%%%%%%%%%%%%%%%%%%%%%%%%%%%%%%%%%%%%%%%%%%%%%%%%
We can relate the above results for the counting of Bethe roots for the decomposed BAE to a more naive diagonalization of the transfer matrix
in sectors with specified magnon number $K$ and momentum $\ell$. To this end, we start from an example to parallel the discussion in
section~\ref{sec:primaryDecQ}.

\paragraph{An example: $M=6$.}
We first classify the states of the six-vertex model transfer matrix $T_M(z)$ given by (\ref{eq:TMz_6v}). We work in the particle picture corresponding
to the right panel of Figure~\ref{fig:lattice} and we let $1$ (resp.\ $0$) denote the presence (resp.\ the absence) of a particle at a given lattice site.
To access the momentum information, we pick a single representative state for each orbit of the cyclic group ${\cal C}_M$, and we denote its
orbit length by $g_K$.

With $K=0$ there is just one state, $|000000\rangle$, so we set $g_0 = 1$.
With $K=1$ there is one orbit with $g_1 = 6$ having the representative state $|100000\rangle$.
With $K=2$ there are three orbits with representative states
\begin{align}
 |110000\rangle\ (g_2^A= 6)\,, \quad |101000\rangle\ (g_2^B = 6)\,, \quad |100100\rangle\ (g_2^C = 3)\,.
\end{align}
And finally, with $M=3$ there are four orbits with representative states
\begin{align}
 |111000\rangle\ (g_3^A = 6)\,, \quad |110100\rangle\ (g_3^B = 6)\,, \quad |101100\rangle\ (g_3^C = 6)\,, \quad |101010\rangle\ (g_3^D = 2)\,.
\end{align}
Let the momentum label $\ell = 1,2,\ldots,M$ be defined as before. In general, a given orbit $s$ of length $g_s$ is compatible with $\ell$ if and only if
\begin{align}
\label{eq:congr_crit}
 \ell g_s = 0 \mbox{ mod } M \,.
\end{align}
Since obviously $g_s | M$, it is not hard to see from (\ref{eq:congr_crit}) that the number of compatible orbits must be constant on each
class identified above on by Galois theory, except that we do not have the $\ell=\infty$ class in the present case. The classes for $M=6$ then
follow from (\ref{eq:15}), namely $\{1,5\},\{2,4\},\{3\},\{6\}$. In table form the numbers of compatible orbits now read:
\begin{center}
\begin{tabular}{c|c|c|c|c}
\hline
  & $\{1,5\}$ & $\{2,4\}$ & $\{3\}$ & $\{6\}$ \\
\hline
  $K=0$ & $0$ & $0$ & $0$ & $1$ \\
\hline
  $K=1$ & $1$ & $1$ & $1$ & $1$ \\
\hline
$K=2$ & $2$ & $3$ & $2$ & $3$ \\
\hline
$K=3$ & $3$ & $3$ & $4$ & $4$ \\
\hline
\end{tabular}
\end{center}

By particle conservation, the full transfer matrix $T_M(z)$, of dimension $2^M$, is a direct sum of blocks $T_{M,K}(z)$,
each having dimension $d_{M,K} = {M \choose K}$. One may now further block-diagonalize the $T_{M,K}(z)$
into momentum sectors $T_{M,K,\ell}(z)$, having the dimensions $d_{M,K,\ell}$ given by
the above table, by using a procedure similar to Appendix A.4 of \cite{JS2018}. To this end we write
\begin{align}
\label{eq:TKMell}
 T_{M,K,\ell}(z) = S_{\rm out} T_{M,K}(z) S_{\rm in} \,.
\end{align}
Here $S_{\rm in}$ is a $d_{M,K,\ell} \times d_{M,K}$ matrix that maps each compatible orbit into its representative state, with weight $g_s$.
And $S_{\rm out}$ is a $d_{M,K} \times d_{M,K\ell}$ matrix that maps each state into a representative (and hence into an orbit),
and attributes a weight $(\xi_M)^{\ell \cdot k} / g_s$ if a state from
orbit $s$ (not necessarily its representative) has to be shifted cyclically through $k$ lattice steps (say, towards the right) in order to make it coincide
with the representative state of $s$. We recall that $\xi_M = \exp(2 \pi i / M)$, as before.
One may now verify by explicit construction of the matrices $T_{M,K,\ell}(z)$ that the spectrum of $T_{M,K}(z)$ is indeed
the union of the spectra of the momentum blocks $T_{M,K,\ell}(z)$.

As it stands, this method does not yet take into account the $\mathfrak{su}(2)$ symmetry of the XXX spin chain. This means that each $T_{M,K,\ell}(z)$
contains all the highest-weight states with equal or higher spin (i.e., $K' \le K$) in its spectrum. To correct this, on the level on the counting,
it suffices to subtract from each row in the table the one just above it, and we arrive at:
\begin{center}
\begin{tabular}{c|c|c|c|c}
\hline
  & $\{1,5\}$ & $\{2,4\}$ & $\{3\}$ & $\{6\}$ \\
\hline
  $K=0$ & $0$ & $0$ & $0$ & $1$ \\
\hline
  $K=1$ & $1$ & $1$ & $1$ & $0$ \\
\hline
$K=2$ & $1$ & $2$ & $1$ & $2$ \\
\hline
$K=3$ & $1$ & $0$ & $2$ & $1$ \\
\hline
\end{tabular}
\end{center}
This can finally be compared with Table~\ref{BAE_decomposition_M6}. It is seen that the two tables are identical, in so far as they assign the same
dimensions to the same $(K,\ell)$ sectors. Notice that the present approach does not particularize the singular case denoted
$\{\infty\}$ in Table~\ref{BAE_decomposition_M6}, but assigns to it straight away the correct regularized momentum, namely
$\{\infty\} \to \{3\}$ for $K=2$, and $\{\infty\} \to \{6\}$ for $K=3$.%
\footnote{The first of these identifications agrees as well with Table~\ref{tab:ABCD} for the subspace called ${\rm A}$ there.}
The counting is consistent with the sum of numbers outside and inside the brackets in Table~\ref{BAE_decomposition_M6}.

\paragraph{General case.}
The case of general $M$ can be treated in the same way. To recover the results corresponding to
Tables~\ref{BAE_decomposition_M6}--\ref{BAE_decomposition_M18}, we only need to know the number
of compatible orbits under ${\cal C}_M$ for each set of $(K,\ell)$. In even simpler terms, pick a divisor
$g_s | M$, and let $N(g_s;M,K)$ be the number of orbits of length $g_s$ with $K$ magnons.
For instance, we have $N(g_s;6,2) = 0,0,1,2$ and $N(g_s;6,3) = 0,1,0,3$ for $g_s = 1,2,3,6$, respectively.
From this data, the dimensions $d_{M,K,\ell}$ is the sum over those $N(g_s;M,K)$ that respect
the criterion (\ref{eq:congr_crit}), and the highest-weight combinations $d_{M,K,\ell}-d_{M,K-1,\ell}$
provide precisely the numbers of Tables~\ref{BAE_decomposition_M6}--\ref{BAE_decomposition_M18},
up to the assignment of a definite momentum to the $\{\infty\}$ classes.

We have written a simple algorithm that carries out this computation. It produces the tables for $M \le 24$
in less than one minute. For $M \le 18$ these are in full agreement with Tables~\ref{BAE_decomposition_M6}--\ref{BAE_decomposition_M18},
after assigning to each $\{\infty\}$ case the corresponding regularized momentum. We note that for
$M$ even, this assignment appears to obey a simple rule: $\{\infty\} \to \{M/2\}$ when $K$ is even,
and $\{\infty\} \to \{M\}$ when $K$ is odd. When $M$ is prime, the decomposition of the BAE is very simple, and there is no $\{\infty\}$.
It remains to discuss the cases of odd non-prime $M \le 18$, namely $M=9$ and $M=15$. For $M=9$ we find that $\{\infty\} \to \{3,6\}$ when $K=3$
(see Table~\ref{BAE_decomposition_M9}), and for $M=15$ we have $\{\infty\} \to \{5,10\}$ when $K=3$ (see Table~\ref{BAE_decomposition_M15}).
But we do not presently know how to establish such assignments for general $M,K$, without going through the actual computations
of regularization via the $TQ$-relations.

To summarize, the computations described in this subsection appear to be an efficient short-cut for obtaining the decomposition
dimension counting of Tables~\ref{BAE_decomposition_M6}--\ref{BAE_decomposition_M18}, without ever actually using the integrability
of the XXX chain, analysing the BAE, or doing any algebraic geometry. This suggests that the solutions of the
BAE simply decompose in a way that respects the conservation of spin and momentum, and the $\mathfrak{su}(2)$ symmetry of the XXX chain,
with no extra hidden structure. But obviously the decomposition of the BAE goes much further than the mere counting of dimensions; in particular
the explicit results for the Gr\"obner bases make possible the efficient computations of the partition functions, as we have seen.

\section{Conclusions and discussions}
\label{sec:concDisc}
%%%%%%%%%%%%%%%%%%%%%%%%%%%%%%%%%%%%%%%%%%%%%%%%%%%%%%%%%%%%%%%%%%%%%%%%
In this paper we developed a method to compute the torus partition function of the six-vertex model exactly and analytically. The method is based on an algebro-geometrical approach to the BAE, together with new ingredients that include the rational $Q$-system, primary decomposition, algbraic extension and Galois theory.\par

Using this approach, we probed new structures in the solution space of the BAE. We found that the solution space can be decomposed into subspaces $\rI_{M,K,\ell}$ on an algebraically extended field $\mathbb{F}_M=\mathbb{Q}(i,\xi_M)$, where $M,K$ are the length, magnon number, and $\ell=1,\cdots,M$ is related to the total momentum $2\pi\ell/M$ of the solutions of the BAE. We classified the subspaces that are related by the action of the Galois group and form an orbit. For each orbit, we computed the Gr\"obner basis, quotient ring and the companion matrices of the transfer matrix $\mathbf{T}_{M,K,\ell}(z)$ and Baxter's $Q$-operator $\mathbf{Q}_{M,K,\ell}(z)$ up to $M=14$. This decomposition can also be understood in terms of a naive momentum-sector decomposition described in section~\ref{sec:naiveMSD}. The decomposition of the solution space into $\rI_{M,K,\ell}$ and the use of Galois theory led to a huge boost in the computation of the partition function. For $M\le 6$, we have closed-form expressions for the partition function for any $N$. For larger $M$ up to $M=14$, the partition functions for fixed $N$ can be computed straightforwardly from the companion matrices $\mathbf{T}_{M,K,\ell}(z)$.\par

The exact partition functions are polynomials in the spectral parameter $z$ of order $MN$ with rational coefficients. When $M$ and $N$ become large, we obtain polynomials with high orders and large coefficients. Since polynomials are essentially specified by their zeros, we solved for the zeros of the partition functions numerically to high accuracy and studied their behavior in the partial thermodynamic limit where $N\gg M$ and $M$ is fixed. We observed that the zeros accumulate on some curves in this limit and gave a numerical method to generate the limiting curves of accumulation points. These curves exhibit some universal behaviors for different values of even and odd $M$ which led us to formulate several observations and conjectures.\par

There are many open questions and new directions that one can pursue in the near future. We discuss some of them in what follows.\par

An immediate interesting direction is to generalize the current work to the the quantum deformed case. In this paper, we focussed on the six-vertex model at the isotropic point where the model is equivalent to the Heisenberg XXX spin chain. Away from the isotropic point, the six-vertex model is still integrable and is equivalent to the XXZ spin chain. In the XXZ spin chain, we have a new parameter $q$ which is related to the anisotropy. The isotropic point corresponds to $q=1$. Usually the BAE of the XXZ spin chain are written in terms of hyperbolic or trigonometric functions, and one might wonder how our approach, which seems to be restricted to rational functions, can be applied to this case. It is actually quite simple to perform a change of variables to recast the BAE in terms of rational functions. To study the solution space of the BAE and the torus partition function as in this paper, we will however have to deal with several very interesting new features.\par

\begin{itemize}
\item First of all, one needs to distinguish between the cases where $q$ takes a generic complex value and the cases where $q$ is a root of unity. It is well known that the latter case is much more subtle than the former, in terms of solutions of BAE. The completeness problem for the generic $q$ case is a straightforward generalization of the XXX case, namely the physical solutions consist of regular and physical singular solutions \cite{JNZ:toAppear}. On the other hand, when $q$ is a root of unity, due to the presence of the so-called exact $K$-strings, there are seemingly infinitely many solutions and the situation for the completeness problem is less clear. It is therefore not clear what are the physical solutions. Before we can compute the torus partition function, it seems that we need to sort out clearly the completeness problem first, which is an interesting question in its own right. Some preliminary calculations show that algebro-geometric methods in these cases are again very useful. For example, we observe that the Gr\"obner bases exhibit singularities when $q$ is a root of unity and the quotient rings become affine varieties with positive dimensions (instead of a collection of points).

\item The six-vertex model is closely related to another famous model, namely the Potts model. The latter can be represented in terms of the affine Temperley-Lieb (TL) algebra (see, \emph{e.g.}, \cite{BGJSV17,JS2018} for a recent overview). The dimensions $\mathcal{N}_{M,K}$ given in (\ref{binomialdiff}) appear naturally as the dimension of standard modules of the affine TL algebra, which we denote by ${\cal W}_{j,\rz}$ (for the isotropic case, we take $\rz=1$). The representations of the affine TL algebra take the graphical form of ``link patterns'' with pairwise connections (arcs) and defect lines (through-lines) where the total number of lines and the number of defect lines $j$ play the role of length $M$ and magnon number $K$, respectively, in our context. To compute the torus partition function of such models,%
\footnote{Notice that these have previously been studied by more combinatorial methods in the special case of the chromatic polynomial \cite{Pottszerostorus}.}
one can perform a decomposition within the standard module with respect to the lattice momentum \cite{JS2018}, which is essentially the same as what we did in section~\ref{sec:naiveMSD}. Let us recall that in our case the decomposition comes completely from studying the solution space of the BAE (or the rational $Q$-system) using algebraic geometry and the dimensions of the subspaces come from counting the number of solutions; while in the Potts model case, these come from studying the representation theory of the affine TL algebra. These similarities are quite remarkable and imply that the physical solutions of the BAE, studied here using the algebro-geometric approach, actually know a lot about the representation theory of affine TL algebra. It will be interesting to see to which extent these connections carry over to the $q$-deformed case.\par

\item For generic $q$, we need to consider the standard module ${\cal W}_{j,\rz}$ with non-trivial $\rz$, namely $\rz\ne 1$. In this case, there is another quantum number that appears which is associated with $\rz$. This is related to the momentum with which the defects spiral around the periodic direction. It will be intriguing to see how such a new quantum number can appear in our context by studying the solutions space of the BAE.\par

\item The case when $q$ is a root of unity is even more interesting. In that case, representations of affine TL should be reducible, but indecomposable. In practice, this will mean that the ${\cal W}_{j,\rz}$ will have to be ``glued'' in various ways \cite{ReadSaleur07,Logreview}. The complexity of these gluings and the expected appearance of Jordan cells will challenge the algebro-geometric approach. It will be very exciting to see how these structures carry over to the solution space of the BAE.
\end{itemize}

In the current work, we considered periodic boundary condition in both directions for the lattice. This corresponds to the torus partition function. It is also interesting to consider the partition function on other topologies, such as an annulus \cite{Pottszeros4}. For this topology, we need to compute the transfer matrix of the spin chain with open boundary conditions. One nice starting point for this case is the quantum group invariant XXZ spin chain \cite{PasquierSaleur}. This spin chain is invariant under the quantum group $U_q(\mathfrak{sl}(2))$. It has several nice properties. In particular, the completeness problem has been studied systematically in \cite{Gainutdinov:2015vba} both for generic $q$, and $q$ at roots of unity. The relation with TL algebra has also been established. Working out this simpler example should also shed light on the more challenging periodic boundary conditions mentioned above.\par

For the zeros of partition function, it will be desirable to find an analytic approach to understand or even predict the condensation curves in the partial thermodynamic limit. It is also interesting to see how the $q$-deformation affects the distribution of the partition function zeros.

\section*{Acknowledgements}
Y.~Jiang is partially supported by the Swiss National Science Foundation through the NCCR SwissMap.
J.L.~Jacobsen acknowledges support from the European Research Council
through the advanced grant NuQFT. Y.~Zhang is supported by
funding from Swiss National Science Foundation (Ambizione grant PZ00P2
161341) and the European Research Council (ERC) under
the European Union's Horizon 2020 research and innovation programme
(grant agreement No 725110). We thank Zoltan Bajnok, Janko B\"{o}hm, Matthias Staudacher and Yuwei Zhang for helpful discussions.

\appendix

%%%%%%%%%%%%%%%%%%%%%%%%%%%%%%%%%%%%%%%%%%%%%%%%%%%%%%%%%%%%%%%%%%%%%%%%
\section{Rational $Q$-system}
\label{app:Q-system}
%%%%%%%%%%%%%%%%%%%%%%%%%%%%%%%%%%%%%%%%%%%%%%%%%%%%%%%%%%%%%%%%%%%%%%%%
In this appendix, we review the rational $Q$-system method for solving Bethe ansatz equations (BAE) proposed by Marboe and Volin \cite{Marboe:2016yyn}. One of the main advantages of this method is that unlike the BAE, the solutions of the rational $Q$-system are all physical. In addition, it is by far the most efficient way of finding Bethe roots for fixed length $M$ and particle number $K$.\par

Let us briefly review how the rational $Q$-system works. For more details, we refer to the original papers \cite{Marboe:2016yyn,Marboe:2017dmb}. The method works for a large class of rational spin chains with $\mathfrak{su}(m,n|k)$ symmetry. For our purpose, we restrict to the $\mathfrak{su}(2)$ case. The main procedure is as follows:
\begin{enumerate}
\item For a BAE of length $M$ and magnon number $K$, we draw a Young diagram of two rows $(M-K,K)$ with $M-K\le K$.
\footnote{Since the solution of BAE gives the same eigenstate as the corresponding dual solution, this restriction already covers the whole Hilbert space.}
\item To each node (\emph{i.e.}, a corner of a box of the Young diagram) we associate a $Q$-function and require that all the $Q$-functions are polynomials. The order of a given $Q$-polynomial is given by the number of boxes on the upper-right part of the corresponding node (see figure~\ref{fig:YD42} for an illustration).
\item Some of the $Q$-functions at the boundary are completely fixed and do not need to be solved. The $Q$-functions at the upper right boundary are completely fixed to be 1 since there are no Bethe roots. In addition, the $Q$-polynomial at node $(0,0)$ is given by $u^M$.
\item The rest of the $Q$-functions are determined by the $QQ$-relations
\begin{align}
\label{eq:Qas}
\rQ_{a+1,s}(u)\rQ_{a,s+1}(u)=\rQ_{a+1,s+1}^+(u)\rQ_{a,s}^-(u)-\rQ_{a+1,s+1}^-(u)\rQ_{a,s}^+(u) \,,
\end{align}
where $\rQ_{a,s}$ denotes the $Q$-function associated with the node at position $(a,s)$. The $QQ$-relation (\ref{eq:Qas}) is a relation between the four nodes around a box. Here and in what follows, we introduce the shorthand notations
\begin{align}
\rQ_{a,s}^\pm(u)=\rQ_{a,s}(u\pm i/2),\quad \rQ_{a,s}^{++}(u)=\rQ_{a,s}(u+i),\quad \rQ_{a,s}^{--}(u)=\rQ_{a,s}(u-i).
\end{align}
To determine these polynomials, one makes an ansatz for the unknown $Q$-polynomials along certain path (see the example in the next subsection) and require that all the $Q$-functions on the Young tableaux are polynomials. This leads to a set of algebraic equations for the unknown coefficients. Solving this set of algebraic equations gives the $Q$-polynomials.
\item The Bethe roots of length $M$ and magnon number $K$ are given by the zeros of $\rQ_{0,1}(u)$. This fact will be shown below.
\end{enumerate}
In order to explain the above procedure, we give an explicit example in the next subsection.

\subsection{A simple example}
\label{sec:example}
\begin{figure}[h!]
\begin{center}
\includegraphics[scale=0.5]{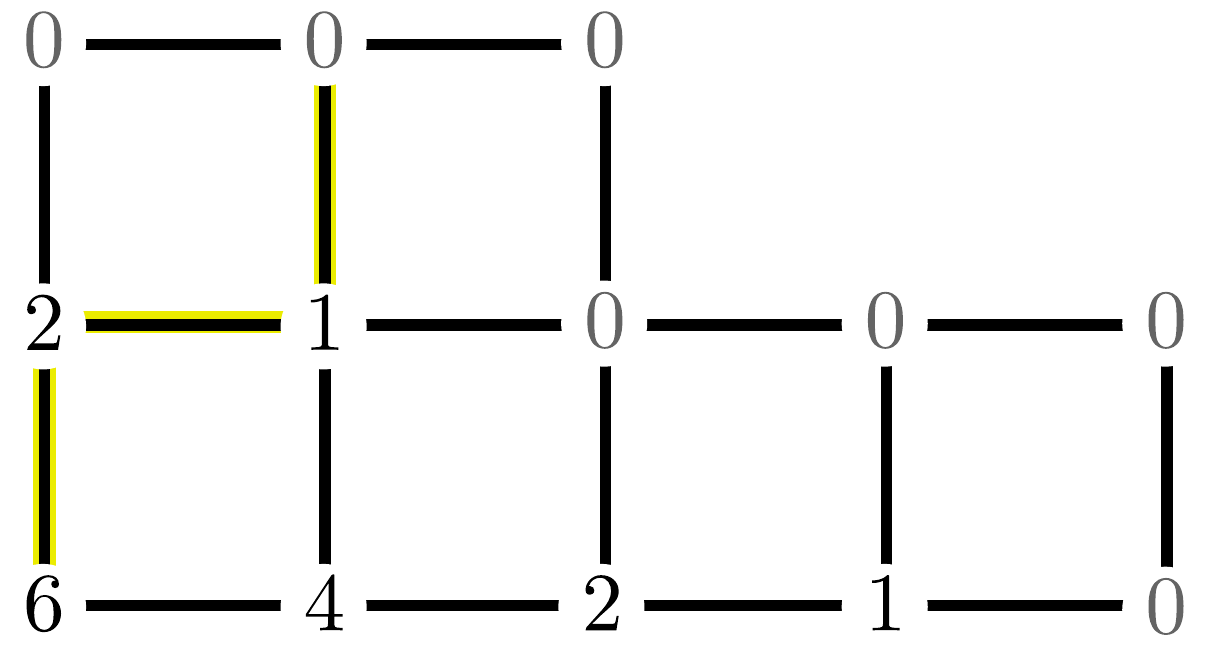}
\caption{The Young diagram for $M=6$, $K=2$. The number on each node denotes the degree of the $Q$-polynomial.}
\label{fig:YD42}
\end{center}
\end{figure}

In order to explain the method, we consider a simple example with $M=6$ and $K=2$, where $M$ is the length of the spin chain and $K$ is the number of magnons. The corresponding Young diagram is given by $(M-K,K)=(4,2)$, and is shown in figure\,\ref{fig:YD42}. The number on each node denotes the degree of the corresponding $Q$-polynomial.
There are thirteen $Q$-functions on the Young tableaux, each $Q$-polynomial being associated with a node of the Young tableaux. The boundary nodes (the ones labeled by 0) are simply taken to be 1. In addition, the $Q$-polynomial at the origin is taken to be
\begin{align}
\label{eq:Qorigin}
\rQ_{0,0}(u)=u^6.
\end{align}
One needs to choose a path from the origin to the upper right boundary. We have chosen the one with yellow color in figure\,\ref{fig:YD42}. Apart from the two boundary $Q$-polynomials, there are two other unknown $Q$-polynomials which are parameterized by
\begin{align}
\label{eq:Qunknown}
\rQ_{0,1}(u)=u^2+c_{0,1}^{(1)}\,u+c_{0,1}^{(0)},\qquad \rQ_{1,1}=u+c_{1,1}^{(0)}.
\end{align}
The remaining task is to use the $QQ$-relations (\ref{eq:Qas}) to determine the remaining unknown $Q$-polynomials as well as the coefficients $c_{0,1}^{(1)}$, $c_{0,1}^{(0)}$ and $c_{1,1}^{(0)}$. In fact, in our case there are only three remaining non-trivial functions to be determined, namely $\rQ_{1,0},\rQ_{2,0}$ and $\rQ_{3,0}$.\par

Let us first determine $\rQ_{1,0}(u)$. Taking $a=s=0$, the $QQ$-relation leads to
\begin{align}
\label{eq:Qexample}
\rQ_{1,0}(u)=c\,\frac{\rQ_{1,1}^+(u)\rQ_{0,0}^-(u)-\rQ_{1,1}^-(u)\rQ_{0,0}^+(u)}{\rQ_{0,1}(u)}
\end{align}
where $c$ is some normalization constant to make the polynomial monic. The quotient and the remainder can be computed straightforwardly. Plugging (\ref{eq:Qorigin}) and (\ref{eq:Qunknown}) into (\ref{eq:Qexample}), we find a non-trivial remainder in terms of the unknown coefficients $c_{0,1}^{(1)}$, $c_{0,1}^{(0)}$ and $c_{1,1}^{(0)}$. Since by definition $\rQ_{1,0}(u)$ is a polynomial, the remainder should be zero. This leads to a set of equations for the unknown coefficients. These relations are called \emph{zero remainder conditions} (ZRC). Repeating this analysis for all the other non-trivial $Q$-functions, we obtain the full set of ZRC, which are the systems of equations that we need to solve.\par

The paths can be chosen in different ways, which result in different forms of ZRC, but finally they lead to the same solution of rational $Q$-systems. The simplest choice of the path is the one that goes from $(0,0)$ to $(0,2)$ and then from $(0,2)$ to the rightmost node $(K,2)$. In this way, we only have one unknown $Q$-function, namely $\rQ_{0,1}$, to determine.\par

After solving the ZRC, we find all the $Q$-polynomials. The Bethe roots are simply given by the zeros of $\rQ_{0,1}(u)$. This is proved in the next subsection.

\subsection{From $Q$-system to BAE: $\mathfrak{su}(2)$ spin chain}
\label{sec:Q_BAESU(2)}
In the previous section, we discussed how to find the solutions of all the $Q$-functions once the boundary conditions are fixed. In this section, we derive the BAE from the $Q$-system, which will demonstrate why the zeros of $\rQ_{0,1}$ are identified with the Bethe roots.\par

\begin{figure}[h!]
\begin{center}
\includegraphics[scale=0.5]{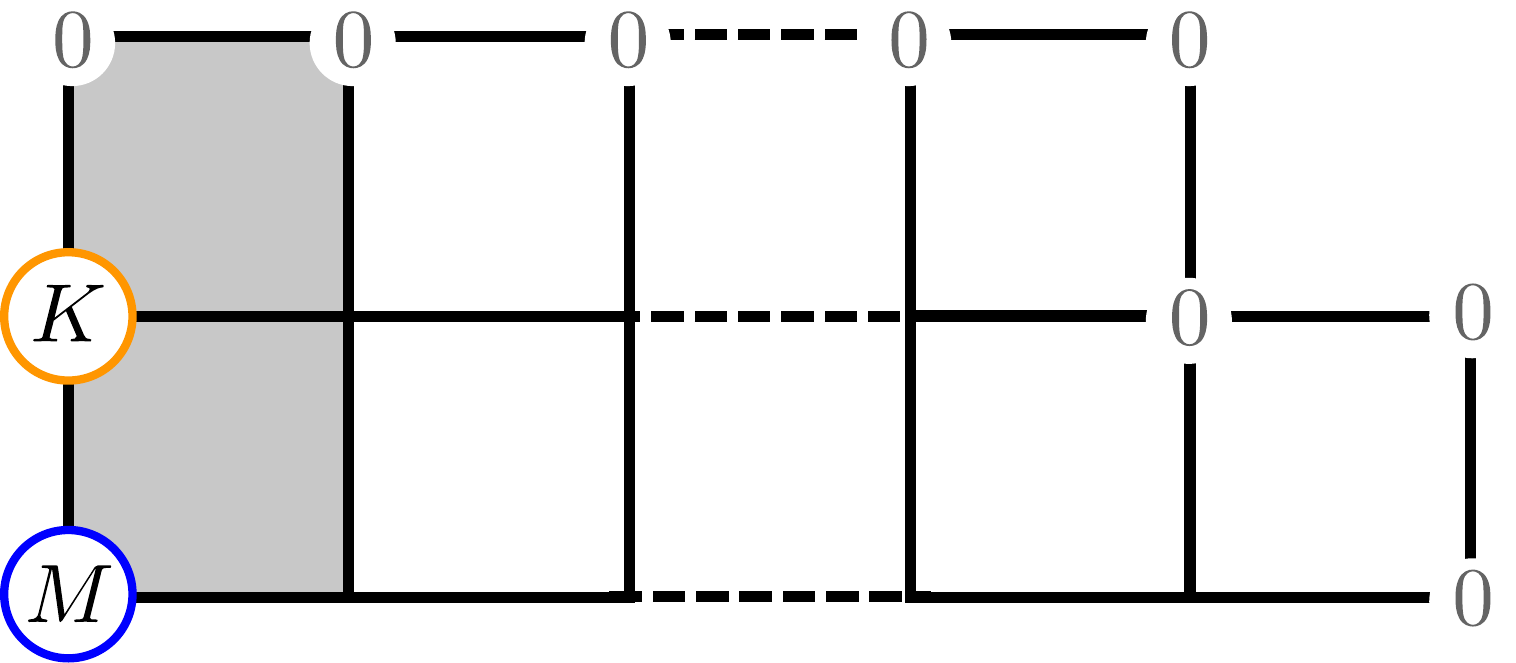}
\caption{The Young tableaux with two rows. The first row has $K$ boxes and the second row has $M-K$ boxes.}
\label{fig:YDMN}
\end{center}
\end{figure}

To this end, let us consider a generic Young tableaux with two rows of $(M-K,K)$ boxes, as shown in figure\,\ref{fig:YDMN}.
We have $K\le M-K$. Let us consider the $Q$-function $\rQ_{0,1}(u)$ (the one at the node with the orange circle) and the related $QQ$-relations. From the power counting we know that $\rQ_{0,1}$ is a monic polynomial of order $K$. We assume that its roots are $u_1,\cdots,u_K$ and write
\begin{align}
\rQ_{0,1}(u)=\prod_{j=1}^K(u-u_j).
\end{align}
We consider the $QQ$-relations in the two shaded boxes, which read
\begin{align}
\rQ_{1,1}\rQ_{0,2}=&\,\rQ_{1,2}^+\rQ_{0,1}^- - \rQ_{1,2}^- \rQ_{0,1}^+ \,, \label{eq:QQ01a} \\
\rQ_{1,0}\rQ_{0,1}=&\,\rQ_{1,1}^+\rQ_{0,0}^- - \rQ_{1,1}^- \rQ_{0,0}^+ \,. \label{eq:QQ01b}
\end{align}
At $u=u_k$, $\rQ_{0,1}(u_k)=0$, and by shifting (\ref{eq:QQ01a}) by $\pm i/2$ we obtain the following relations:
\begin{align}
\rQ_{1,1}^+(u_k) \rQ_{0,2}^+(u_k)=-\rQ_{1,2}(u_k)\rQ_{0,1}^{++}(u_k),\qquad \rQ_{1,1}^-(u_k)\rQ_{0,2}^-(u_k)=\rQ_{1,2}(u_k)\rQ_{0,1}^{--}(u_k).
\end{align}
Using the boundary condition $\rQ_{0,2}=\rQ_{1,2}=1$, we obtain
\begin{align}
\label{eq:Q11_Q01}
\rQ_{1,1}^+(u_k) =-\rQ_{0,1}^{++}(u_k),\qquad \rQ_{1,1}^-(u_k)=\rQ_{0,1}^{--}(u_k).
\end{align}
Meanwhile, evaluating (\ref{eq:QQ01b}) at $u=u_k$ produces
\begin{align}
\label{eq:QQBAE}
\rQ_{1,1}^+(u_k)\rQ_{0,0}^-(u_k) - \rQ_{1,1}^-(u_k) \rQ_{0,0}^+(u_k)=0 \,,
\end{align}
and plugging (\ref{eq:Q11_Q01}) into (\ref{eq:QQBAE}) we obtain
\begin{align}
\rQ_{0,1}^{++}(u_k)\rQ_{0,0}^-(u_k)+\rQ_{0,1}^{--}(u_k)\rQ_{0,0}^+(u_k)=0.
\end{align}
Use the boundary condition and the expression for $Q_{0,1}(u)$,
\begin{align}
\rQ_{0,0}(u)=u^M,\qquad \rQ_{0,1}(u)=\prod_{j=1}^K(u-u_j) \,,
\end{align}
we finally arrive at
\begin{align}
\left(\frac{u_k+i/2}{u_k-i/2}\right)^M=-\prod_{j=1}^K\frac{u_k-u_j+i}{u_j-u_k-i} \,,
\end{align}
which is nothing but the BAE of $SU(2)$ spin chain with length $M$ and magnon number $K$.
It follows in particular that the roots $u_j$ of $Q_{0,1}(u)$ are precisely the Bethe roots, so we can identify the latter with the Baxter polynomial
\eqref{Baxter-polynomial}, viz.\ $Q_{0,1}(u) = Q(u)$, as previously claimed.

%%%%%%%%%%%%%%%%%%%%%%%%%%%%%%%%%%%%%%%%%%%%%%%%%%%%%%%%%%%%%%%%%%%%%%%%
\section{More on algebraic geometry}
\label{sec:Gr}
%%%%%%%%%%%%%%%%%%%%%%%%%%%%%%%%%%%%%%%%%%%%%%%%%%%%%%%%%%%%%%%%%%%%%%%%

In this appendix, we briefly review the important algebraic geometry
technique used in our paper, {\it primary decomposition}. The
mathematics reference is \cite{MR1878556}. For the
concept of {\it Gr\"obner basis} and {\it companion matrix}, we refer
to the introduction in \cite{Jiang:2017phk}.

Let $\mathbb F$ be a field and $A=\mathbb F[z_1, \ldots z_n]$ be a
polynomial. Any ideal $I$ in $A$ decomposes into the intersection of
several primary ideals.
\begin{equation}
  \label{primary_decomposition}
  I=I_1 \cap I_2 \cap \ldots \cap I_{k} \,,
\end{equation}
where each $I_j$ is {\it primary}. A primary ideal $J$ is an ideal such
that if $a,b\in A$, $ab \in J$, then either $a\in J$ or $b^n \in
J$. The decomposition \eqref{primary_decomposition} is an analogy for
the factorization of integers.

Note that with the decomposition \eqref{primary_decomposition},
geometrically, the zero sets of $I$ decompose into the union of several
algebraic sets
\begin{equation}
  \label{eq:5}
  \mathcal Z(I) =\mathcal Z(I_1) \cup \mathcal Z(I_2)  \cup \ldots
  \cup \mathcal Z(I_k) \,.
\end{equation}
This property is useful for the study of the complicated zero set
$\mathcal Z(I)$. Especially, when $Z(I)$ is  zero-dimensional, like
the set of Bethe roots, this decomposition classifies the Bethe roots
and we shall call each $I_j$ \emph{decomposed BAE}.

Note that the definition of primary decomposition depends on the
coefficient field $\mathbb F$. For example, within $\mathbb Q[x,y]$,
\begin{equation}
  \label{eq:2}
  \langle x^2-2 y^2 \rangle
\end{equation}
is already primary and hence cannot be decomposed further. However, within
the algebraic closure $\mathbb{\bar Q}[x,y]$, we have the primary decomposition:
\begin{equation}
  \label{eq:2}
  \langle x^2-2 y^2 \rangle =\langle x-y \sqrt2 \rangle \cap \langle
  x+y \sqrt2\ \rangle\,.
\end{equation}

We refer to \cite{MR1878556} for the introduction of algebraic
extension and Galois theory. The computation of primary decomposition can be carried out by the
computer algebra system {\sc Singular} \cite{DGPS}.

%%%%%%%%%%%%%%%%%%%%%%%%%%%%%%%%%%%%%%%%%%%%%%%%%%%%%%%%%%%%%%%%%%%%%%%%
\section{Power of companion matrices}
\label{sec:largematrix}
%%%%%%%%%%%%%%%%%%%%%%%%%%%%%%%%%%%%%%%%%%%%%%%%%%%%%%%%%%%%%%%%%%%%%%%%
The main method we present in this paper to compute partition function
is to calculate the companion matrices for BAE. Recall that, according to \eqref{eq:ZMN-ag},
the six-vertex model partition function is given by
\begin{align}
Z_{M,N}=\sum_{K=0}^{[M/2]}(M-2K+1)\,\tr\!\left(\mathbf{T}_{M,K}(z)^N\right)\,,
\end{align}
where $\mathbf{T}_{M,K}(z)$ is the companion matrix of the polynomial
$t_{M,K}({\bf s} ,z)$. Recall also that $\mathbf{T}_{M,K}(z)$, calculated
from the Gr\"obner basis, contains only rational numbers. Therefore the
whole computation is manifestly analytic.

In practice, although $\mathbf{T}_{M,K}(z)$ can be calculated from the
straightforward Gr\"obner basis and polynomial division procedure, the
matrix power $\mathbf{T}_{M,K}(z)^N$ computation can be difficult.

We
here present an alternative algorithm which speeds up the computation and saves
RAM usage. The algorithm can be sketched as follows:
\begin{enumerate}
\item For given $M$ and $K$, calculate the Gr\"obner basis $G(I_{M,K})$.
\item Divide $t_{M,K}({\bf s} ,z)$ towards $G(I_{M,K})$; the remainder
  $\overline{t_{M,K}({\bf s} ,z)}$ is called the normal form.
\item Assume $N$ is a power of $2$, \emph{i.e.}, $N=2^c$. Recursively compute
  $t_{M,K}^{(i)}({\bf s} ,z) \equiv (\overline{t_{M,K}^{(i-1)}({\bf s}
    ,z)})^2$, for $i=2, \ldots, c+1$, where $t_{M,K}^{(1)}({\bf s} ,z)\equiv
  \overline{t_{M,K}({\bf s} ,z)}$. Divide $t_{M,K}^{(i)}({\bf s} ,z)$
  towards $G(I_{M,K})$ and define $\overline{t_{M,K}^{(i)}({\bf s}
    ,z)}$ to be the remainder.
\item Calculate the  companion matrix of $\overline{t_{M,K}^{(c+1)}({\bf s}
    ,z)}$. This is the demanded companion matrix $\mathbf{T}_{M,K}(z)^N$.
\end{enumerate}
This strategy avoids the storage and multiplication of dense
matrices. Here we used the algebraic geometry property \eqref{eq:propertyM} that the companion matrix
of a product of two polynomials, equals the product of the
corresponding two companion matrices. If $N$ is not a power of $2$, we
find the binary representation of
  $N$ which is $N=2^{c_1}+2^{c_2} +...$ and repeat the computation
  several times.

In this algorithm, we trim each intermediate polynomial $t_{M,K}^{(i)}({\bf s} ,z)$
  towards $G(I_{M,K})$ via the polynomial division, to get a much
  short polynomial $\overline{t_{M,K}^{(i)}({\bf s}
    ,z)}$. This significantly saves the RAM usage.

This algorithm can also be combined with the decomposition described
in Section $\ref{sec:primaryDec}$. Instead of calculating
$G(I_{M,K})$, we calculate $G(I_{M,K,\ell})$ over the algebraic extension
$\mathbb F_M$. Since $\mathcal Z(I_{M,K,\ell})$ contains much fewer
points than those in $\mathcal Z(I_{M,K})$, the quotient ring $\mathbb
F_M[s_0, \ldots s_{K-1}]/I_{M,K,\ell}$ has much lower dimension. Hence the
polynomial division step trims the polynomial size more dramatically
and further speeds up the computation. By
\eqref{galois_action_partition_function}, we just need to compute
$\sigma_0(M)+1$ distinct $\mathbf{T}_{M,K,l}(z)^N$, and get the rest
by the Galois group action.

Our algorithm is powered by the {\sc Singular} \cite{DGPS} code. For the application
with decomposed BAE, we introduce $\xi_M$ through the \emph{minpoly}
command in {\sc Singular} with the explicit minimal polynomial for
$\xi_M$ (cyclotomic polynomial).

%%%%%%%%%%%%%%%%%%%%%%%%%%%%%%%%%%%%%%%%%%%%%%%%%%%%%%%%%%%%%%%%%%%%%%%%
\section{Details on finding limiting curves}
\label{sec:codes}
%%%%%%%%%%%%%%%%%%%%%%%%%%%%%%%%%%%%%%%%%%%%%%%%%%%%%%%%%%%%%%%%%%%%%%%%

We here describe the numerical procedure employed in Section~\ref{sec:condcurv} to find the limiting curves along which the partition function zeros
$Z_{M,N}(z) = 0$ accumulate, in the limit $N \to \infty$, for a fixed value of $M$.

The key ingredient is obviously to be able to efficiently diagonalize the transfer matrix, so that the loci of equimodularity
($|\Lambda_1(z)| = |\Lambda_2(z)|$) can be identified. To this end, we do not need the entire spectrum of $T_M(z)$ but only
the first few eigenvalues (in principle, just the first two, but see the remarks below). The most efficient means of finding those
are iterative Krylov-subspace methods that depend only on implementing the multiplication of the transfer matrix with a vector,
i.e., to compute $v' = T_M(z) v$ when given some vector $v$. Our method of choice is the Arnoldi algorithm, for which we use
the {\sc Arpack} implementation of Arnoldi's algorithm for complex matrices \cite{Arpack}.%
\footnote{We thank C.R.\ Scullard for technical discussions about the use of {\sc Arpack}.}
This implementation is capable of sorting the eigenvalues in order of decreasing norm.

A huge advantage of such iterative methods is that $T_M(z)$ is given as a product of sparse matrices via (\ref{eq:TMz_6v}),
where each $R$-matrix contains at most two non-zero entries per column. Therefore the computation of $v'$ in each iteration requires at most
$2 M d$ operations, where $d$ is the dimension of the matrix. The trace $\tr_a$ is performed by going
from $M$ to $M+2$ sites when the auxiliary space is inserted, and back to $M$ sites after a row of the lattice
has been completed and the trace operation performed.

If the magnon and momentum labels $(K,\ell)$ for the equimodular eigenvalues $\Lambda_1(z)$ and $\Lambda_2(z)$ were known beforehand,
it would obviously be most efficient to diagonalize the smaller matrices $T_{M,K,\ell}(z)$ constructed in (\ref{eq:TKMell}). But since
the accumulation curves in practice turn out to have multiple branches and T-points where the sector labels may change, this
approach would necessitate a considerable amount of manual intervention. We have thus chosen a more brute-force approach in
which the entire matrix $T_M(z)$ is diagonalized, for all $K=0,1,\ldots,\lfloor M/2 \rfloor$ simultaneously and with decomposing the momentum
with respect to the $\ell$ label. Note that the $\mathfrak{su}(2)$ highest-weight constraint is also not enforced in
this approach, so the eigenvalues with $K < \lfloor M/2 \rfloor$ present degeneracies. We deal with this in practice by imposing
the equimodularity criterion $|\Lambda_1(z)| = |\Lambda_r(z)|$, where $r \ge 2$ is a suitably chosen (small) integer, which may need some adjustment
as we run through the various branches of the equimodular curves.

The first step in our procedure is to acquire some approximate knowledge about where to start the search for the equimodular curves. In the case
of Figure~\ref{fig:curves4-8} we first made a rough plot of the norms of the first few eigenvalues along a few straight lines with constant
${\rm Re}\, z$, or along the real axis, to get a finite list of points close to the equimodular curves. In a second step, we then launched
a direct-search algorithm, taking each of these points as the initial point. The direct-search method is carefully described in \cite{Pottszeros1};
it has the property of first locking onto any close-by equimodular curve and then following it in small steps.%
\footnote{We chose $|\Delta z| = 10^{-2}$ in most cases, but decreased to $10^{-3}$ close to T-points and other fine details.}
The search is stopped whenever a branch of the equimodular wanders off to infinity, or if it starts overlapping with a part of the curve
which is already known. One all starting points have been exploited, we have completed the second step.

The third and final step consists in making sure that the set of equimodular curves is complete. This requires in particular
examining carefully the surroundings of any point where the curves present a discontinuous tangent vector, since this is the sign of a
T-point or a higher-order bifurcation point. As in the first step, we make a rough plot of the norms along a small circle surrounding
any potential bifurcation point. If a branch is identified that has not yet been traced out, we go back to the second step as many times as necessary.

Obviously, if the whole set of equimodular curves has some very small disconnected pieces, they may be missed by this procedure.
However, the fact that the curves in Figure~\ref{fig:curves4-8} present only a single connected component gives appealing evidence
that they are actually complete.

\newpage
%%%%%%%%%%%%%%%%%%%%%%%%%%%%%%%%%%%%%%%%%%%%%%%%%%%%%%%%%%%%%%%%%%%%%%%%%%%%%%%%%%%%%
\section{More results of primary decomposition on $\mathbb{F}_M$}
\label{sec:resultDcom9to18}
%%%%%%%%%%%%%%%%%%%%%%%%%%%%%%%%%%%%%%%%%%%%%%%%%%%%%%%%%%%%%%%%%%%%%%%%%%%%%%%%%%%%%
In this appendix, we list the results of primary decomposition on $\mathbb{F}_M$ for $9\le M\le 18$.
\begin{table}[h!]
  \centering
\begin{tabular}{c|c|c|c|c}
\hline
  & $\{1,2,4,5,7,8\}$ & $\{3,6\}$ & $\{9\}$ & $\{\infty\}$ \\
\hline
  $K=1$ & $1$ & $1$ & $0$ & $0$ \\
\hline
$K=2$ & $3$ & $3$ & $3$ & $0$ \\
\hline
$K=3$ & $5$ & $5\ (1)$ & $6$ & $2$  \\
\hline
$K=4$ & $5$ & $4$ & $4$ & $0$  \\
\hline
\end{tabular}
\caption{Number of Bethe roots for decomposed BAE with $M=9$. }
\label{BAE_decomposition_M9}
\end{table}

\begin{table}[h!]
  \centering
\begin{tabular}{c|c|c|c|c|c}
\hline
  & $\{1,3,7,9\}$ & $\{2,4,6,8\}$ & $\{5\}$ & $\{10\}$ & $\{\infty\}$ \\
\hline
  $K=1$ & $1$ & $1$ & $1$ & $0$ & $0$ \\
\hline
$K=2$ & $3$ & $4$ & $2\ (1)$ & $4$ & $1$\\
\hline
$K=3$ & $8$ & $7$ & $8$ & $6\ (1)$ & $1$ \\
\hline
$K=4$ & $8$ & $10$ & $4\ (4)$ & $10$ & $4$  \\
\hline
$K=5$ & $5$ & $3$ & $6$ & $0\ (4)$ & $4$ \\
\hline
\end{tabular}
\caption{Number of Bethe roots for decomposed BAE with $M=10$. }
\label{BAE_decomposition_M10}
\end{table}

\begin{table}[h!]
  \centering
\begin{tabular}{c|c|c|c}
\hline
  & $\{1,2,3,4,5,6,7,8,9,10\}$ & $\{11\}$ & $\{\infty\}$ \\
\hline
  $K=1$ & $1$ & $0$ & $0$ \\
\hline
$K=2$ & $4$ & $4$ & $0$ \\
\hline
$K=3$ & $10$ & $10$ & $0$ \\
\hline
$K=4$ & $15$ & $15$ & $0$ \\
\hline
$K=5$ & $12$ & $12$ & $0$ \\
\hline
\end{tabular}
\caption{Number of Bethe roots for decomposed BAE with $M=11$. }
\label{BAE_decomposition_M11}
\end{table}

\newpage

\begin{table}[h!]
  \centering
\begin{tabular}{c|c|c|c|c|c|c|c}
\hline
  & $\{1,5,7,11\}$ & $\{2,10\}$ & $\{3,9\}$ & $\{4,8\} $ & $\{6\}$&
                                                                    $\{12\}$ & $\{\infty\}$ \\
\hline
  $K=1$ & $1$ & $1$ & $1$ & $1$ & $1$ & $0$ & $0$ \\
\hline
$K=2$ & $4$ & $5$ & $4$ & $5$ & $4\ (1)$ & $5$ & $1$\\
\hline
$K=3$ & $13$ & $12$ & $14$ & $12$ & $13$ & $12\ (1)$ & $1$ \\
\hline
$K=4$ & $22$ & $24$ & $21$ & $25$ & $18\ (5)$ & $24$ & $5$  \\
\hline
$K=5$ & $26$ & $24$ & $26$ & $23$ & $24$ & $18\ (5)$ & $5$ \\
\hline
$K=6$ & $9$ & $12$ & $10$ & $12$ & $4\ (10)$ & $14$ & $10$ \\
\hline
\end{tabular}
\caption{Number of Bethe roots for decomposed BAE with $M=12$. }
\label{BAE_decomposition_M12}
\end{table}

\begin{table}[h!]
  \centering
\begin{tabular}{c|c|c|c}
\hline
  & $\{1,2,3,4,5,6,7,8,9,10,11,12\}$ & $\{13\}$ & $\{\infty\}$ \\
\hline
  $K=1$ & $1$ & $0$ & $0$ \\
\hline
$K=2$ & $5$ & $5$ & $0$ \\
\hline
$K=3$ & $16$ & $16$ & $0$ \\
\hline
$K=4$ & $33$ & $33$ & $0$ \\
\hline
$K=5$ & $44$ & $44$ & $0$ \\
\hline
$K=6$ & $33$ & $33$ & $0$ \\
\hline
\end{tabular}
\caption{Number of Bethe roots for decomposed BAE with $M=13$. }
\label{BAE_decomposition_M13}
\end{table}
\begin{table}[h!]
  \centering
\begin{tabular}{c|c|c|c|c|c}
\hline
  & $\{1,3,5,9,11,13\}$ & $\{2,4,6,8,10,12\}$ & $\{7\}$ & $\{14\}$ & $\{\infty\}$ \\
\hline
  $K=1$ & $1$ & $1$ & $1$ & $0$ & $0$ \\
\hline
$K=2$ & $5$ & $6$ & $4\ (1)$ & $6$ & $1$\\
\hline
$K=3$ & $20$ & $19$ & $20$ & $18\ (1)$ & $1$ \\
\hline
$K=4$ & $44$ & $47$ & $38\ (6)$ & $47$ & $6$  \\
\hline
$K=5$ & $73$ & $70$ & $73$ & $64\ (6)$ & $6$ \\
\hline
$K=6$ & $69$ & $74$ & $54\ (15)$ & $74$ & $15$ \\
\hline
$K=7$ & $33$ & $28$ & $34$ & $14\ (15)$ & $15$ \\
\hline
\end{tabular}
\caption{Number of Bethe roots for decomposed BAE with $M=14$. }
\label{BAE_decomposition_M14}
\end{table}

\newpage
\begin{table}[h!]
  \centering
\begin{tabular}{c|c|c|c|c|c}
\hline
  & $\{1,2,4,7,8,11,13,14\}$ & $\{3,6,9,12\}$ & $\{5,10\}$ & $\{15\}$ & $\{\infty\}$ \\
\hline
  $K=1$ & $1$ & $1$ & $1$ & $0$ & $0$ \\
\hline
$K=2$ & $6$ & $6$ & $6$ & $6$ & $0$\\
\hline
$K=3$ & $23$ & $24$ & $22\ (1)$ & $24$ & $2$ \\
\hline
$K=4$ & $61$ & $60$ & $61$ & $60$ & $0$  \\
\hline
$K=5$ & $109$ & $109$ & $110$ & $110$ & $0$ \\
\hline
$K=6$ & $133$ & $135$ & $132$ & $134$ & $0$ \\
\hline
$K=7$ & $96$ & $94$ & $96$ & $94$ & $0$\\
\hline
\end{tabular}
\caption{Number of Bethe roots for decomposed BAE with $M=15$. }
\label{BAE_decomposition_M15}
\end{table}
\begin{table}[h!]
  \centering
\begin{tabular}{c|c|c|c|c|c|c}
\hline
  & $\{1,3,5,7,9,11,13,15\}$ & $\{2,6,10,14\}$ & $\{4,12\}$ & $\{8\}$&$\{16\}$ & $\{\infty\}$ \\
\hline
  $K=1$ & $1$ & $1$ & $1$ & $1$ & $0$ & $0$ \\
\hline
$K=2$ & $6$ & $7$ & $7$ & $6\ (1)$ & $7$ & $1$\\
\hline
$K=3$ & $28$ & $27$ & $27$ & $27$ & $26\ (1)$ & $1$ \\
\hline
$K=4$ & $77$ & $80$ & $81$ & $74\ (7)$ & $81$ & $7$  \\
\hline
$K=5$ & $161$ & $158$ & $157$ & $157$ & $150\ (7)$ & $7$ \\
\hline
$K=6$ & $224$ & $231$ & $231$ & $210\ (21)$ & $231$ & $21$ \\
\hline
$K=7$ & $218$ & $211$ & $211$ &  $211$  & $190\ (21)$ & $21$ \\
\hline
$K=8$ & $85$ & $93$ & $94$ & $60\ (35)$ & $95$ & $35$ \\
\hline
\end{tabular}
\caption{Number of Bethe roots for decomposed BAE with $M=16$. }
\label{BAE_decomposition_M16}
\end{table}

\begin{table}[h!]
  \centering
\begin{tabular}{c|c|c|c}
\hline
  & $\{1, 2, 3, 4, 5, 6, 7, 8, 9, 10, 11, 12, 13, 14, 15, 16\}$ & $\{17\}$ & $\{\infty\}$ \\
\hline
  $K=1$ & $1$ & $0$ & $0$ \\
\hline
$K=2$ & $7$ & $7$ & $0$ \\
\hline
$K=3$ & $32$ & $32$ & $0$ \\
\hline
$K=4$ & $100$ & $100$ & $0$ \\
\hline
$K=5$ & $224$ & $224$ & $0$ \\
\hline
$K=6$ & $364$ & $364$ & $0$ \\
\hline
$K=7$ & $416$ & $416$ & $0$ \\
\hline
$K=8$ & $286$ & $286$ & $0$ \\
\hline
\end{tabular}
\caption{Number of Bethe roots for decomposed BAE with $M=17$. }
\label{BAE_decomposition_M17}
\end{table}

\begin{table}[h!]
  \centering
\begin{tabular}{c|c|c|c|c|c|c|c}
\hline
  & $\{1, 5, 7, 11, 13, 17\}$ & $\{2, 4, 8, 10, 14, 16\}$ & $\{3,
                                                            15\}$ &
                                                                    $\{6,
                                                                    12\}$&$\{9\}$
  &$\{18\}$ & $\{\infty\}$ \\
\hline
$K=1$ & $1$ & $1$ & $1$ & $1$ & $1$ & $0$ & $0$\\
 \hline
$K=2$ & $7$& $8$ & $7$ & $8$& $6\ (1)$& $8$& $1$\\
 \hline
$K=3$& $37$& $36$ & $38$ & $37$ & $38$& $36\ (1)$ & $1$\\
 \hline
$K=4$& $123$ & $127$ & $122$& $126$& $114\ (8)$& $126$& $8$\\
 \hline
$K=5$& $308$& $304$& $308$& $304$& $308$& $296\ (8)$& $8$\\
 \hline
$K=6$& $550$& $559$& $552$& $562$& $524\ (28)$& $562$& $28$\\
 \hline
$K=7$& $742$& $733$& $740$& $730$& $740$& $702\ (28)$& $28$\\
 \hline
$K=8$&$656$& $670$& $656$& $670$& $600\ (56)$& $670$& $56$\\
 \hline
$K=9$& $276$ & $262$& $279$& $265$& $280$& $210\ (56)$& $56$\\
 \hline
\end{tabular}
\caption{Number of Bethe roots for decomposed BAE with $M=18$. }
\label{BAE_decomposition_M18}
\end{table}

%\bibliographystyle{JHEP}
%\bibliography{yunfeng}

\end{document}